\documentclass[preprint]{aastex}

\shorttitle{Disk-galaxy lenses in the SDSS}
\shortauthors{Chlo\'e F\'eron et al.}

\begin{document}

\title{A Search for Disk-Galaxy Lenses in the Sloan Digital Sky Survey}

\author{Chlo\'e F\'eron\altaffilmark{1}, Jens Hjorth\altaffilmark{1}, John P.\ McKean\altaffilmark{2} and Johan Samsing\altaffilmark{1}}

\altaffiltext{1}{Dark Cosmology Centre, Niels Bohr Institute, University of Copenhagen, Juliane Maries Vej 30, DK-2100 Copenhagen, Denmark}
\altaffiltext{2}{Max-Planck-Institut f\"{u}r Radioastronomie, Auf dem H\"{u}gel 69, D-53121 Bonn, Germany}

\begin{abstract}

We present the first automated spectroscopic search for disk-galaxy lenses, using the Sloan Digital Sky Survey database. We follow up eight gravitational lens candidates, selected among a sample of $\sim40000$ candidate massive disk galaxies, using a combination of ground-based imaging and long-slit spectroscopy. We confirm two gravitational lens systems: one probable disk galaxy, and one probable S0 galaxy. The remaining systems are four promising disk-galaxy lens candidates, as well as two probable gravitational lenses whose lens galaxy might be an S0 galaxy. The redshifts of the lenses are $z_{\rm{lens}}~\sim~0.1$.  The redshift range of the background sources is $z_{\rm{source}}~\sim~0.3 - 0.7$. The  systems presented here are (confirmed or candidate) galaxy-galaxy lensing systems, that is, systems where the multiple images are faint and extended, allowing an accurate determination of the lens galaxy mass and light distributions without contamination from the background galaxy. Moreover, the low redshift of the (confirmed or candidates) lens galaxies  is favorable for measuring rotation points to complement the lensing study. We estimate the rest-frame total mass-to-light ratio within the Einstein radius for the two confirmed lenses: we find  $M_{\rm{tot}}/L_I=5.4~\pm1.5$ within $3.9~\pm0.9$ kpc for SDSS J081230.30+543650.9, and  $M_{\rm{tot}}/L_I=1.5~\pm0.9$ within $1.4~\pm0.8$ kpc for SDSS J145543.55+530441.2 (all in solar units). \textit{Hubble Space Telescope}  or Adaptive Optics imaging is needed to further study the systems.

\end{abstract}

\keywords{dark matter --- galaxy: fundamental parameters (M/L) --- galaxy: spiral --- galaxy: structure --- gravitational lensing}

%%%%%%%%%%%%%%%%%%%%%%%%%%%%%%%%%%%%%%%%%%%%%%%%%%%%%%%%%%%%
\section{Introduction}

The Lambda-Cold Dark Matter ($\Lambda$CDM)  model has gained the place of cosmological paradigm to describe our Universe, yet its success at explaining the large-scale universe is not reproduced at galactic scales.  The central cusps of simulated dark matter (DM) halos are in disagreement with the observed cores of spiral galaxy DM halos \citep{gentile04,kassin06}, and the disk galaxies formed in $\Lambda$CDM cosmological simulations appear to be offset from the Tully-Fisher relation \citep[TF;][]{tully77} observed for distant disk galaxies  \citep{navarro00a, navarro00b, portinari07, dutton07}. Gravitational lensing provides a promising way of constraining the mass distribution of disk galaxies and measuring their mass-to-light ($M/L$) ratio\footnote{The $M/L$ ratios are expressed in solar units throughout the paper. }. These are key components for comparing  the mass of disks grown in cosmological simulations to the luminosity of observed galaxies. 

It is still unclear whether  the TF relation zero point problem originates in the simulations (\citet{dutton07}; but see \citet{governato07}), in the $\Lambda$CDM cosmology \citep{vdbosch03}, or in the generalization of the Milky Way (MW)  initial mass function (IMF) to other disk  galaxies \citep{flynn06}. However, recent studies found the MW  to be offset from  the observed TF relation by  about the same amount as disk-galaxies simulated using Solar Neighborhood IMFs \citep{flynn06, portinari07}. This suggests that the simulations may be consistent with the model they reproduce, but the Galactic IMF may not be representative of all disk galaxies (see \citet{portinari04} for a review of  IMFs). 

The $M/L$ ratio of disk galaxies can be used to constrain  the shape of the IMF \citep{dejong07}, as well as to test  the maximum disk hypothesis \citep{valbada86}. Yet, to date, detailed individual measurements of $M/L$ ratios exist for three disk galaxies only. These being  the MW (\citet{flynn06}, $M_*/L_I=1.20$ for the stellar matter in the Galactic disk), the Sc galaxy NGC~4414 (\citet{vallejo02}, $M_*/L_I \sim 1$), and the Sab spiral lens 2237+0305, i.e., the Einstein Cross (\citet{trott02}, $M_*/L_I=1.1$ for the disk; \citet{trott08}, $M_*/L_B=1.2$ for the disk). There are several methods used to determine indirectly the $M/L$ ratio of disk galaxies, for example, stellar population models (assuming the shape of the IMF), and relations between the color and $M/L$ ratio derived from the maximum rotation curves of spiral galaxies \citep{bell01, salucci08}.

Gravitational lensing can provide an independent measurement of the $M/L$ ratio of disk galaxies. The lensing geometry is sensitive to the total mass enclosed within the Einstein radius, and to the total projected ellipticity, giving information on the combination of the bulge, disk and DM halo. However, a disk-halo degeneracy remains due to the lack of constraints on the halo ellipticity. The study of the spiral lens B1600+434 \citep{jaunsen97} by \citet{maller00} proved the feasibility of breaking this degeneracy  assuming an independent mass measurement which would constrain the disk and halo contributions at larger radii.  Following this method, \citet{trott02} and \citet{trott08} studied the Einstein Cross \citep{huchra85} and measured the bulge and disk $M/L$ ratios of the spiral lens, using  in addition to the lensing constraints two \ion{H}{1} rotation points, as well as rotation curve and velocity dispersion profiles. These three studies found evidence for a sub-maximum disk\footnote{The theoretical study of \citet{shin07} also points towards spiral lens galaxies having sub-maximum disks.}, in contradiction with rotation curve studies \citep{salucci99}.

%ooooooooooooooooooooooooooooooooooooooooooooooooooooooo
\begin{deluxetable}{lccccccc}
\tabletypesize{\scriptsize}
\tablecaption{ Disk-galaxy lenses \label{tab_disklenses}}
\tablewidth{0pt}
\tablehead{
\colhead{Name} & \colhead{N\tablenotemark{a}} & \colhead{$z_{\rm{lens}}$} & \colhead{$z_{\rm{source}}$} & \colhead{$r_{\rm{Einstein}}$\tablenotemark{b}}
 & \colhead{disk $M/L$} & \colhead{Comments}   & \colhead{ref}
} 
\startdata
B0218+357 & 2 & 0.685 & 0.944 & 0.17 & \nodata & Too crowded by the quasar images & 1, 2 \\
B1600+434 & 2 & 0.41 & 1.59 & 0.70 & \nodata & \nodata & 3, 4, 5 \\
CXOCY J220132.8-320144 & 2 & 0.32 & 3.90 & 0.41 & $M_*/L_V=4$  & One [O II] rotation point & 6 \\
OAC-GL J1223-1239 & 2 & 0.4656 & unknown & 0.42 & \nodata & \nodata & 7 \\
PKS 1830-211 & 2 & 0.886 & 2.507 & 0.491 & \nodata & System near to galactic plane & 8  \\   
PMN J2004-1349 & 2 & unknown & unknown & 0.56 & \nodata & \nodata & 9 \\
Q2237+0305 & 4 & 0.0394 & 1.695 & 0.9 & $M_*/L_I=1.1$  & Two H I rotation points and & 10, 11, 12\\
  &  &  &  &  & $M_*/L_B=1.2$ & stellar-kinematic information & \\
SDSS J0841+3824 & \nodata &  0.1159 &  0.6567 &  4.21 & \nodata & \nodata &  13\\
SDSS J1432+6317 & \nodata &  0.1230 &  0.6643 &  5.85 & \nodata & \nodata &  13\\
SDSS J2141-0001 & \nodata &  0.1380 &  0.7127 &  1.81 & \nodata & \nodata &  13\tablenotemark{c}
\enddata
\tablenotetext{a}{Number of lensed images.}
\tablenotetext{b}{In arcsec.}
\tablenotetext{c}{Three other late-type galaxy lenses are published in Bolton et al.\ (2008), but whether the lens galaxy is a  disk or an irregular galaxy is not confirmed as yet.}
\tablerefs{
(1) Patnaik et al.\ 1992; (2) York et al.\ 2005; (3) Jaunsen \& Hjorth 1997; (4) Koopmans et al.\ 1998; (5) Maller et al.\ 2000; (6) Castander et al.\ 2006; (7)  Covone et al.\ 2008; (8) Winn et al.\ 2002; (9) Winn et al.\ 2003; (10) Huchra et al.\ 1985; (11) Trott \& Webster 2002; (12) Trott et al.\ 2008;  (13) Bolton et al.\ 2008}
\end{deluxetable}
%oooooooooooooooooooooooooooooooooooooooooooooooooooooooooooooooooooooo

Unfortunately, only  ten confirmed disk-galaxy lenses, and three late-type galaxy lenses which might have a disk, are known to date  (see Table \ref{tab_disklenses}). In comparison, about a hundred elliptical-galaxy lenses were already known\footnote{CASTLES, \url{http://cfa-www.harvard.edu/castles/}}  before the large surveys of recent years \citep{bolton08,faure08}. This difference is due in part to the fact that disk galaxies are less massive than elliptical galaxies, and so have a lower multiple image cross-section. Theoretical models predict that only 10\% to 20\% of lenses are due to spiral galaxies \citep{keeton98,moller07}. Moreover, their detection by optical imaging can be made more challenging due to the small separation of the lensed images and by dust extinction in the disk.

Fortunately, spectroscopic selection of candidates opens new perspectives for finding disk-galaxy lens systems. The Sloan Lens Advanced Camera for Surveys (SLACS) project has pioneered this new technique to discover strong gravitational lenses \citep{bolton04,bolton06}, using the large spectroscopic database of the Sloan Digital Sky Survey \citep[SDSS;][]{york00}. They selected luminous red galaxies with absorption-line dominated spectra that also showed at least three emission lines from a background galaxy along the line-of-sight. Imaging with the \textit{Hubble Space Telescope} (\textit{HST}) has confirmed over 70 new   gravitational lenses\footnote{We note that   the full SLACS sample of confirmed lenses presented in \citet{bolton08} contains six systems classified as late-type (Sa or later).}, with a survey efficiency over 65 \%. All of the  confirmed systems had measured lens and source redshifts, which is required for determining the mass of the lens \citep{bolton08}. The Optimal Line-of-Sight (OLS) lens survey \citep{willis05, willis06} extended the search to those lens candidates showing only one emission line from a background galaxy. This allowed lensed galaxies at higher redshifts to be investigated. They found seven new early-type gravitational lenses from the SDSS, which they confirmed using a combination of ground-based imaging and spectroscopic observations. 

The field of strong-lens searching has experienced a dramatic surge in recent years, from the discovery of new radio-loud or optical-loud lensed quasars \citep[e.g., ][]{browne03,myers03,oguri06,kayo07,ghosh08,inada08,oguri08a,oguri08b}, to seredenpitous galaxy-galaxy lens discoveries \citep[e.g., ][]{fassnacht06,belokurov07}, strong lensing optical and spectroscopic surveys \citep[e.g., ][]{willis05,willis06,bolton07,bolton08,faure08,limousin08}, as well as the study of the methodology and biases of strong-lens automatic searches \citep[e.g., ][]{moller07,oguri07,dobler08,mandelbaum08,marshall08,newton08}.

Spectroscopic selection can   be used to efficiently find disk-galaxy lenses in the SDSS,  building on techniques developed by the SLACS and OLS-Lens surveys. However, the foreground galaxy as well as the background galaxy can have strong emission lines, making the search more difficult. Also, the small number of disk-galaxies discovered to date indicates that this project can be challenging. Yet, here we show that it is feasible. 

In addition, galaxy-galaxy lenses bring better constraints than lensed quasars on the mass distribution of the lens galaxy. The lensed images are typically extended, containing much more information than point sources, and they are faint, allowing accurate photometry and astrometry of the disk galaxy to be carried out. Moreover, the low redshift of  $z\sim0.1$ for SDSS galaxies is favorable for adding rotation curve measurements to the lensing constraints in the future.

The purpose of this paper is to prove the feasibility of efficiently discovering disk-galaxy lenses by an automated spectroscopic search, using the large spectroscopic database of the SDSS. This is not a statistical study. We present a method to find low redshift galaxy-galaxy lenses where the lens galaxy is a disk, these systems being particularly well suited for measuring the $M/L$ ratios of disk galaxies and the mass distribution in their central parts. 

The paper is organized as follows. The selection of the lens candidates is explained in \S~\ref{sec_selection}. In \S~\ref{sec_obs} we present the ground-based follow-up of the lens candidates with imaging and long-slit spectroscopy; with the combined results we identify the genuine disk-galaxy lenses. In \S~\ref{sec_ml} we estimate the $M/L$ ratio of the confirmed lenses. Improvements and perspectives are discussed in \S~\ref{sec_discussion}. Conclusions follow in \S~\ref{sec_conclusion}.

Throughout the paper we assume a $\Lambda$CDM cosmology with the following cosmological parameter values: $\Omega_m=0.3$, $\Omega_{\Lambda}=0.7$, $H_0=70~\rm{km~s^{-1}~Mpc^{-1}}$. In this cosmology, $1\arcsec$ corresponds to $1.84~\rm{kpc}$ at a redshift $z=0.1$.

\section{Lens candidates selection}\label{sec_selection}

\subsection{Massive disk galaxies in the SDSS}

The SDSS provides a large optical imaging and spectroscopic database of astronomical objects, covering more than a quarter of the sky \citep{york00}. Defining proper selection criteria to constitute a sample of massive disk galaxies out of the SDSS database is not trivial. Disk galaxies have very diverse spectral types, ranging from the absorption dominated spectra of S0 and Sa galaxies to the emission line spectra of Sb and Sc galaxies. However, it is possible to identify disk galaxies using color criteria. The study of \citet{strateva01}  shows that galaxies in the SDSS are distributed in two groups divided by the color separator $u-r=2.22$ around redshift $z=0.1$, with a redshift evolution roughly parallel to the separator line. For each population, the peak density corresponds to elliptical and spiral galaxies, respectively. This provides a good criterion to select a sample of spiral galaxies in the SDSS. However, because gravitational lensing depends highly on the mass of the lensing galaxy, and spiral galaxies are on average less massive than elliptical galaxies, we want to find massive disk galaxies to increase the probability that the background galaxies will be multiply imaged. That is, try to select preferentially S0 and Sa galaxies, which are more massive than Sb and Sc galaxies.  

The study of \citet{fukugita95} gives more constraints on the color of different galaxy types in the SDSS passband system. Tables are available for various galaxy redshifts ($z=0.0, 0.2, 0.5, 0.8$); we use the $z=0.2$ galaxy color values, as we impose on our sample the redshift range $0.1<z<0.3$. The upper redshift limit is fixed by the redshift selection in the Main Galaxy Sample of the SDSS \citep{percival07}. The lower redshift limit is chosen to increase the efficiency of our survey. Indeed, about half of the SDSS Main Galaxy Sample lies below $z=0.1$, but these galaxies have a lower probability of being gravitational lenses due to their low redshift. We compare the color values from \citet{fukugita95} to the color distribution of SDSS galaxies found by \citet[see their Fig.\ 1]{strateva01}. The color of galaxies at $z=0.2$ calculated in the SDSS passband system by \citet{fukugita95} have a higher (redder) $u-g$ when compared to those galaxies actually observed by the SDSS in the same passband system \citep{strateva01}. Therefore, we decided to use the color of Sab and Sbc galaxies at $z=0.2$ from \citet{fukugita95}, which cover part of the region  of color-space situated between the density peaks for the elliptical and spiral galaxy populations in the SDSS \citep{strateva01}. This is where we expect to find the most massive disk galaxies in color-space.

This leads us to define a color selection criterion\footnote{We used a combination of $u-g$ and $g-r$ color rather than $u-r$ color in order to match the search criteria available in the SDSS SkyServer Spectroscopic Query Form, which we used to retrieve and download our sample of galaxies.}
 $1.4<u-g<2$ and $0.7<g-r<1$, in addition to the $r<17.7$ selection intrinsic to the SDSS Main Galaxy Sample. We find a sample of 41625 candidate massive disk galaxies, using the SDSS Fifth Data Release \citep{adelman07}. Being based only on color, this selection  will include some elliptical galaxies. Visual inspection and galaxy modeling will be necessary to identify the disk galaxies among the gravitational lens candidates.

%oooooooooooooooooooooooooooooooooooooooooooooooooooooooooooo
\begin{figure*}
\centering
\includegraphics[scale=0.35]{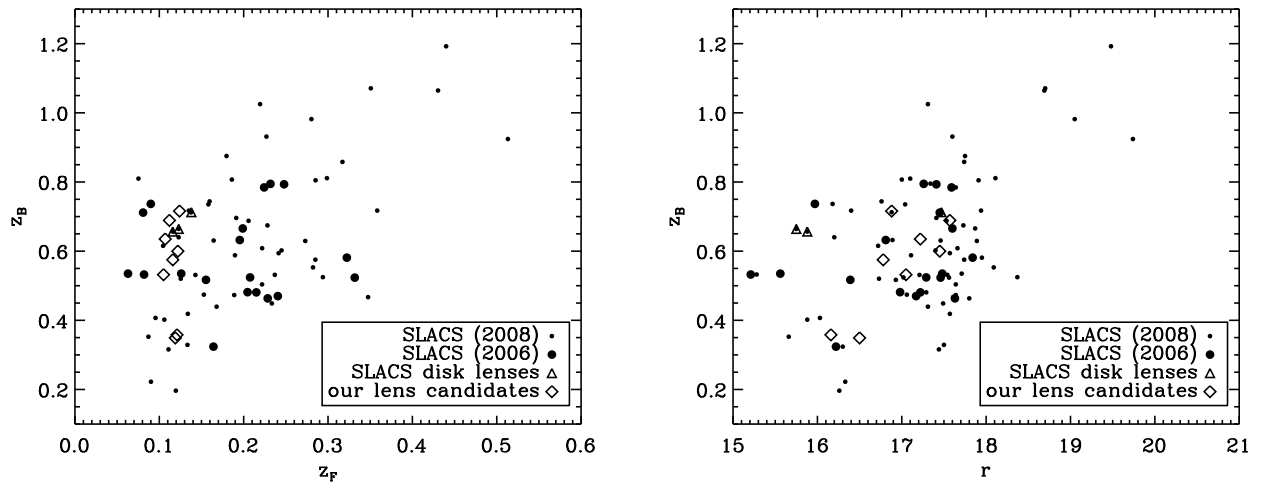}
\caption{Distribution of the foreground galaxy redshift $z_F$, the background galaxy redshift $z_B$ and the $r$ magnitude of our eight final lens candidates compared to the SLACS confirmed strong lenses. The sample of SLACS strong lenses from  \citet{bolton06} was the one we actually used to select our final lens candidates. We show also the final sample of SLACS strong lenses published in \citet{bolton08}, as well as the three disk-galaxy lenses found in it. 
\label{fig_redshifts}}
\end{figure*}
%ooooooooooooooooooooooooooooooooooooooooooooooooooooooooooooooo

%ooooooooooooooooooooooooooooooooooooooooooooooooooooooooooooo
\begin{deluxetable}{lcccccc}
\tabletypesize{\scriptsize}
\tablecaption{Lens candidates \label{tab_lenscandidates}}
\tablewidth{0pt}
\tablehead{
\colhead{Name} & \colhead{R.A. (J2000)} & \colhead{Decl. (J2000)} & \colhead{Plate-MJD-Fiber} & \colhead{$z_F$} & \colhead{$z_B$} &
\colhead{$r$\tablenotemark{a}}
}
\startdata
J0812+5436 & 08 12 13.30 & +54 36 50.9 & spSpec-53384-1871-114 & 0.121 & 0.358 & 16.16 \\
J0903+5448 & 09 03 15.62 & +54 48 56.4 & spSpec-51908-0450-226 & 0.112 & 0.689 & 17.57 \\
J0942+6111 & 09 42 49.08 & +61 11 15.5 & spSpec-51910-0486-155 & 0.124 & 0.716 & 16.88 \\
J1150+1202 & 11 50 19.56 & +12 02 57.3 & spSpec-53142-1609-141 & 0.105 & 0.532 & 17.05 \\
J1200+4014 & 12 00 46.90 & +40 14 00.0 & spSpec-53401-1976-183 & 0.116 & 0.575 & 16.78 \\
J1356+5615 & 13 56 16.53 & +56 15 06.1 & spSpec-52797-1323-531 & 0.122 & 0.600 & 17.45 \\
J1455+5304 & 14 55 43.55 & +53 04 41.2 & spSpec-52674-1164-270 & 0.107 & 0.635 & 17.22 \\
J1625+2818 & 16 25 51.95 & +28 18 21.4 & spSpec-52822-1408-417 & 0.119 & 0.349 & 16.50 \\
\enddata
\tablenotetext{a} {De Vaucouleurs model SDSS (AB) magnitude.}
\end{deluxetable}
%oooooooooooooooooooooooooooooooooooooooooooooooooooooooooooooooooooooo

%oooooooooooooooooooooooooooooooooooooooooooooooooooooooooooooooooooooo
\begin{figure*}
\includegraphics[scale=0.45]{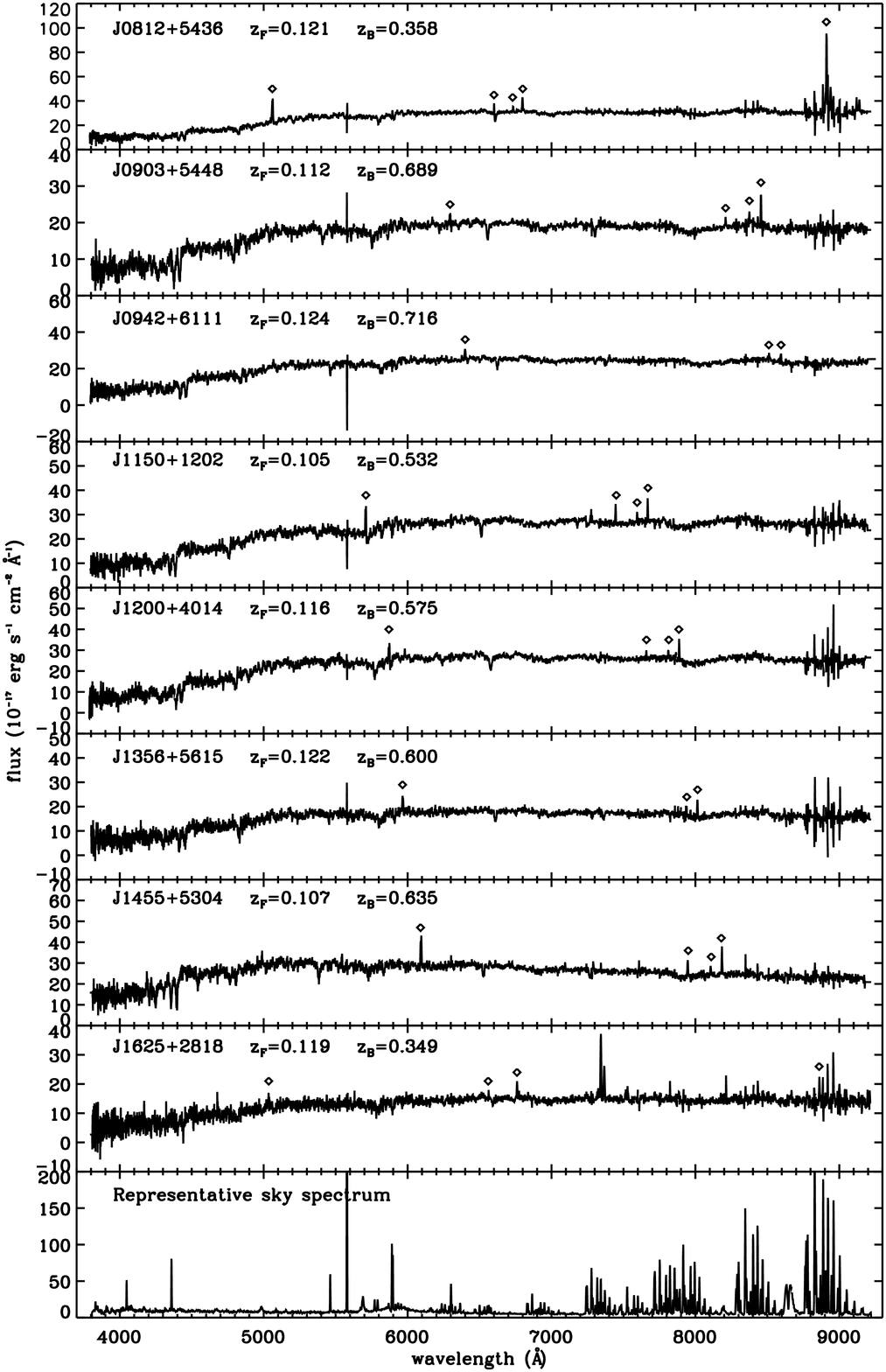}
\caption{Full SDSS spectra of the selected gravitational lens candidates (from Princeton/MIT SDSS Spectroscopy).  The background galaxy's emission lines  are shown by marks in these spectra, and are presented in detail in Fig.\ \ref{fig_emlines}.  Information on the gravitational lens candidates is given in Table \ref{tab_lenscandidates}.
\label{fig_spectra}}
\end{figure*}

\begin{figure*}
\includegraphics[scale=0.3]{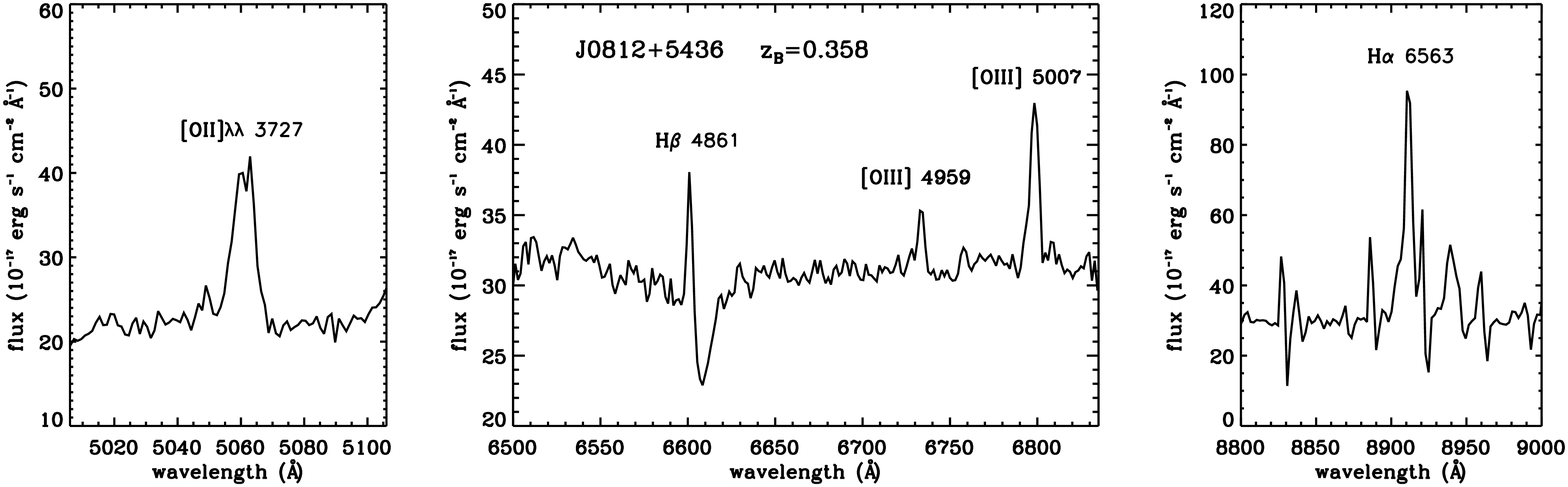}
\includegraphics[scale=0.3]{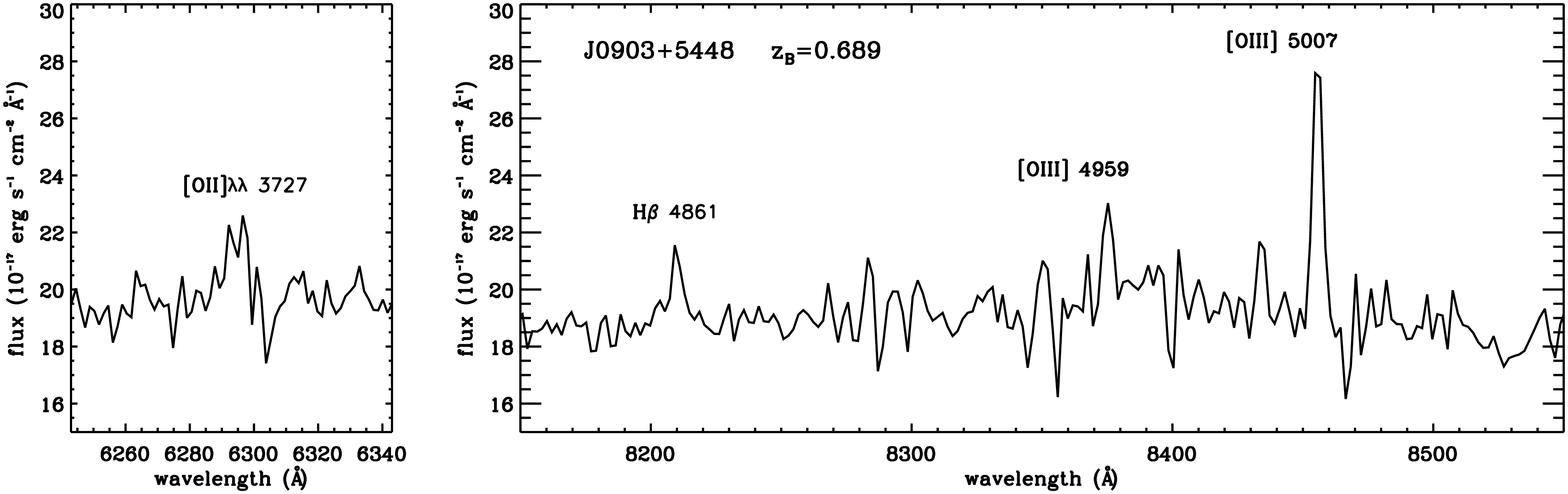}
\includegraphics[scale=0.3]{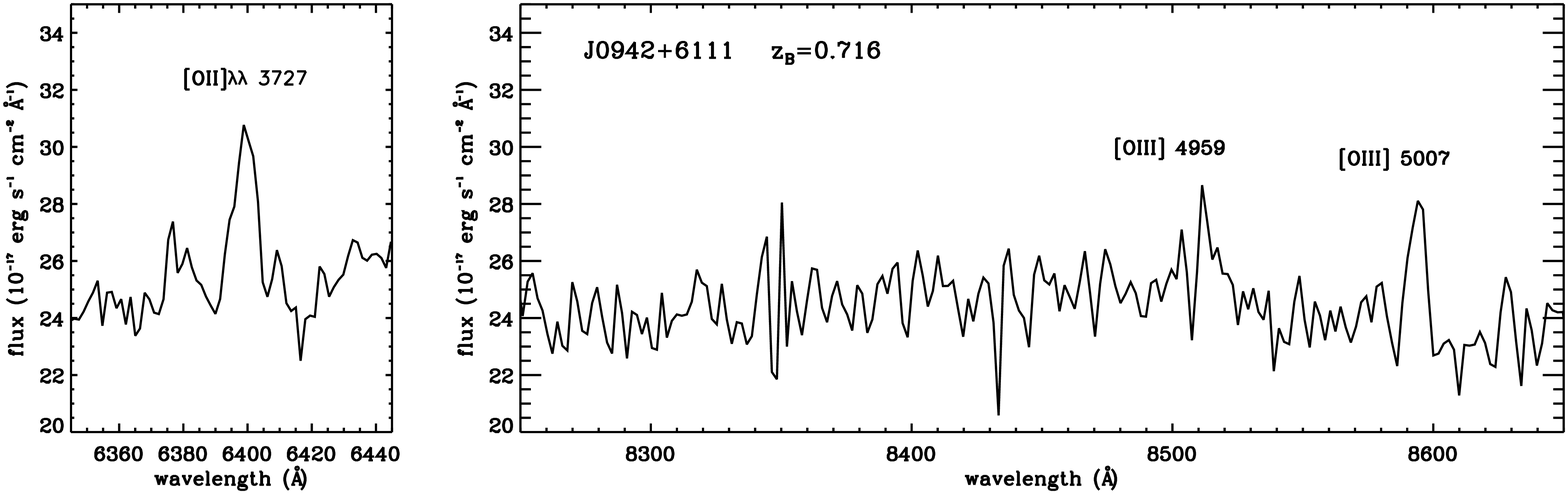}
\includegraphics[scale=0.3]{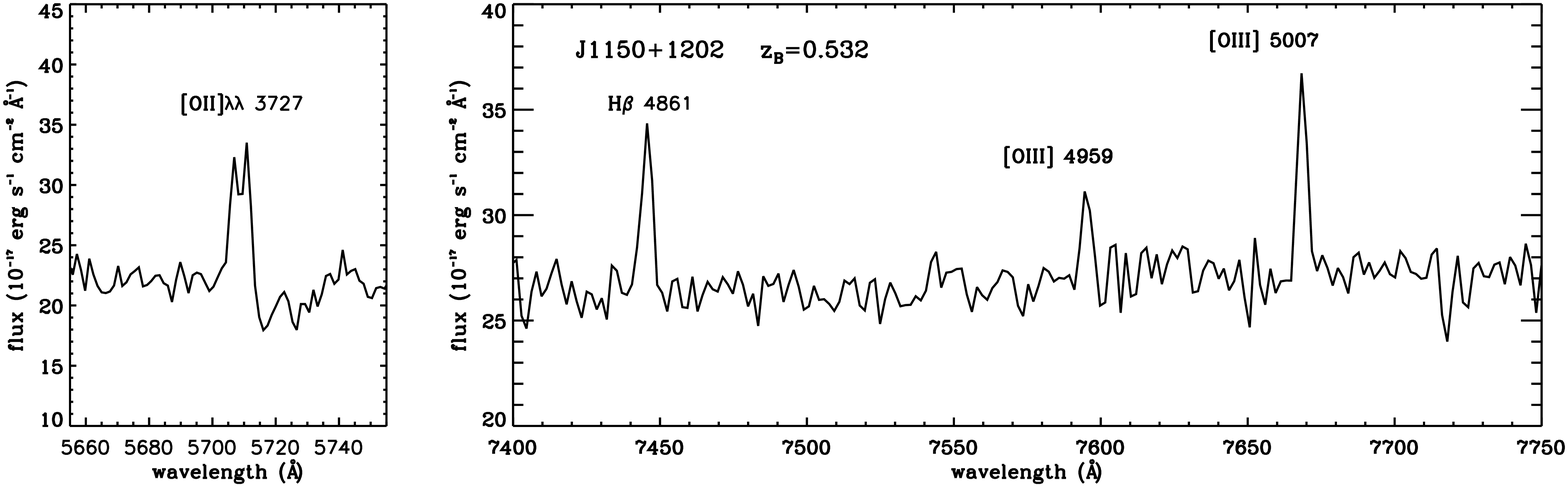}
\caption{Emission lines of the background galaxies in the SDSS spectra for the selected gravitational lens candidates (from Princeton/MIT SDSS Spectroscopy). Although the SDSS spectra are in vacuum wavelengths, we indicate the emission lines in their usual air wavelengths for clarity. The full SDSS spectra are presented in Fig.\ \ref{fig_spectra}.  Information on the gravitational lens candidates is given in Table \ref{tab_lenscandidates}.
\label{fig_emlines}}
\end{figure*}

\addtocounter{figure}{-1}

\begin{figure*}
\includegraphics[scale=0.3]{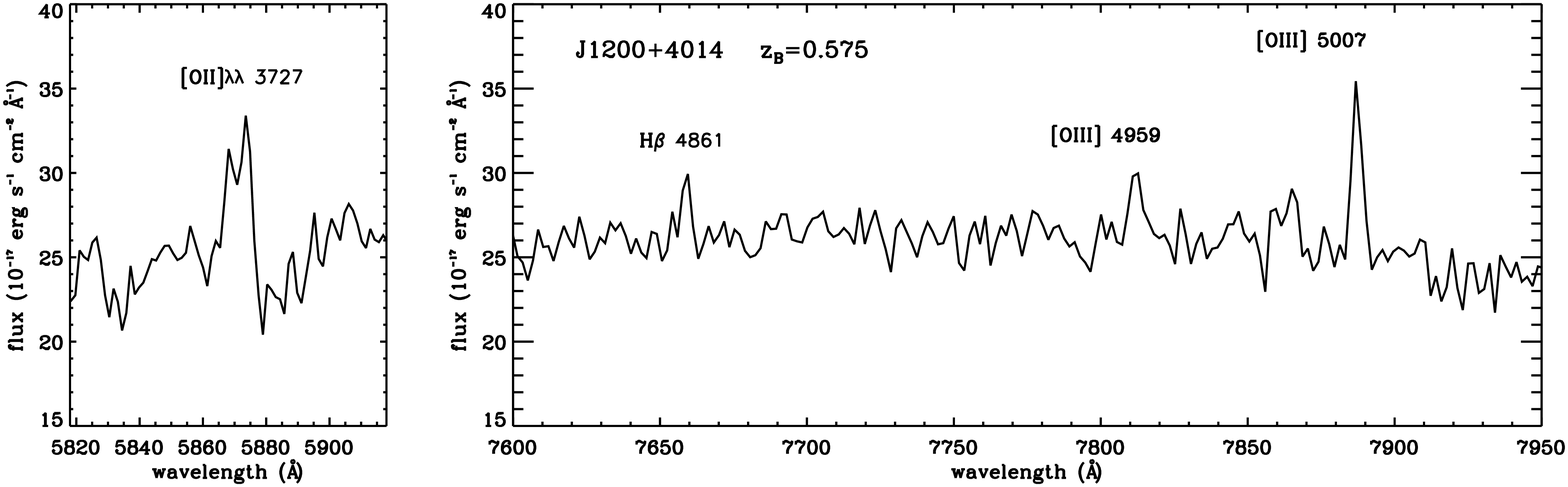}
\includegraphics[scale=0.3]{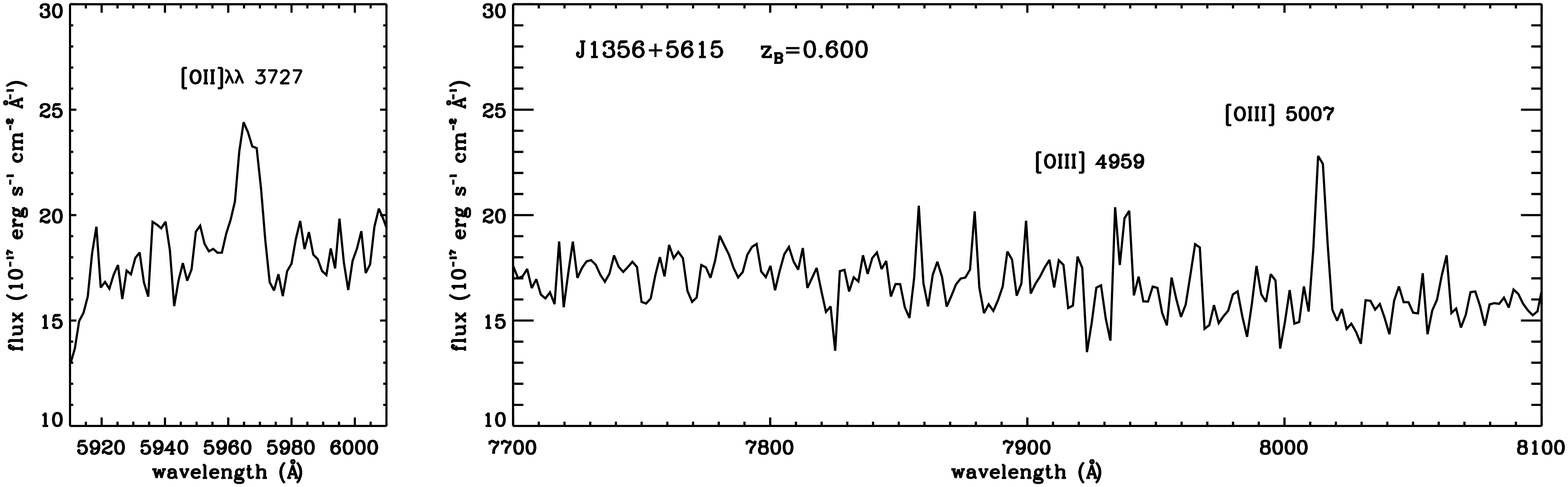}
\includegraphics[scale=0.3]{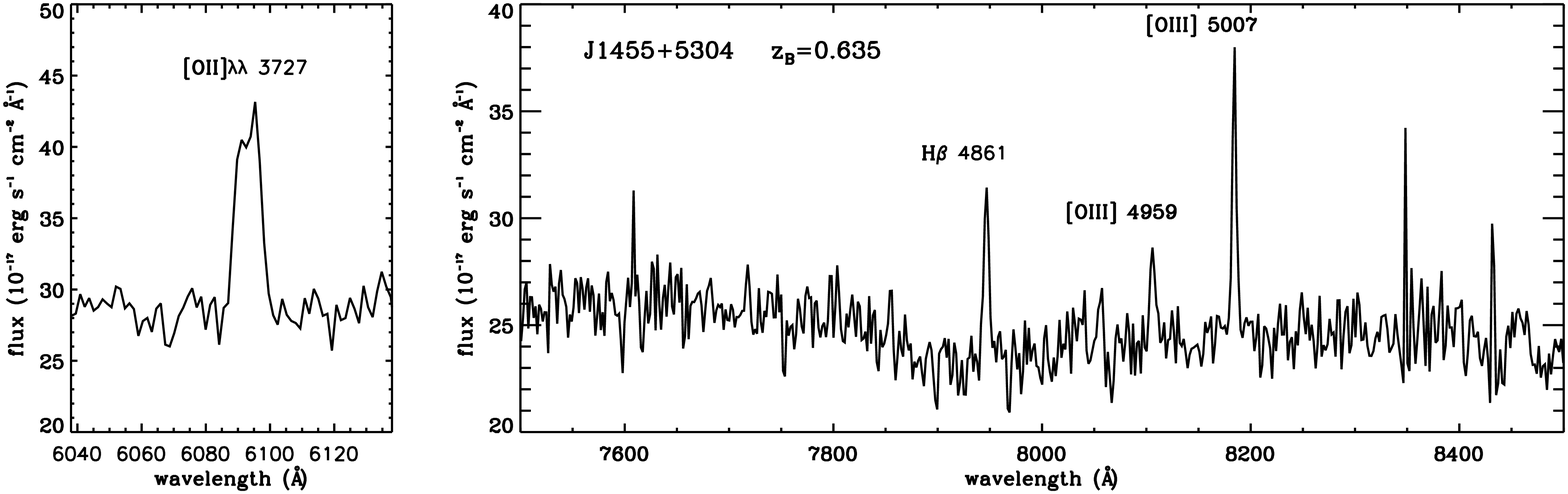}
\includegraphics[scale=0.3]{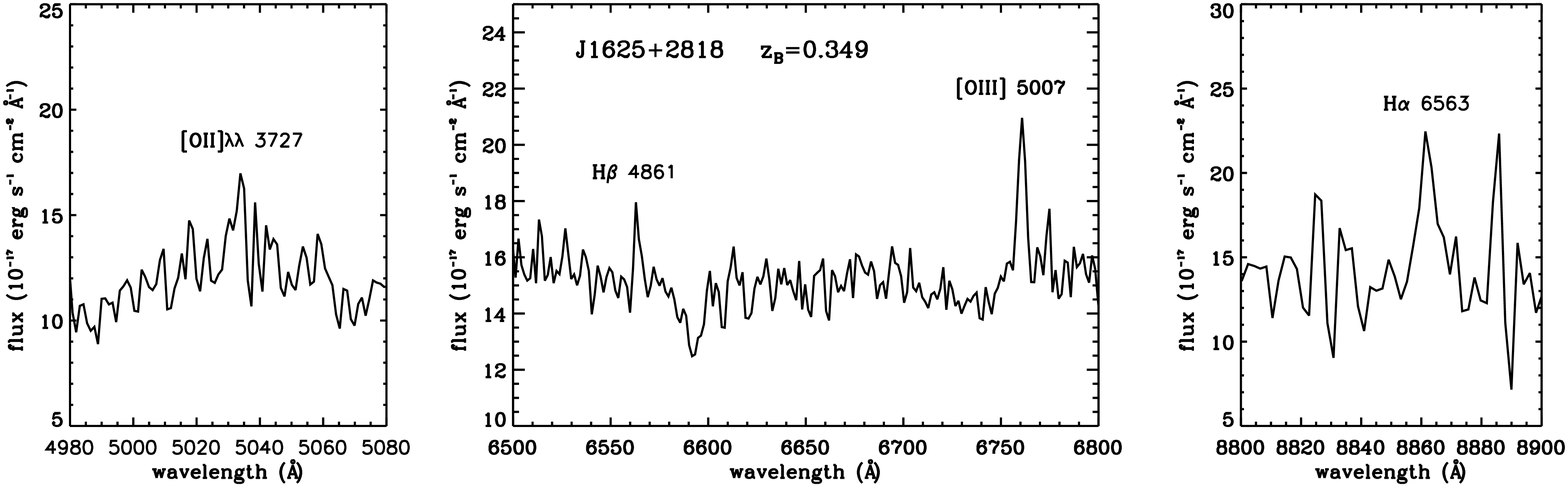}
\caption{(continued)}
\end{figure*}
%oooooooooooooooooooooooooooooooooooooooooooooooooooooooooooooooooooo

\subsection{Spectroscopic selection}

The selection of lens candidates among our sample of massive disk galaxies is two-stepped. First, we select spectra presenting evidence for at least three emission lines from a background galaxy. Second, we filter these candidates based on redshift and magnitude criteria to keep only those candidates with the highest lensing probabilities. 

We developed an automated method to select the spectra presenting evidence of at least three emission lines from a background galaxy.  The method used here is similar to that used in the SLACS project \citep[see][]{bolton04} and the OLS-Lens survey \citep[see][]{willis05}, but it differs in the details. While these two works performed independent processing and background subtraction on the SDSS spectra \citep[a detailed comparison of the methods is presented in][]{willis05}, we used directly the processed spectra distributed by the SDSS, particularly the continuum-subtracted and noise spectra.  The main features of the continuum-subtracted spectra are skyline residuals left from skyline subtraction, galaxy emission lines (if the foreground galaxy has an emission spectrum), and anomalous emission lines which are none of the above. Knowing the redshift of the foreground galaxy from the SDSS database, we can flag the emission lines belonging to the foreground galaxy as well as any skyline residuals.  The remaining, anomalous emission lines are matched with each other to search for combinations of wavelength intervals which would correspond to the emission lines of a galaxy at a different redshift than the foregroung galaxy. For the selected spectra presenting evidence of anomalous emission lines from another galaxy along the line of sight, we inspect visually that the anomalous emission lines are not residuals from the subtraction of bright skylines, which were not flagged as such because they are too far from the central wavelength of the bright skyline. We eliminate in this way the false detections, and keep only the robust candidates. We select as lens candidates those spectra showing at least three emission lines among [\ion{O}{2}] $\lambda\lambda$3727, H$\beta$ 4861, [\ion{O}{3}] 4959, [\ion{O}{3}] 5007 and H$\alpha$ 6563, that belong to a higher redshift galaxy.

In this pilot study, we looked only for bright lenses, that is, candidates we can easily follow-up with ground-based observations. Therefore, we decided to use a signal-to-noise ($S/N$) ratio per pixel selection to find emission lines, which is directly obtained from the ratio of the SDSS  continuum-subtracted and noise spectra. This method is sufficient to detect bright emission lines, and saves the computation time that an integrated $S/N$ ratio selection would require. The limits we used for the $S/N$ ratio per pixel were, $S/N >$~5 for the first emission line peak and then $S/N >$~3 for the other emission line peaks (that were at the same redshift as the first emission line peak). This first step of the selection results in 20 spectra presenting a higher redshift galaxy aligned along the line-of-sight.

We apply a second selection based on lensing probability: systems with a large strong lensing cross-section will have a higher probability for the background galaxy to be multiply imaged, and not only magnified. Studies of gravitational lens systems have found that isothermal mass models are a good approximation for the lensing mass distribution  of early-type galaxies \citep{koopmans06}. As we target massive disk-galaxy lenses,  all our candidates are likely to have a non-negligible bulge component, which accounts for the first order lensing potential, and is well described by an isothermal sphere mass model.  For a singular isothermal sphere (SIS) model, the strong lensing cross-section is $\sigma_{SL}~=~\pi~r_E^2$ with
$r_E~=~4~\pi~(\sigma^2/c^2)~(D_{LS}/D_S)$, where $r_E$ is the Einstein radius, $\sigma$ is the velocity dispersion of the lens galaxy, $D_{LS}$ is the angular diameter distance between the lens and the source, which depends on the lens redshift, the source redshift and the cosmological parameters, and $D_S$ is the source angular diameter distance. The lensing cross-section increases as a function of the lens galaxy mass and of the lens-source angular diameter distance, that is, of the redshift interval between the lens and the source. Unfortunately, we do not have measurements of $\sigma$, which provides a prior on the mass for the candidate galaxies in our sample.  However, we can establish a qualitative comparison of our lens candidates to the confirmed SLACS lenses, which were selected using a SIS model, to determine the highest lensing cross-section candidates. In addition to the redshift interval between the lens and the source, we can use as a comparison criterion  the observed $r$-band magnitude from the SDSS database to obtain a rough estimate of the old stellar population luminosity and hence the mass of each galaxy (that is, the most massive galaxies are  those with brighter $r$ magnitudes).

We select a final sample of eight gravitational lens candidates as those systems with the same range in $r$ magnitude, foreground galaxy redshift $z_F$ and background galaxy redshift $z_B$ as the confirmed SLACS lenses (see Fig.\ \ref{fig_redshifts}). The final sample has the following characteristics. All of the lens candidates have $z_F\sim0.1$. The background galaxy redshifts are distributed in two subsamples. The first in a group at $0.5<z_B<0.8$ with $16.7<r<17.7$, and the second in a group of two lens candidates with specially low background galaxy redshifts at $0.3<z_B<0.4$, but more massive lens galaxies with $16<r<16.7$. The list of final lens candidates with redshift and $r$ magnitude information is presented in Table \ref{tab_lenscandidates}.  The full SDSS spectra of the lens candidates are presented in Fig.\ \ref{fig_spectra}, and the detected background emission lines are shown in Fig.\ \ref{fig_emlines}. The twelve lens candidates which were discarded in the second selection step are shown in Appendix A.

\section{Follow-up of lens candidates}\label{sec_obs}

\subsection{Strategy}
 
We  used a combination of $u'$-band imaging and spatially resolved long-slit spectroscopy, observed at the 2.5-m Nordic Optical Telescope (NOT), at the Observatorio del Roque de los Muchachos (Canary Islands, Spain), to confirm or reject the lensing hypothesis for the eight disk-lens candidates.  A combination of both methods  was required as the multiple images we want to detect are faint (they are not visible in the SDSS images) and are expected to be close to the galaxy center (disk galaxies have smaller Einstein radii than elliptical galaxies, due to their lower central concentration). 

Imaging in the $u'$ band provided, after modeling and subtraction of the foreground  galaxy, an indication of the position of the potential lensed images, which guided us for positioning the slit for spectroscopic observations. However, the residuals in the image could also be non-smooth parts from the galaxy  \citep[an interesting discussion on that topic can be found in][]{marshall08}; therefore we considered them only as potential  lensed images of a background source, and upgraded them to confirmed lensed images only when finding in the long-slit spectroscopy a background galaxy emission line at the same position. The $u'$ band was chosen over other optical bands because it minimizes the contamination from the foreground massive disk galaxy, which is faint in the blue end of the spectrum, while the background source is likely to be bright in the $u'$ band, as indicated by its strong emission lines tracing the presence of star-forming regions.

 The SDSS spectra, which were taken with a $3\arcsec$ diameter fiber, have already shown that there is a background source near to the line-of-sight of the candidate lens galaxy. Spatially resolved long-slit spectroscopy allows us to detect spatially resolved multiple images of the emission lines of the background galaxy. As the slit can probe only a limited region around the foreground galaxy (e.g., the north-south axis, or the east-west axis), we placed it according to the galaxy-subtracted $u'$-band image. If we detected one potential  lensed image we  placed the slit in a position going through the image and the center of the galaxy; if we  found multiple images around the galaxy, we  took spectra in two different positions of the slit which  optimized the number of potential  lensed images covered by our observations.

This strategy  gave us a high chance of detecting the presence of multiple images of  the same background source around the foreground  lensing galaxy.

Confirming the presence of a disk in the  candidate lensing galaxies  was achieved by taking additional optical images in the $g'$, $r'$ and $i'$ bands. Four of the candidate lensing galaxies were visually identified as disk galaxies.  For the other galaxies, we made a fit of the main galaxy component in the $u'$, $g'$, $r'$ and $i'$ bands and  deduced a first-order estimate about the morphology of the galaxy.

 We now present the optical imaging and spectroscopy of the eight gravitational lens candidates.

%oooooooooooooooooooooooooooooooooooooooooooooooooooooooooooooooooo
\begin{deluxetable}{lcccccc}
\tabletypesize{\scriptsize}
\tablecaption{Seeing (FWHM) conditions during observations (in arcsec)\label{tab_observations}}
\tablewidth{0pt}
\tablehead{
\colhead{Name} & \colhead{$u'$} &\colhead{$g'$} & \colhead{$r'$} & \colhead{$i'$} & \colhead{spectroscopy:} & \colhead{spectroscopy:}\\
\colhead{ } & \colhead{exptime 1800 s} & \colhead{exptime 300 s} & \colhead{exptime 300 s} & \colhead{exptime 300 s} & \colhead{east-west slit} & \colhead{north-south slit}
}
\startdata
J0812+5436 & 1.15 & 1.43 & 1.39 & 1.65 & 1.08  & \nodata \\
   
J0903+5448 & 1.13 & 1.3 & 1.32 & 1.50 & \nodata  & 1.14 \\
      
J0942+6111 & 1.30 & 1.39 & 1.24 & 1.52 & 0.70  & \nodata \\
         
J1150+1202 & 1.17 & 2.02 & 1.24 & 1.71 & 0.89 & \nodata \\
          
J1200+4014 & 1.17 & 1.11 & 1.17 & 1.02 & 0.76 & 1.14 \\
          
J1356+5615 & 1.11 & 1.17 & 1.15 & 0.85 & 0.76 & 0.91 \\
          
J1455+5304 & 1.02 & 1.28 & 1.04 & 0.95 & \nodata  & 1.00 \\
          
J1625+2818 & 1.00 & \nodata  & \nodata & \nodata & 0.57  & 0.86  \\
\enddata
\end{deluxetable}
%oooooooooooooooooooooooooooooooooooooooooooooooooooooooooooooooooooooo

\subsection{Imaging}
\subsubsection{Observations}

Optical images of the lens candidates were taken at the  NOT  with the MOSaic CAmera (MOSCA) during 2007 April 14-17 under variable weather conditions. We used the MOSCA instrument to take advantage of its good sensitivity in the $u'$ band. As the size of our targets does not require a mosaic camera, we centered our observations on one of the four CCDs of the mosaic, the CCD10 which showed the best properties regarding flux gradient and bad pixels  on the chip. We used the SDSS filters $u'$, $g'$, $r'$ and $i'$ so as to be able to directly calibrate our observations with the SDSS database. The $u'$-band observations were performed during dark sky conditions and when the seeing conditions were  favorable, in order to resolve the multiple images of the background sources. Exposure times of 1800~s were used due to the expected faintness of the lensed images. The $g'$-, $r'$- and $i'$-band observations were performed during grey sky conditions with exposure times of 300~s in each band.  Seeing conditions during the observations are detailed in Table \ref{tab_observations}; the mean seeing during the imaging run was  $1.26\arcsec$.

\subsubsection{Data reduction}

The optical images of the lens candidates were processed in IDL using standard data reduction techniques. The data were bias-subtracted and flat-fielded, the cosmic rays were removed using L.A.Cosmic \citep{vdokkum01}. The dark current was negligible. 

%oooooooooooooooooooooooooooooooooooooooooooooooooooooooooooooooooooooo

%FIGURE J0812+5436

\begin{figure}
\centering
\includegraphics[scale=0.4]{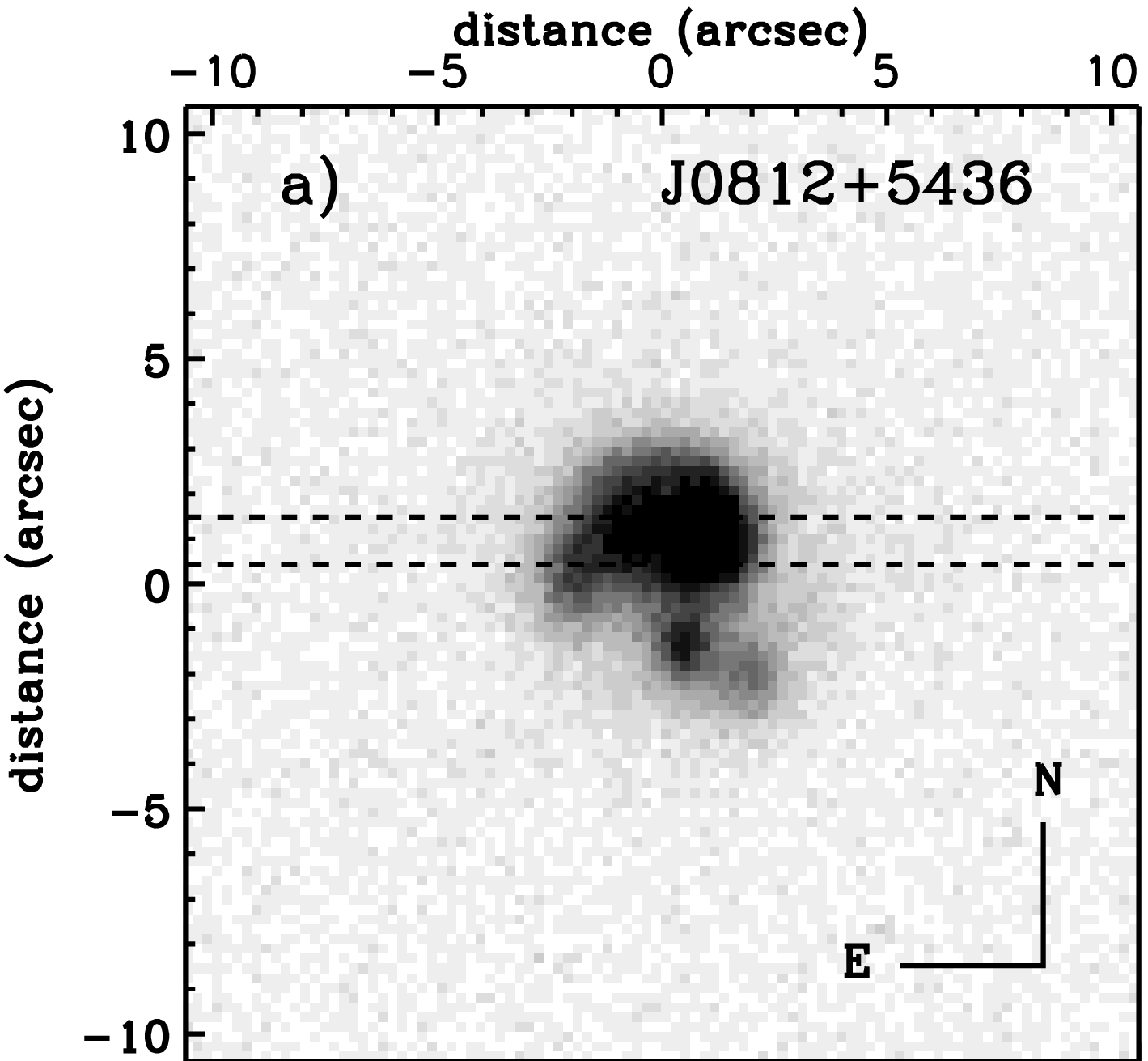}\\
\includegraphics[scale=0.4]{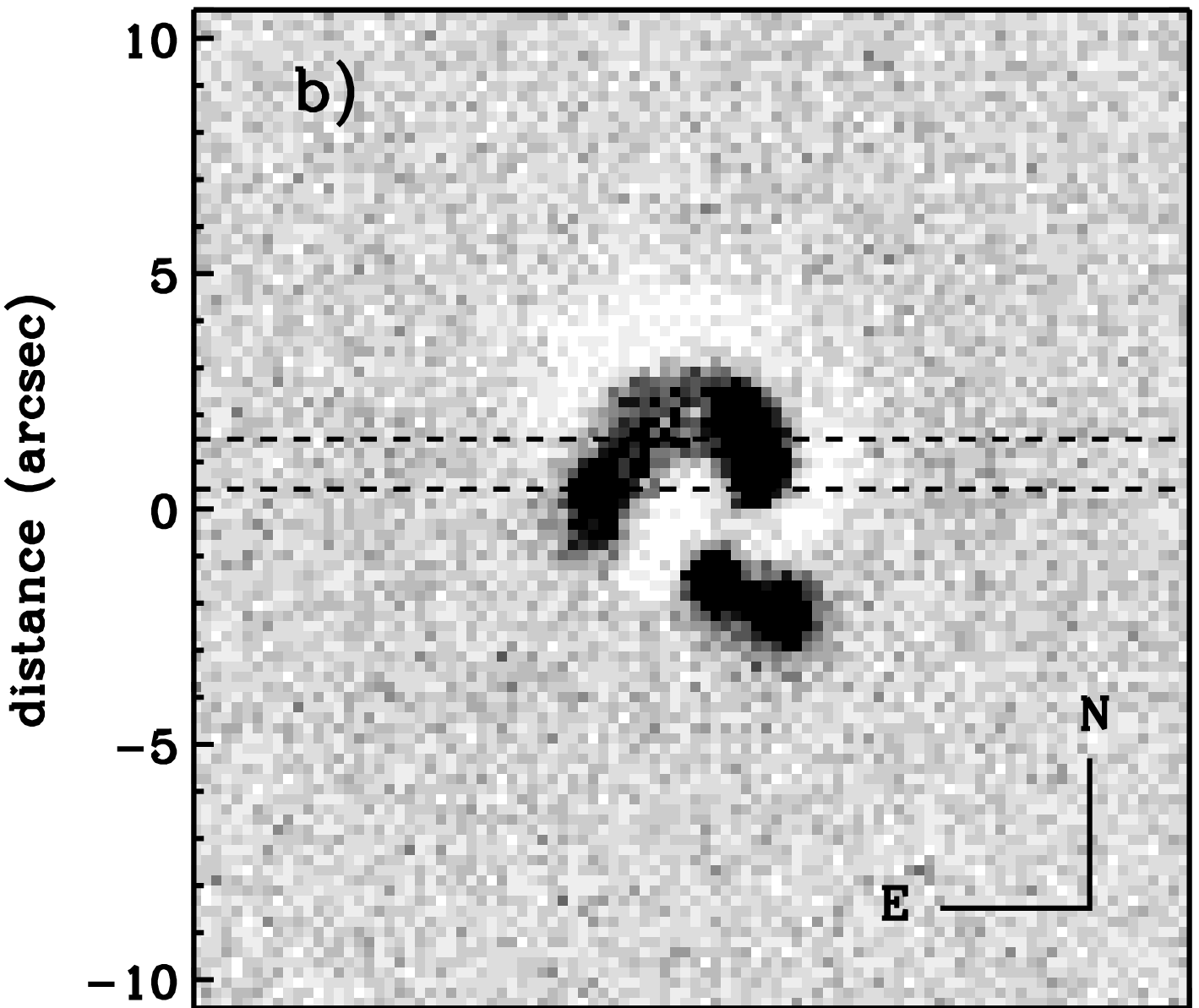}\\
\includegraphics[scale=0.4]{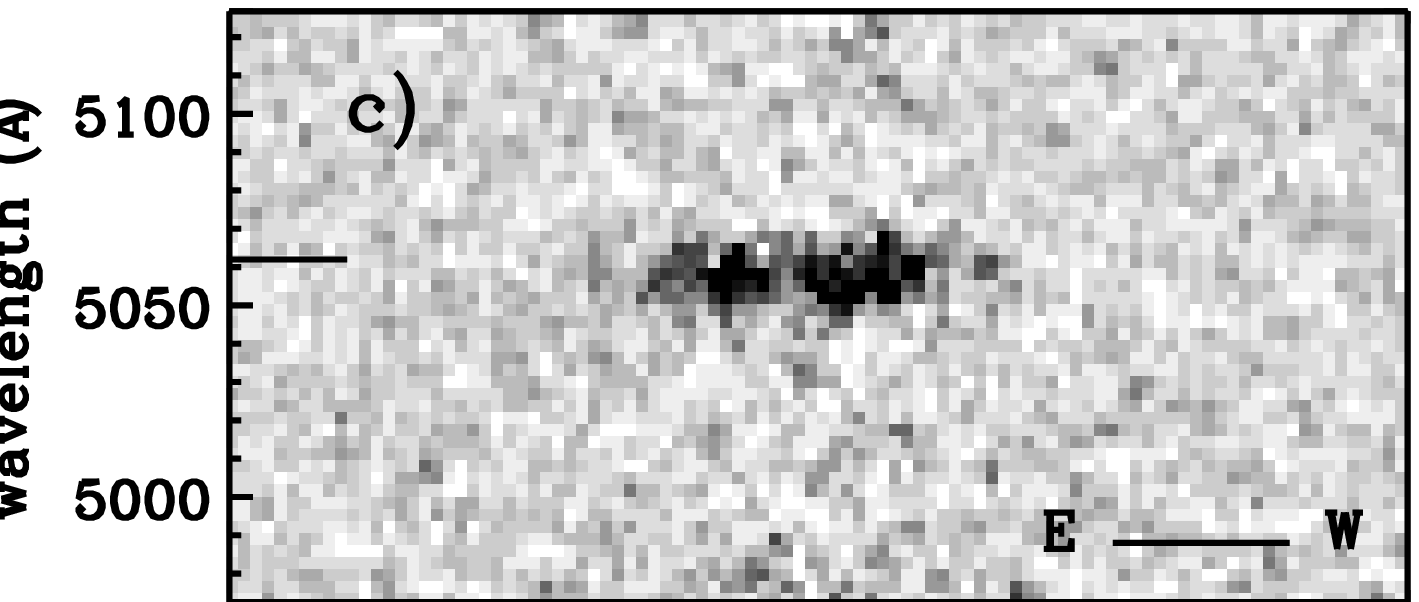}\\
\includegraphics[scale=0.4]{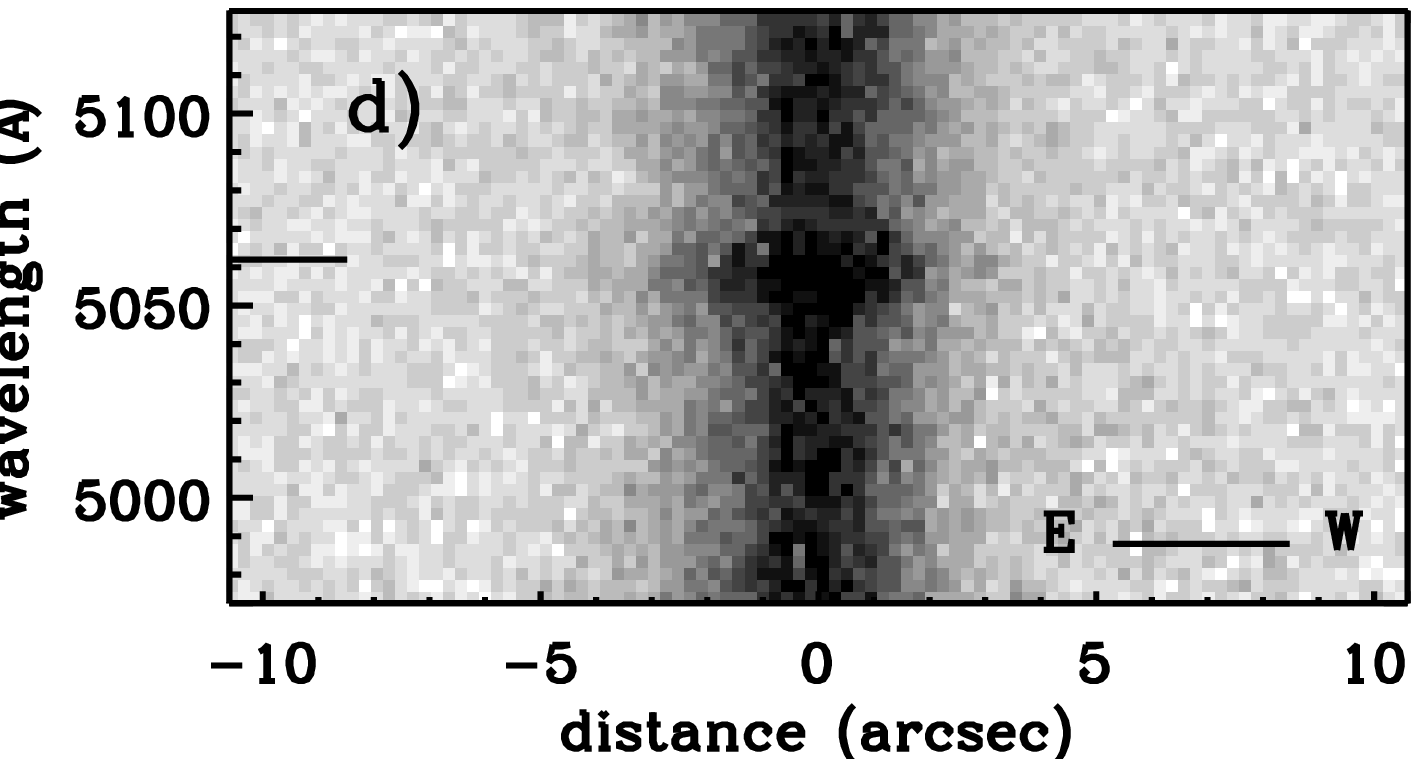}
\caption{\small{Panel a) shows the $u'$-band image of J0812+5436, and panel b) the $u'$-band  residual image after galaxy subtraction. The dashed lines indicate the position and width of the slit used to take long-slit spectra of the target. Panel c) shows a section of the 2D spectrum of J0812+5436 with the [\ion{O}{2}] $\lambda\lambda$3727 emission line of the background galaxy, at $\lambda=5062$~\AA, after skyline and galaxy subtraction, convolved with a gaussian of $\sigma=1$ pixel for display purposes ; panel d) shows the same part of spectrum after skyline subtraction only.}  
\label{fig_0812}}
\end{figure}

%FIGURE J0903+5448

\begin{figure}
\centering
\includegraphics[scale=0.4]{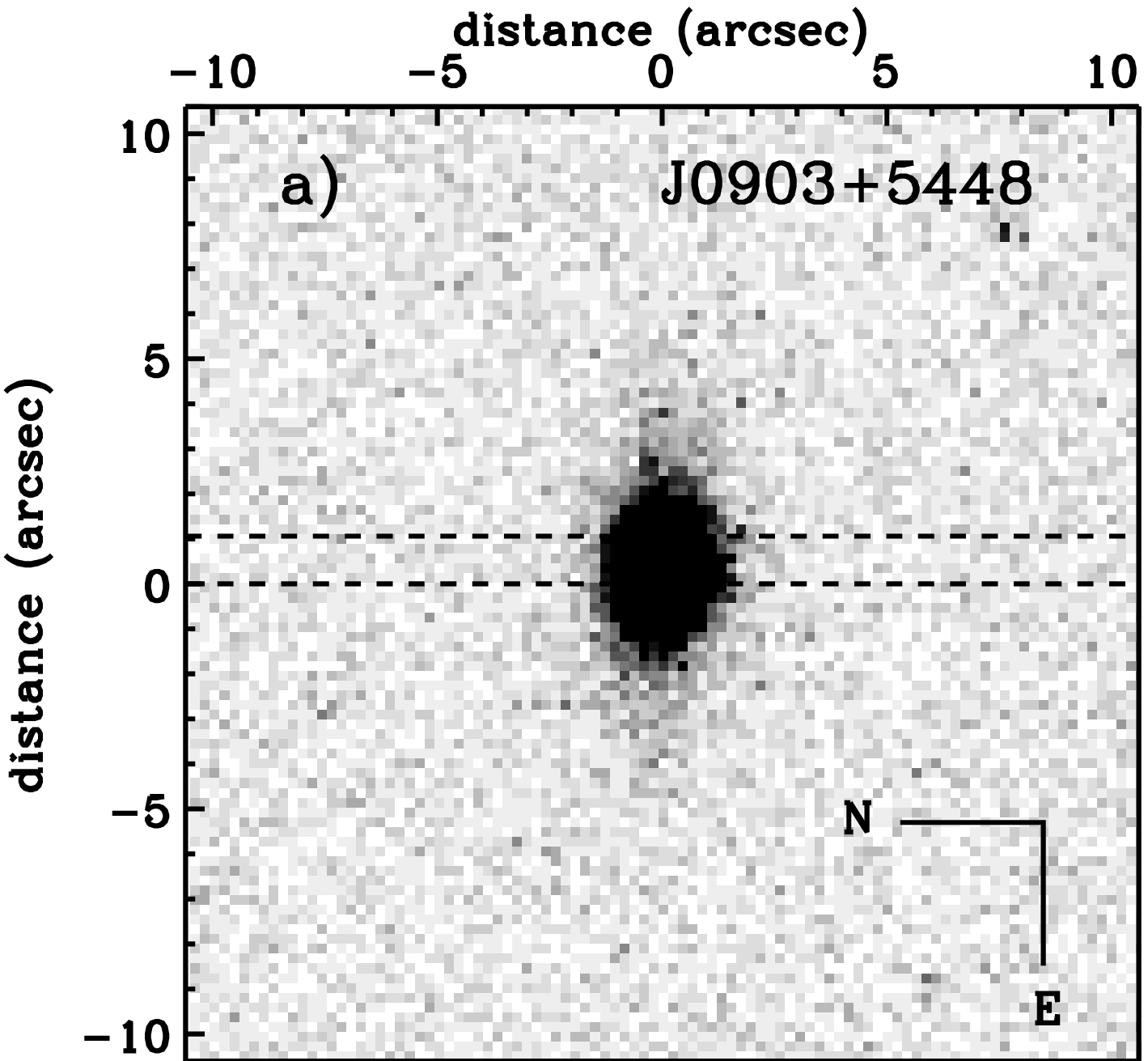}\\
\includegraphics[scale=0.4]{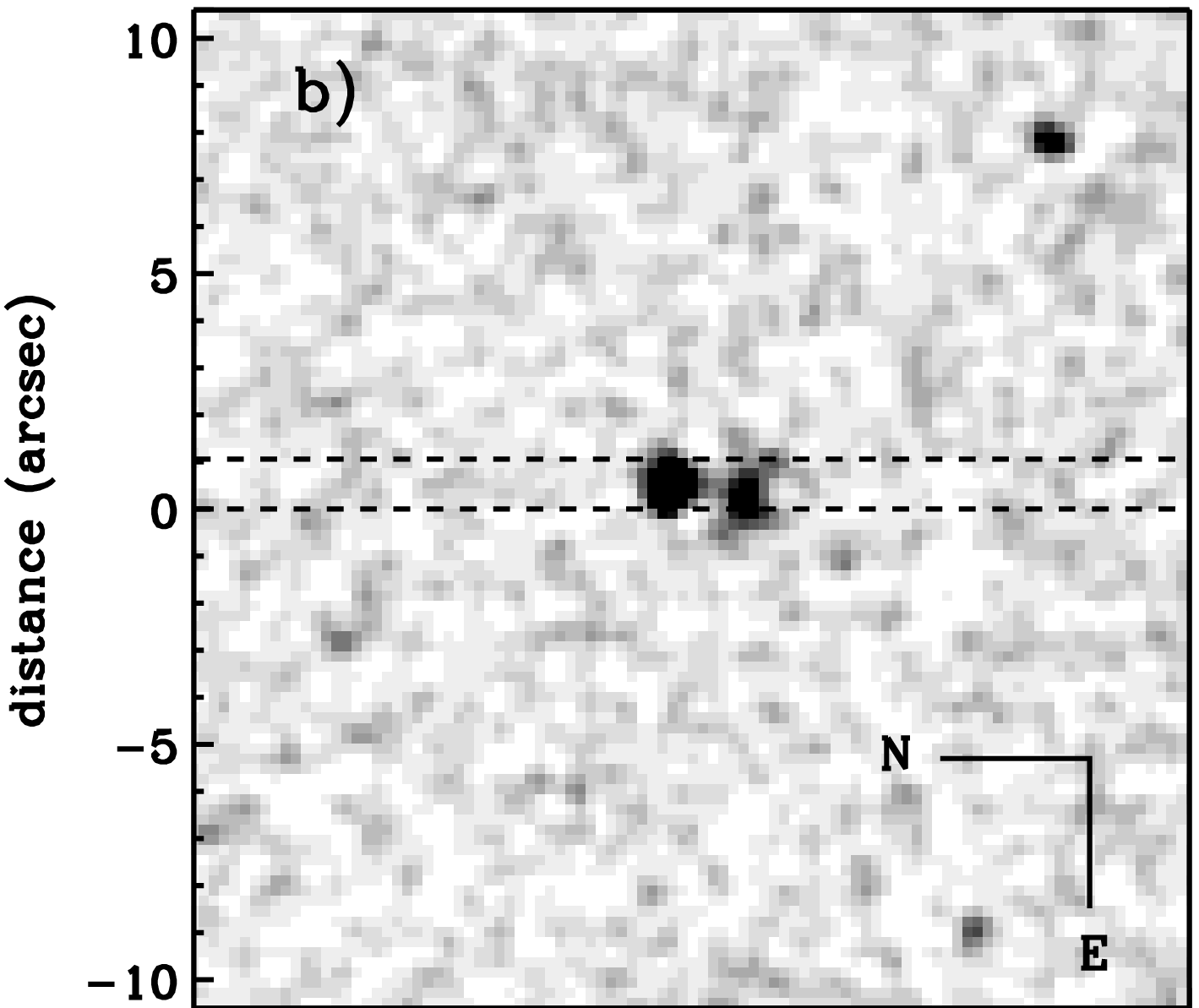}\\
\includegraphics[scale=0.4]{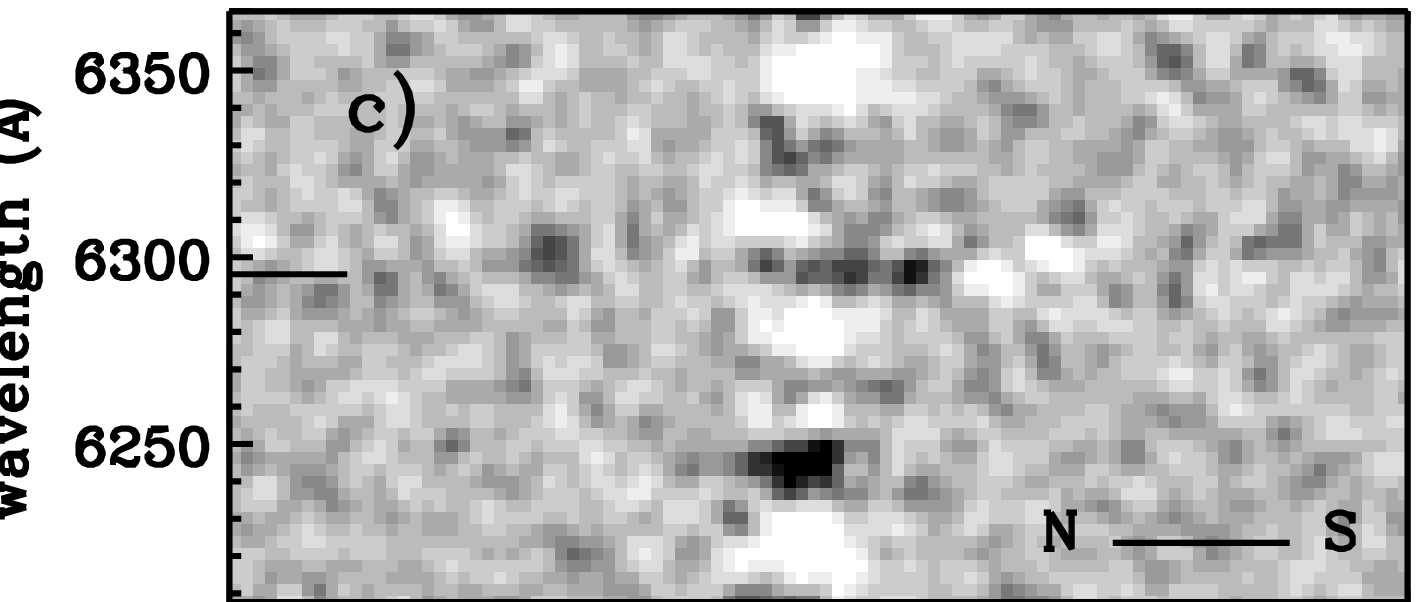}\\
\includegraphics[scale=0.4]{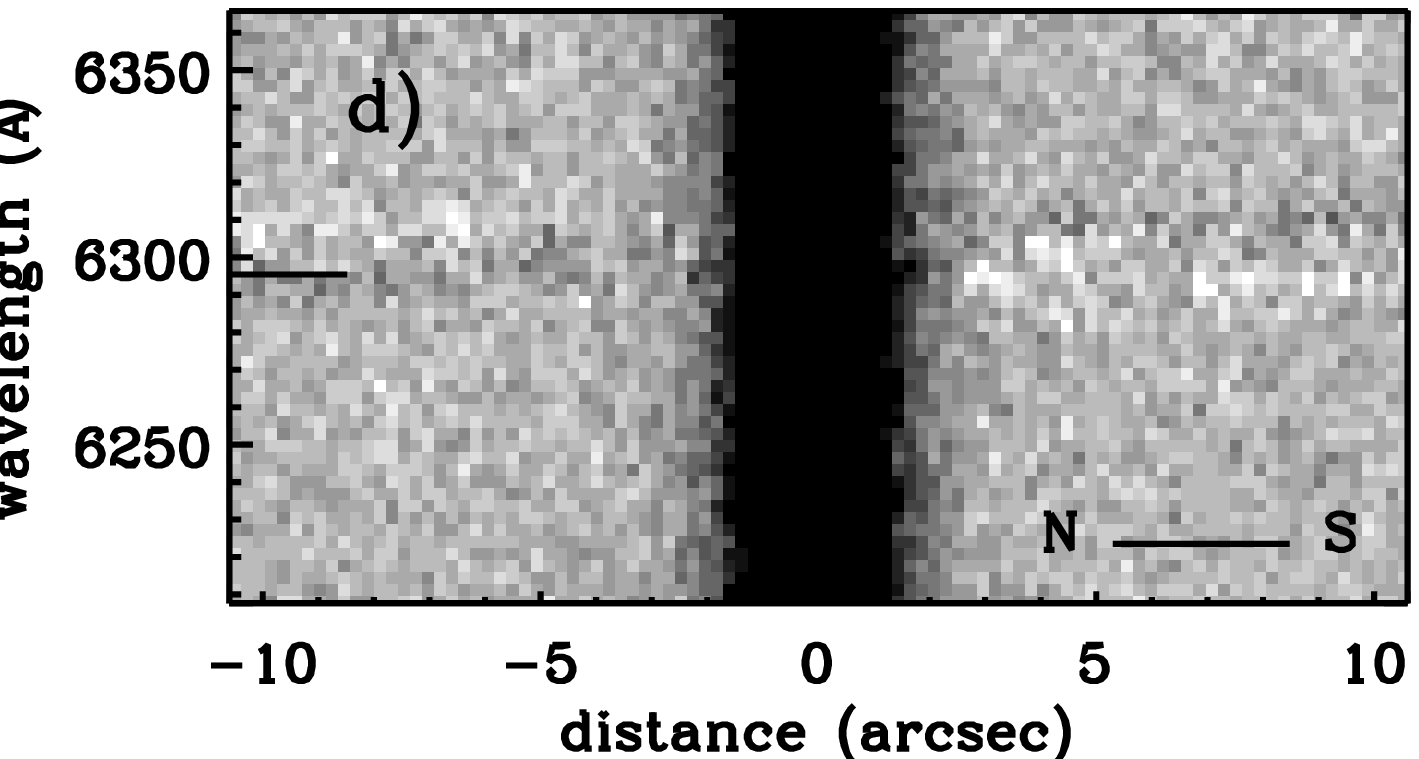}
\caption{\small{Panel a) shows the $u'$-band image of J0903+5448, and panel b) the $u'$-band  residual image after galaxy subtraction, convolved with a gaussian of $\sigma=2$ pixels for display purposes. The dashed lines indicate the position and width of the slit used to take long-slit spectra of the target. Panel c) shows a section of the 2D spectrum of J0903+5448 with the [\ion{O}{2}] $\lambda\lambda$3727 emission line of the background galaxy, at $\lambda=6295$~\AA, after skyline and galaxy subtraction, convolved with a gaussian of $\sigma=2$ pixels for display purposes; panel d) shows the same part of spectrum after skyline subtraction only.}  
\label{fig_0903}}
\end{figure}

%FIGURE J0942+6111

\begin{figure}
\centering
\includegraphics[scale=0.4]{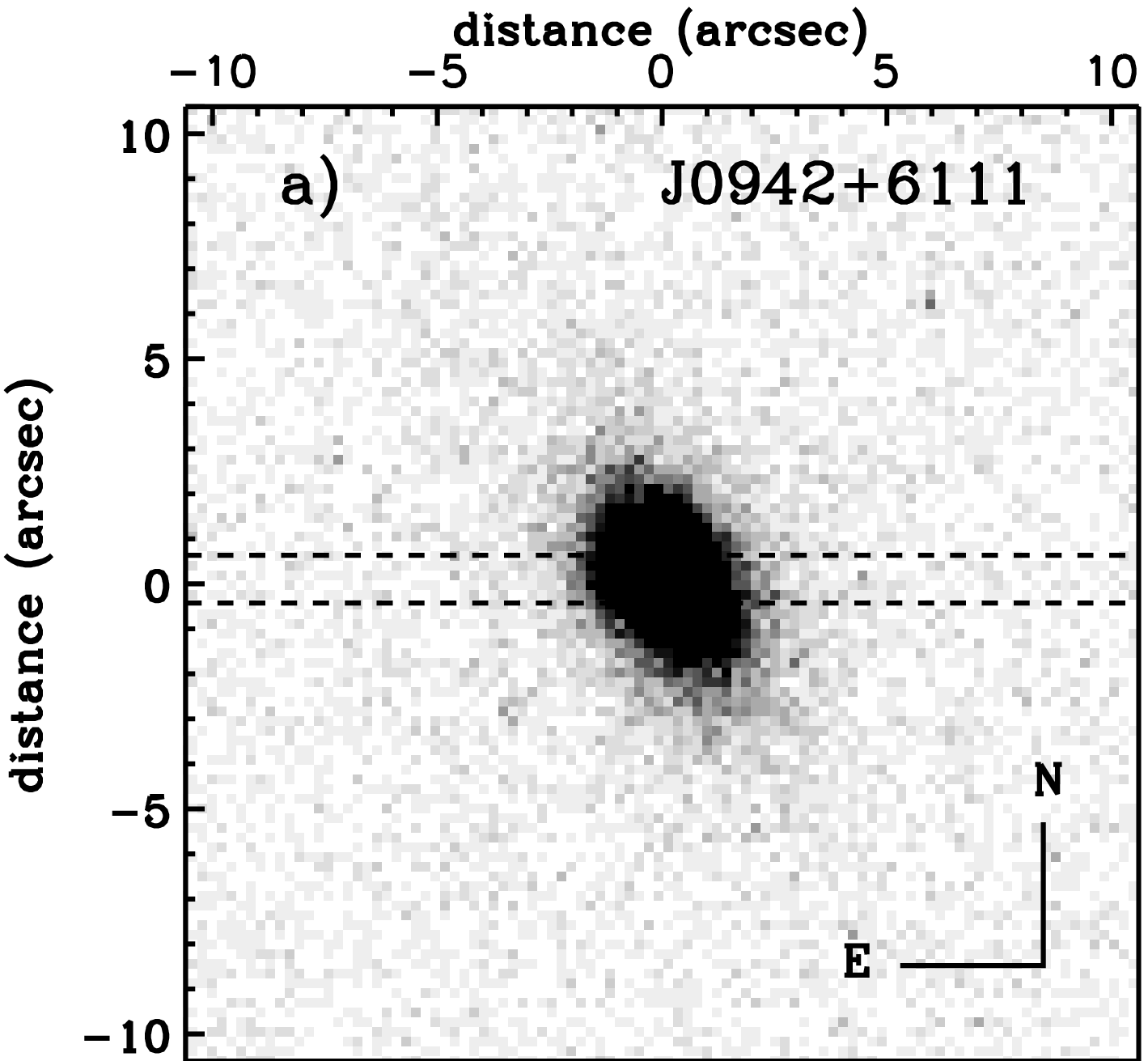}\\
\includegraphics[scale=0.4]{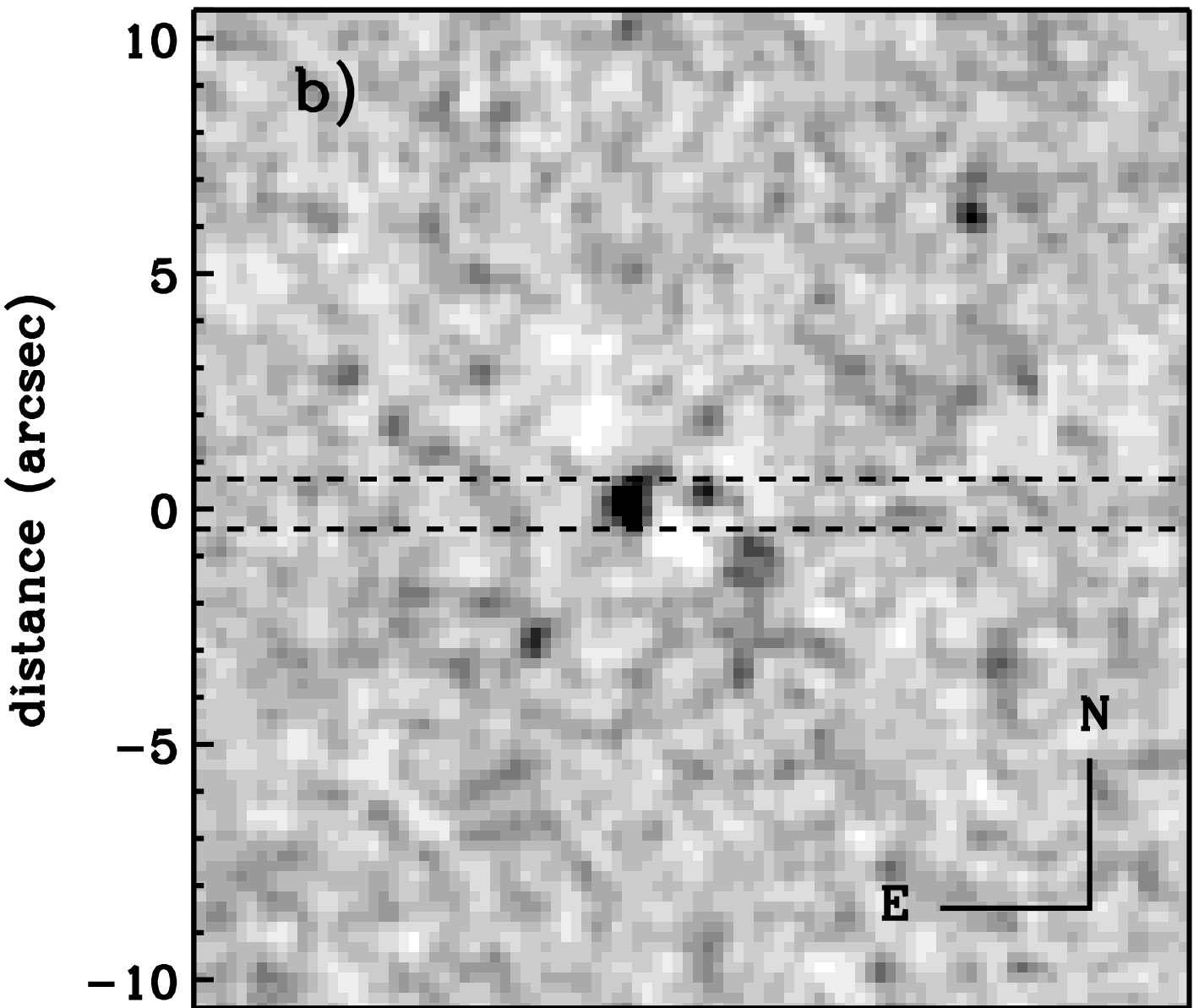}\\
\includegraphics[scale=0.4]{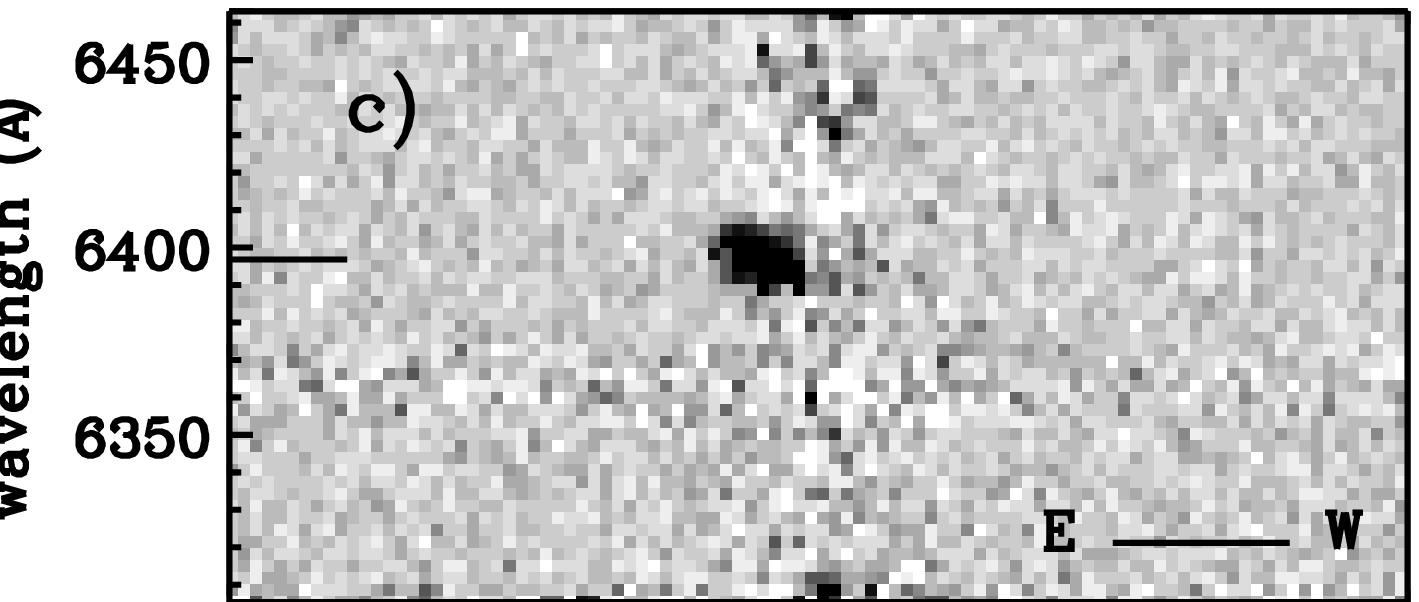}\\
\includegraphics[scale=0.4]{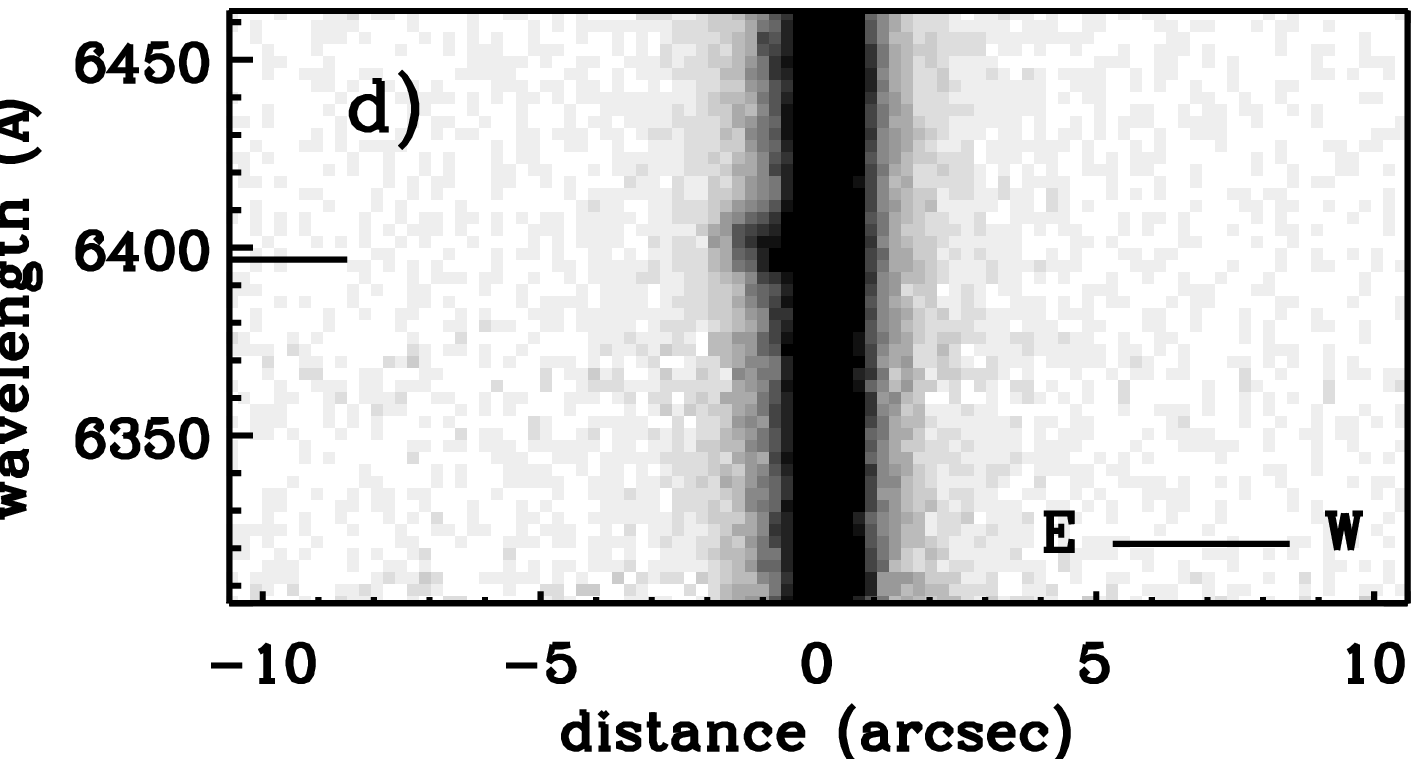}
\caption{\small{Panel a) shows the $u'$-band image of J0942+6111, and panel b) the $u'$-band  residual image after galaxy subtraction, convolved with a gaussian of $\sigma=2$ pixels for display purposes. The dashed lines indicate the position and width of the slit used to take long-slit spectra of the target. Panel c) shows a section of the 2D spectrum of J0942+6111 with the [\ion{O}{2}] $\lambda\lambda$3727 emission line of the background galaxy, at $\lambda=6397$~\AA, after skyline and galaxy subtraction; panel d) shows the same part of spectrum after skyline subtraction only.} 
\label{fig_0942}}
\end{figure}

%FIGURE J1150+1202

\begin{figure}
\centering
\includegraphics[scale=0.4]{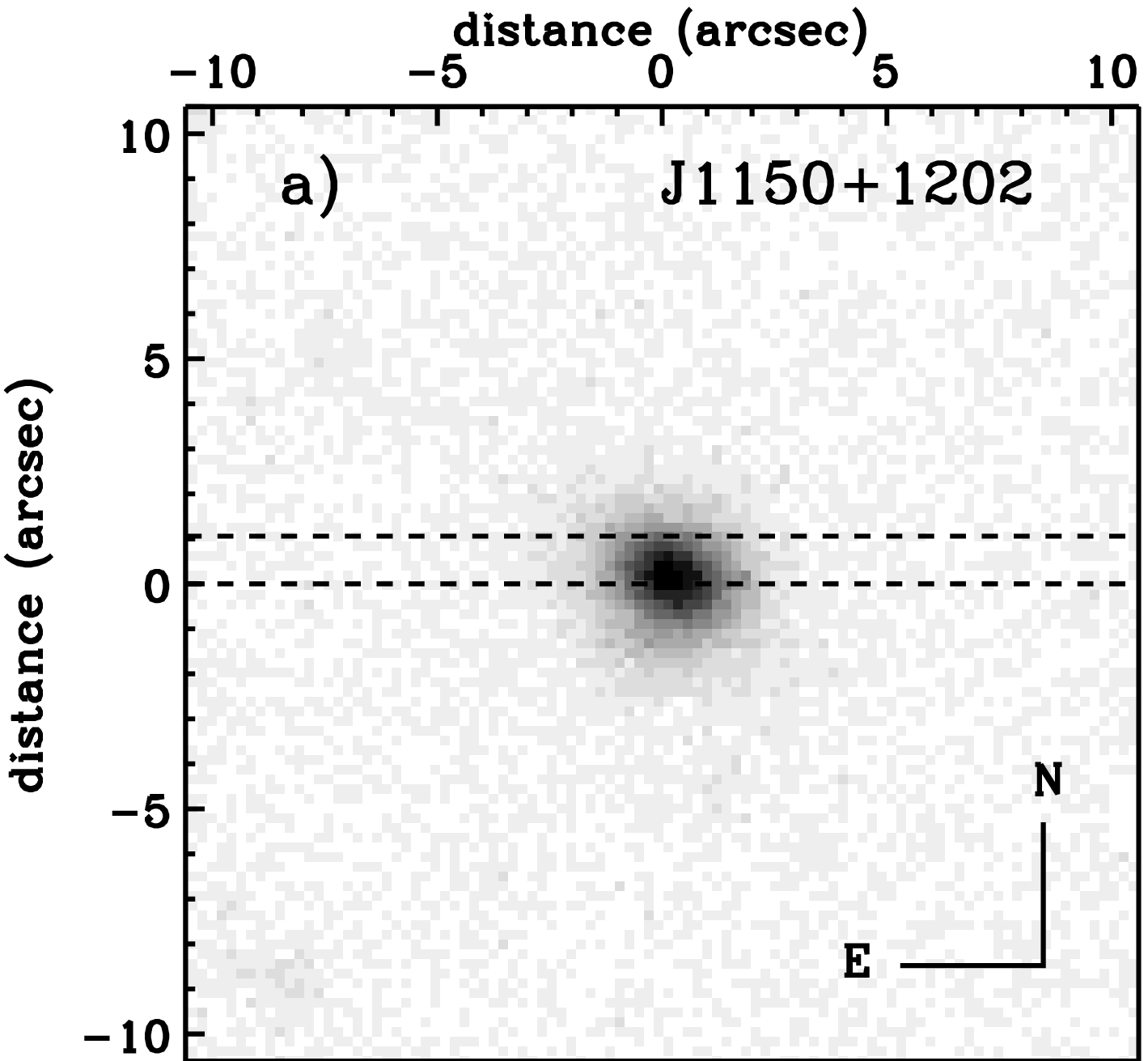}\\
\includegraphics[scale=0.4]{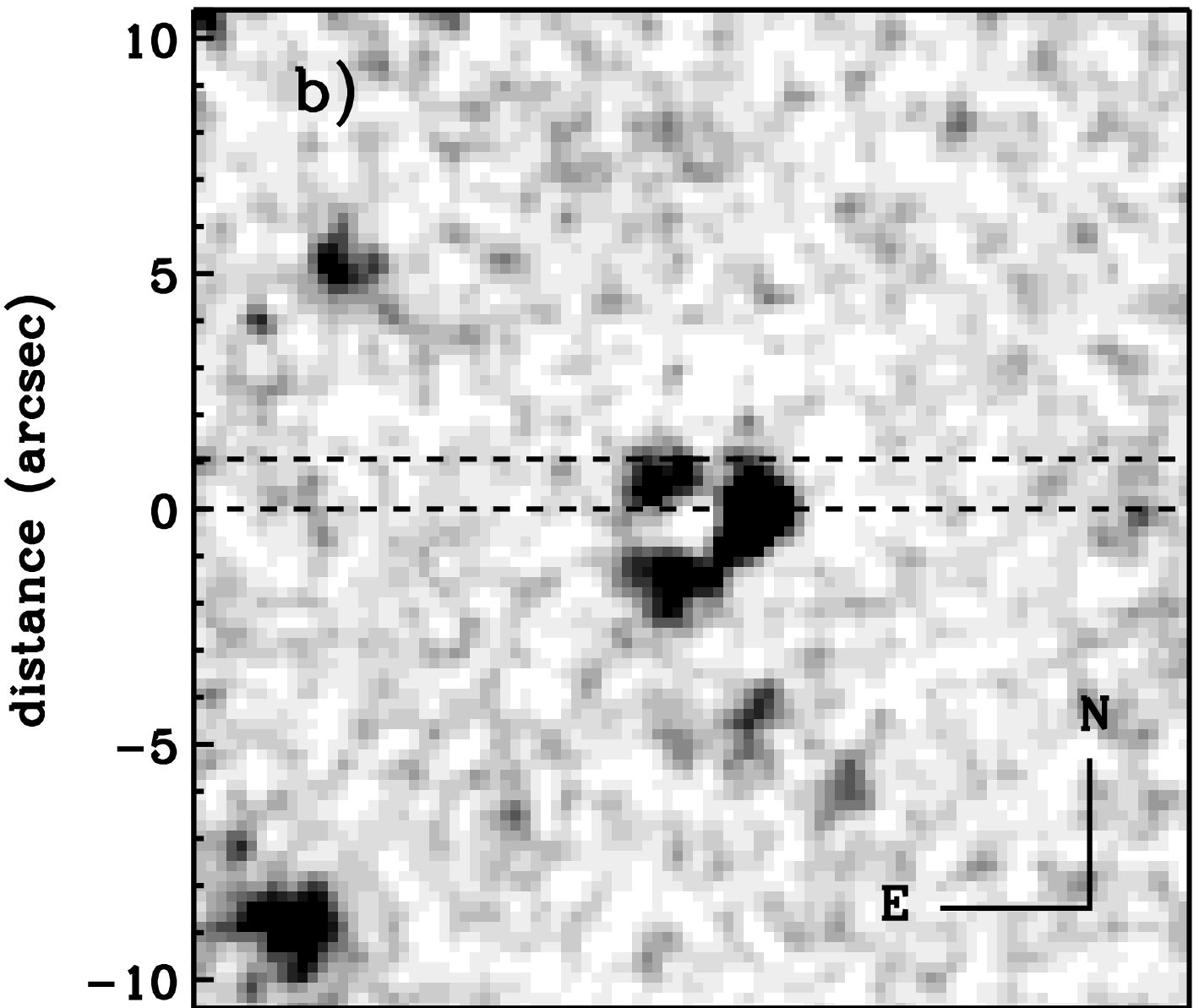}\\
\includegraphics[scale=0.4]{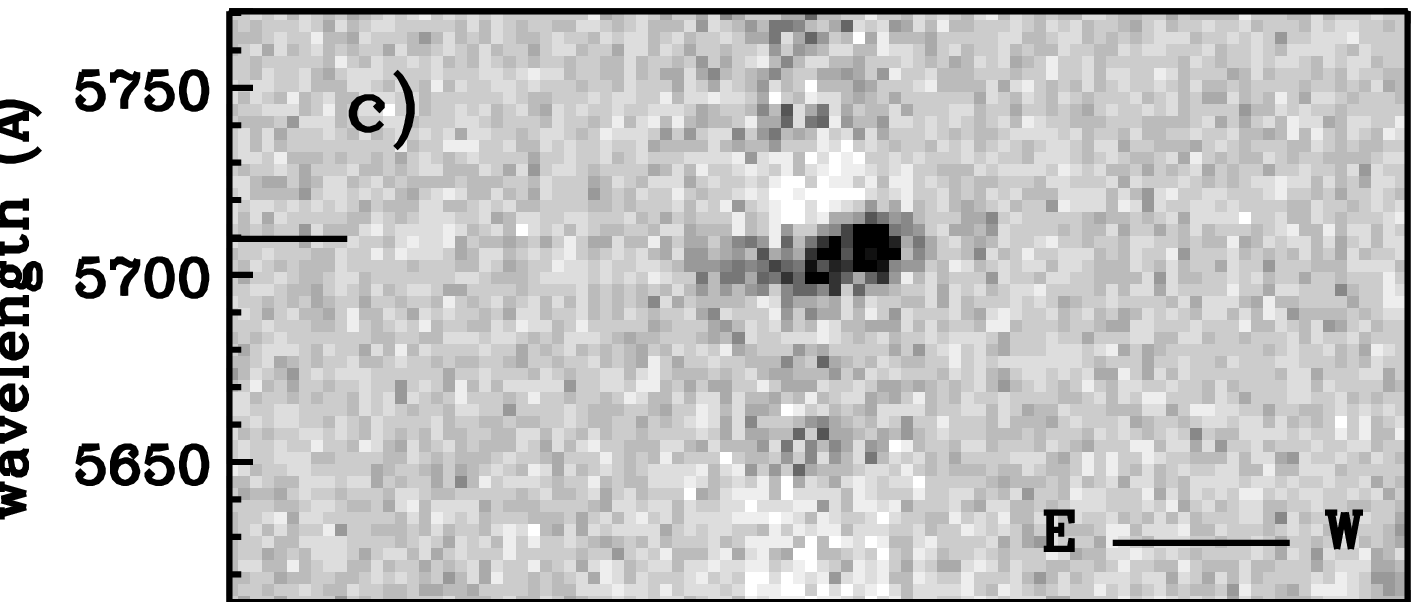}\\
\includegraphics[scale=0.4]{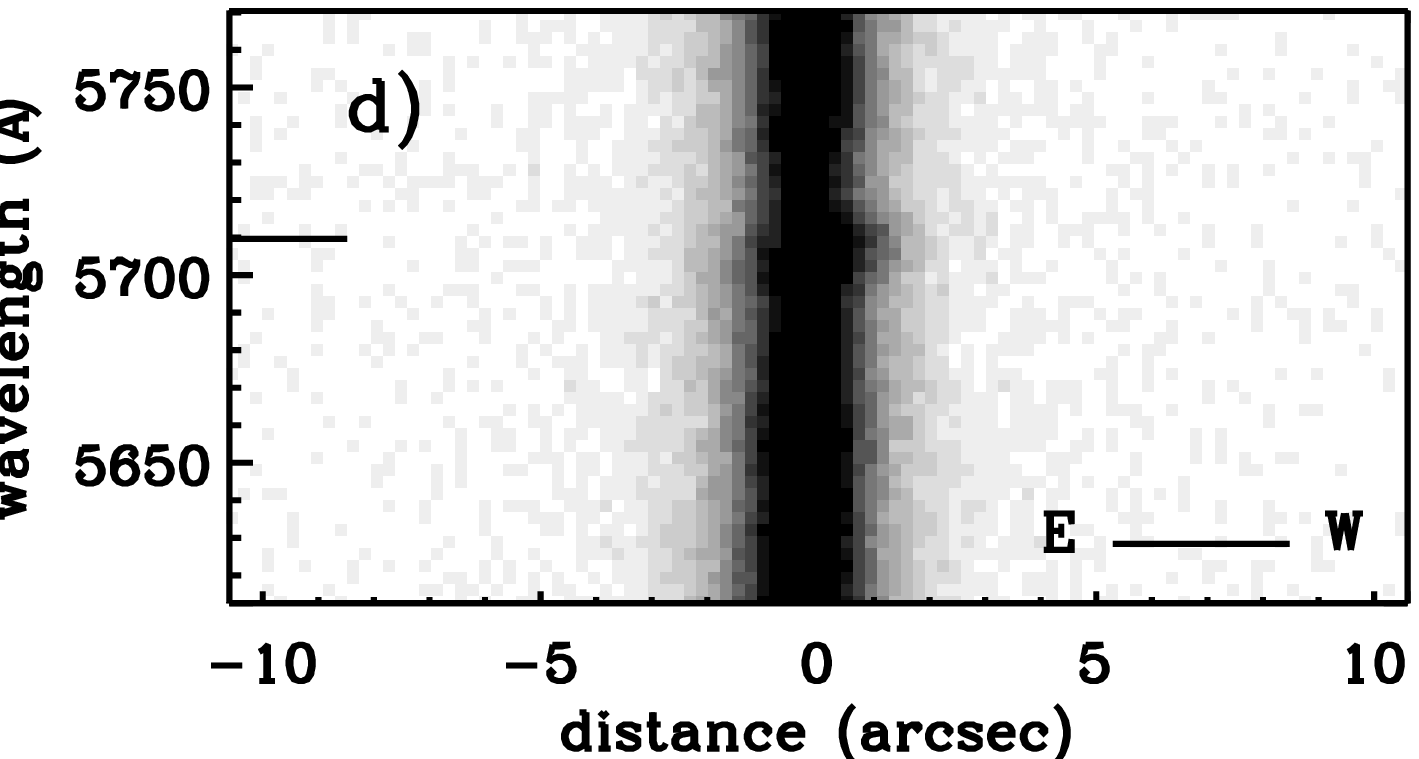}
\caption{\small{Panel a) shows the $u'$-band image of J1150+1202, and panel b) the $u'$-band  residual image after galaxy subtraction, convolved with a gaussian of $\sigma=1$ pixel for display purposes. The dashed lines indicate the position and width of the slit used to take long-slit spectra of the target. Panel c) shows a section of the 2D spectrum of J1150+1202 with the [\ion{O}{2}] $\lambda\lambda$3727 emission line of the background galaxy, at $\lambda=5710$~\AA, after skyline and galaxy subtraction; panel d) shows the same part of spectrum after skyline subtraction only.} 
\label{fig_1150}}
\end{figure}

%FIGURE J1200+4014

\begin{figure}
\centering
\includegraphics[scale=0.4]{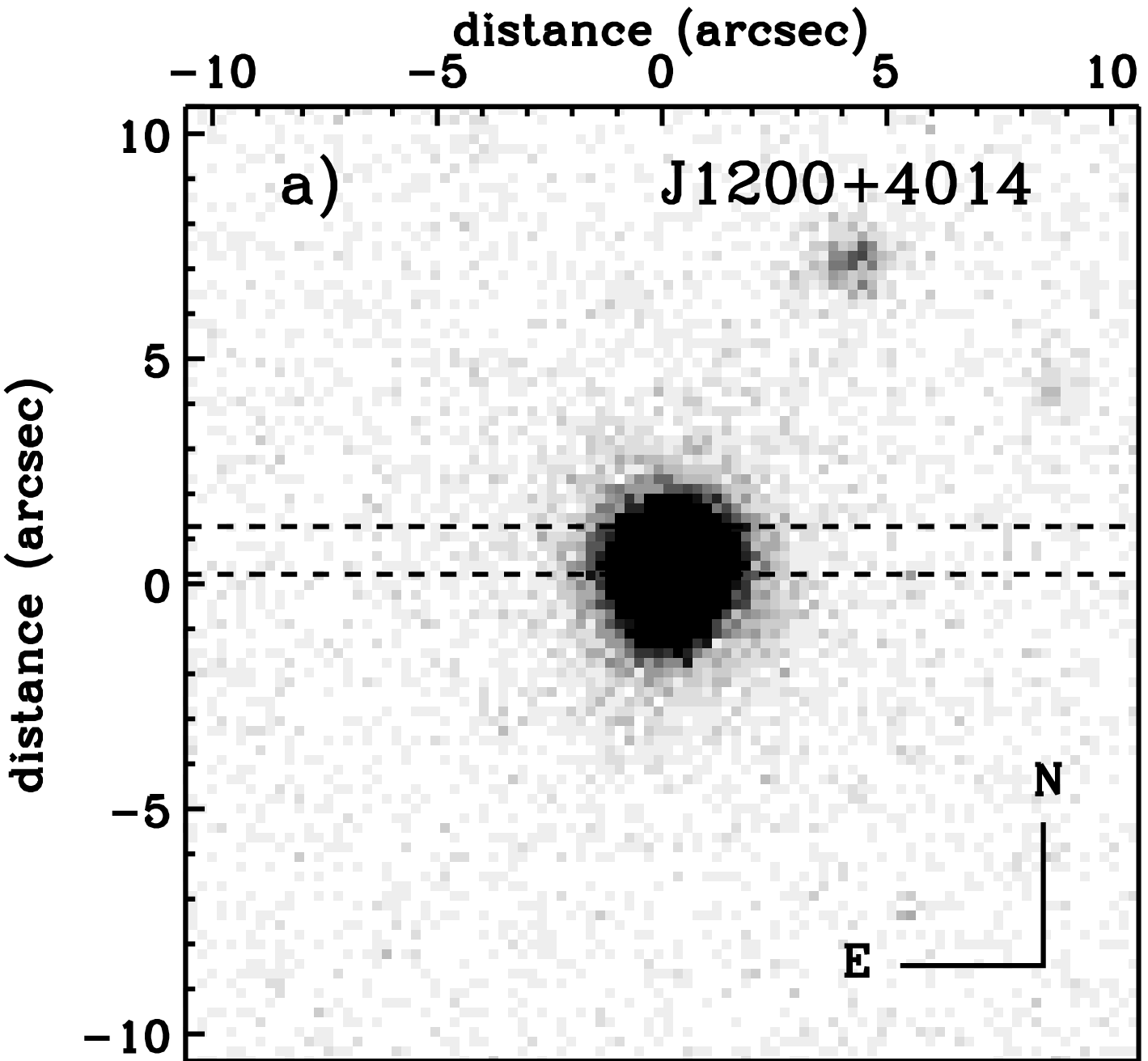} \hspace{1.cm}
\includegraphics[scale=0.4]{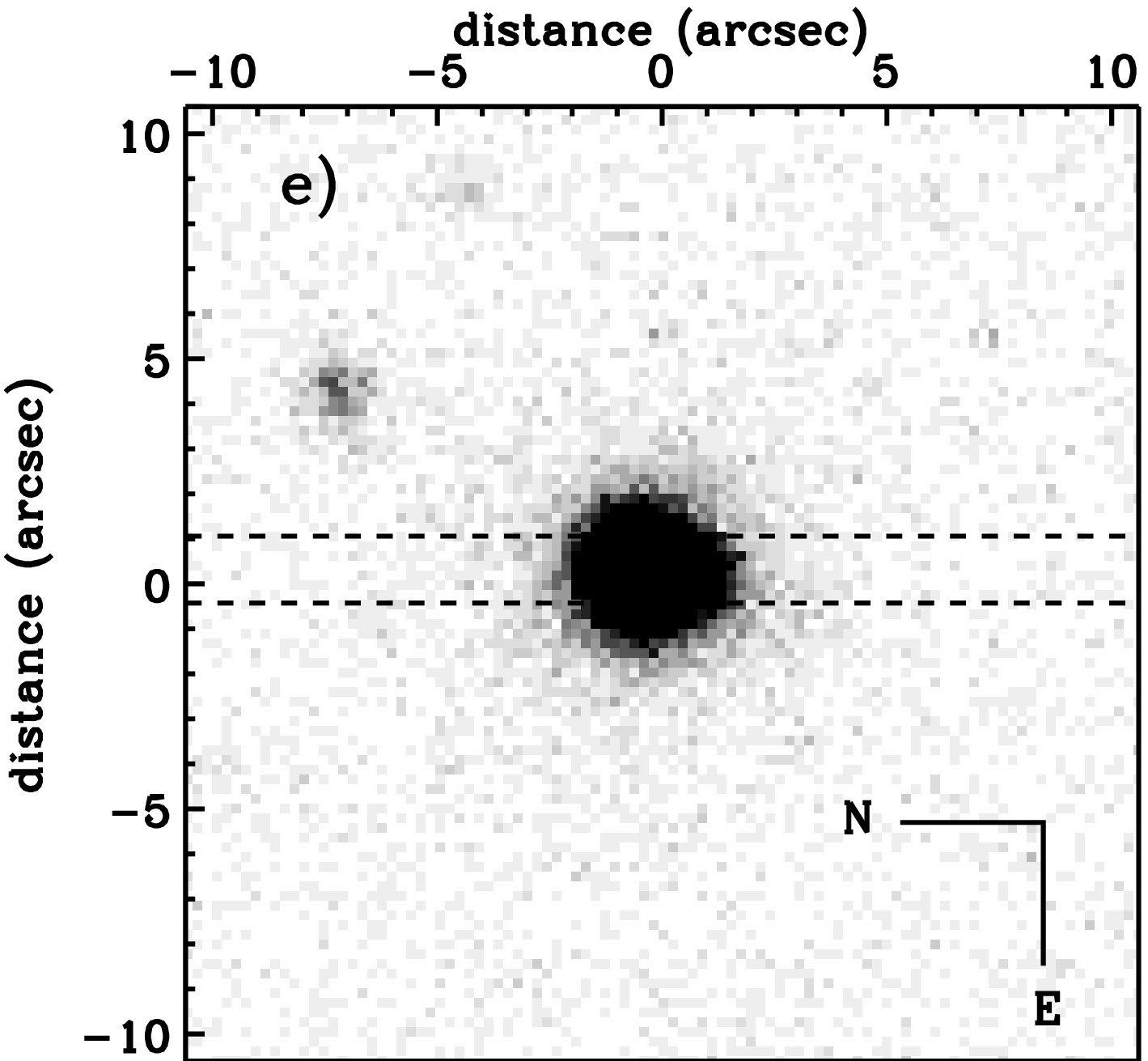}\\
\includegraphics[scale=0.4]{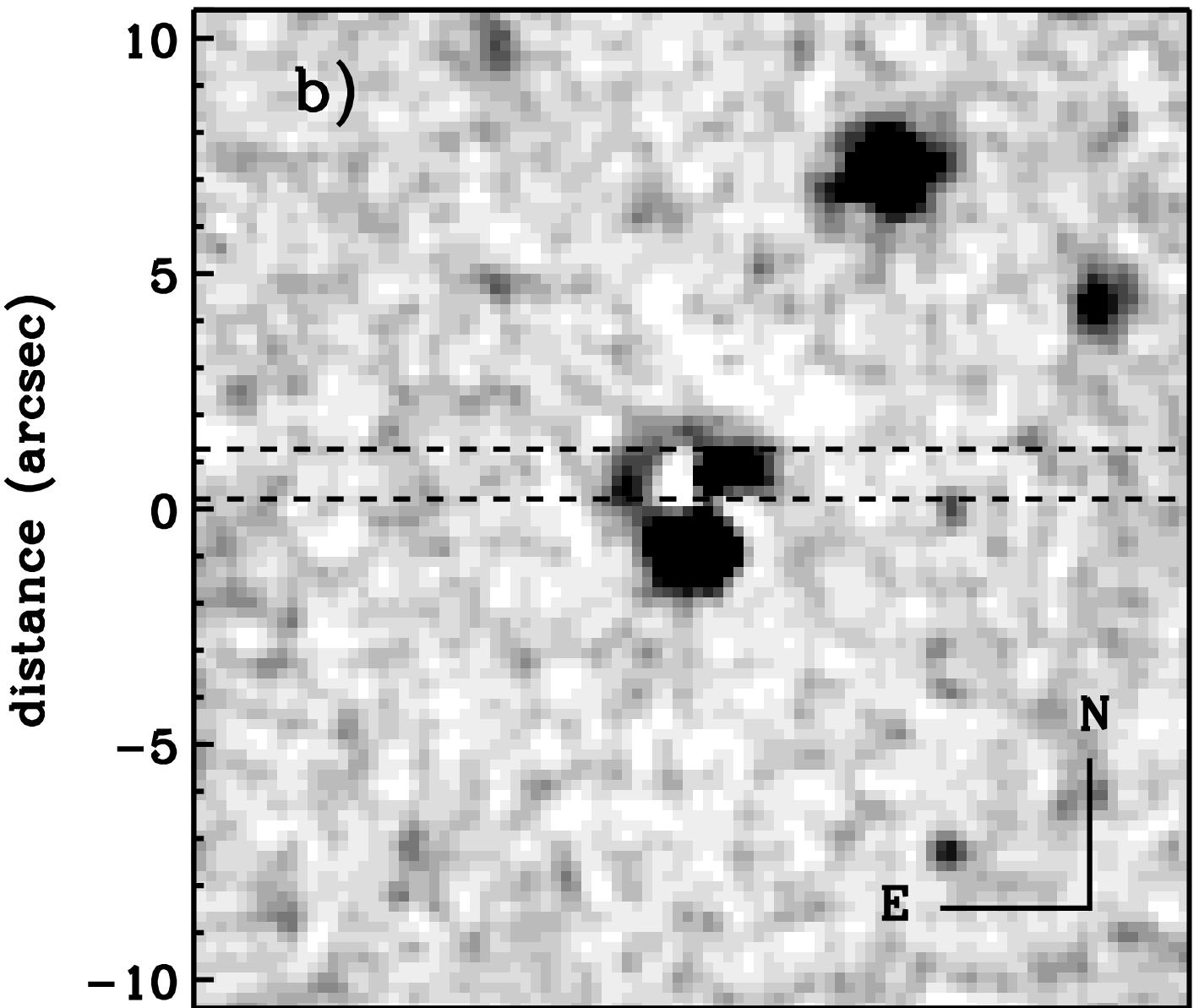} \hspace{1.cm}
\includegraphics[scale=0.4]{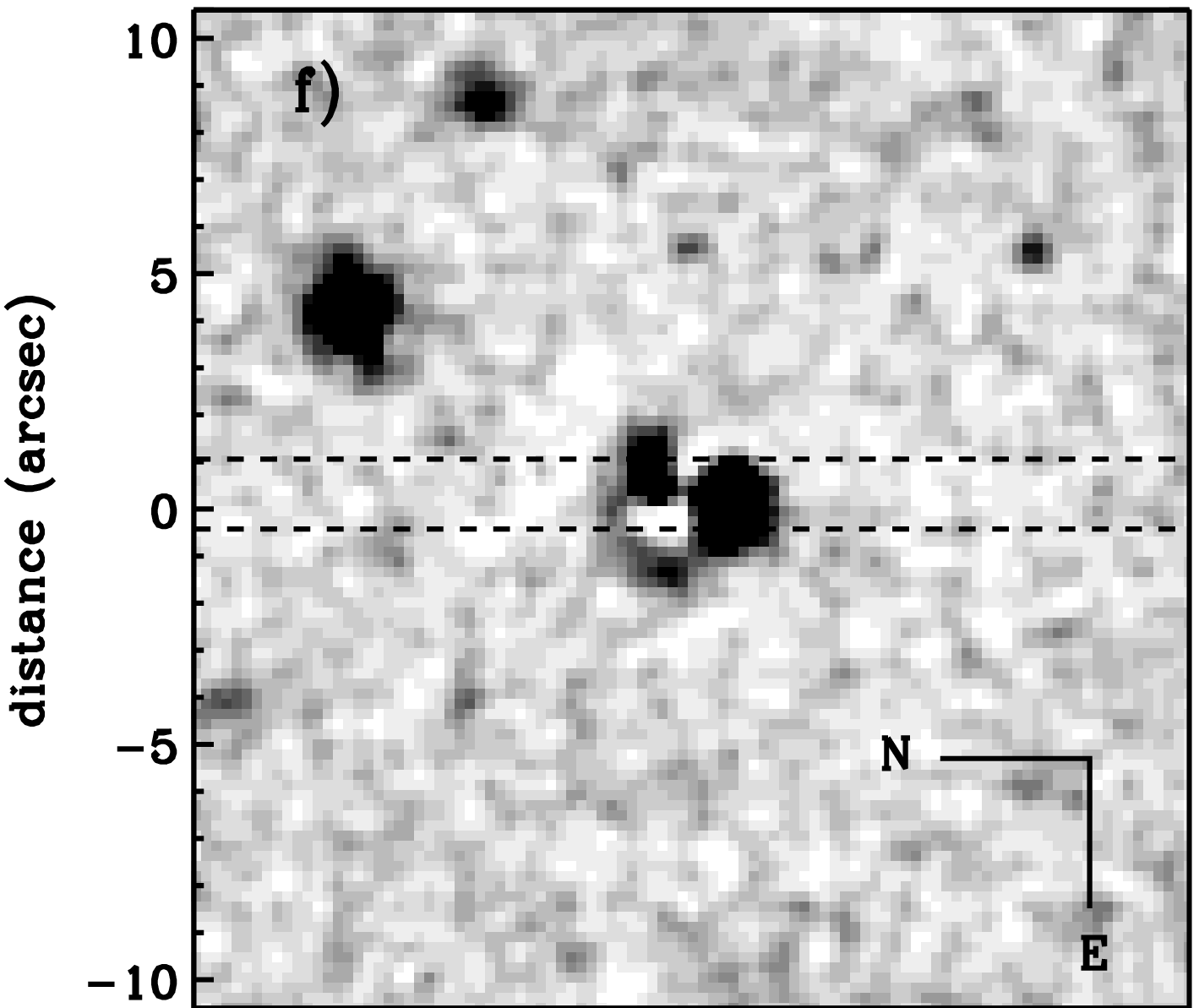}\\
\includegraphics[scale=0.4]{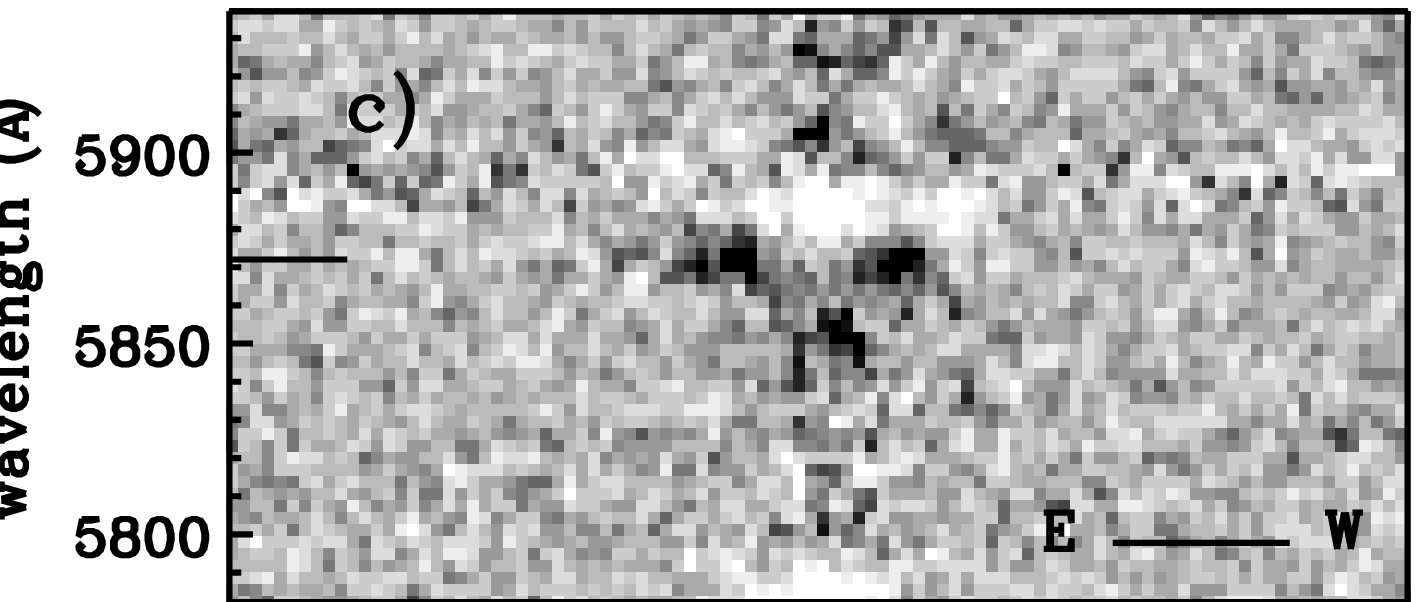} \hspace{1.cm}
\includegraphics[scale=0.4]{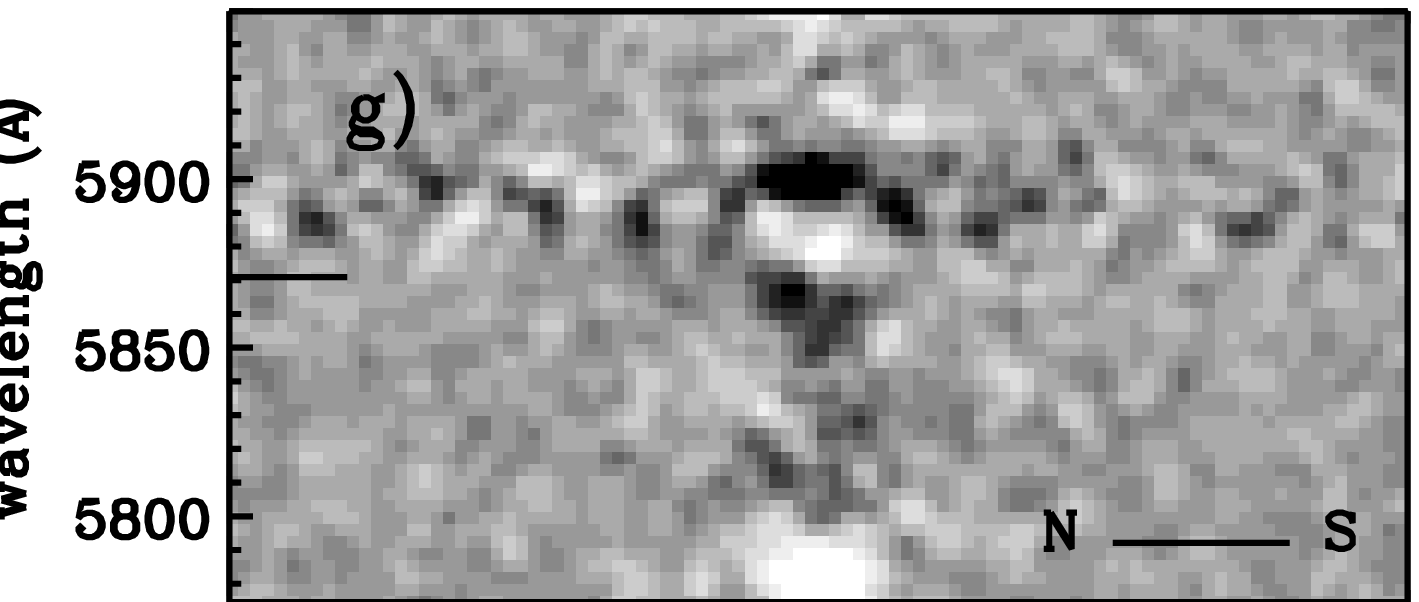}\\
\includegraphics[scale=0.4]{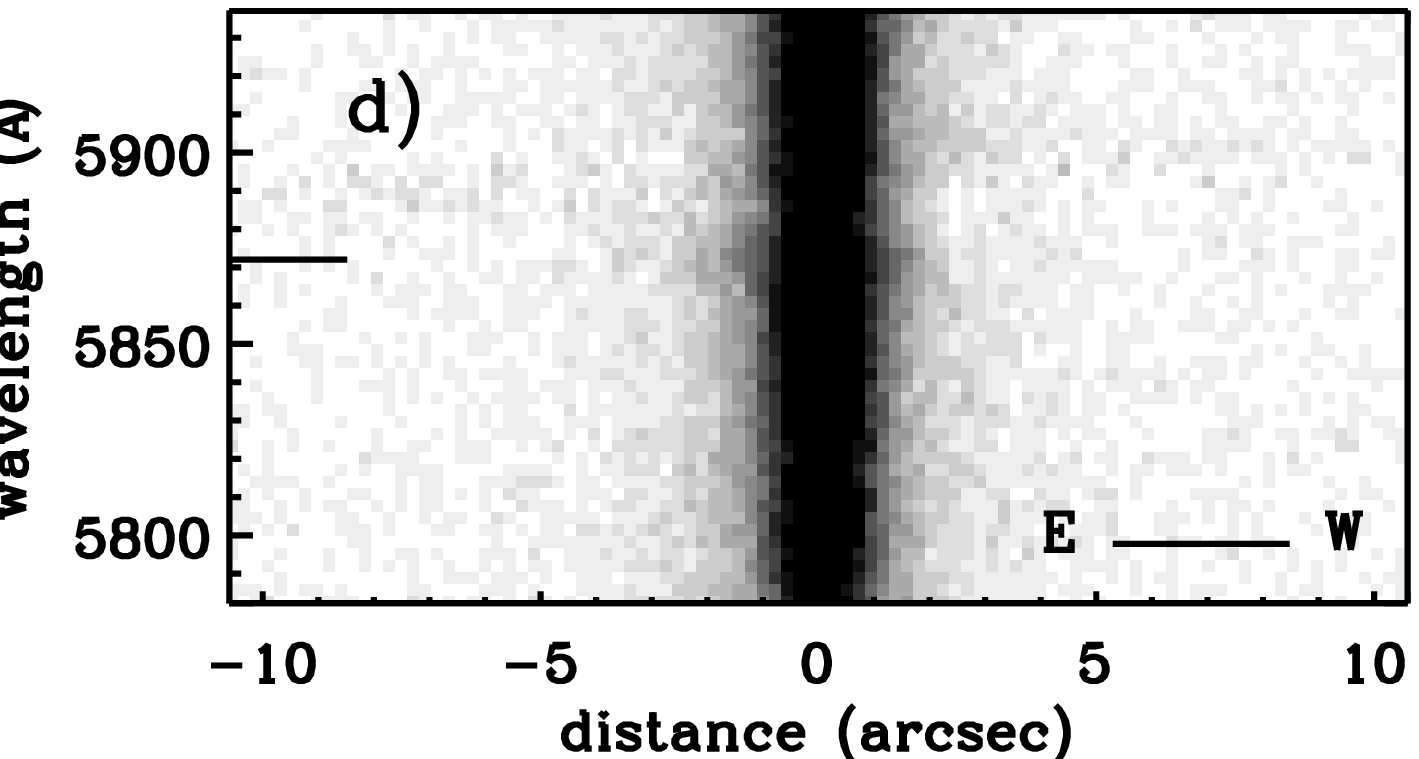} \hspace{1.cm}
\includegraphics[scale=0.4]{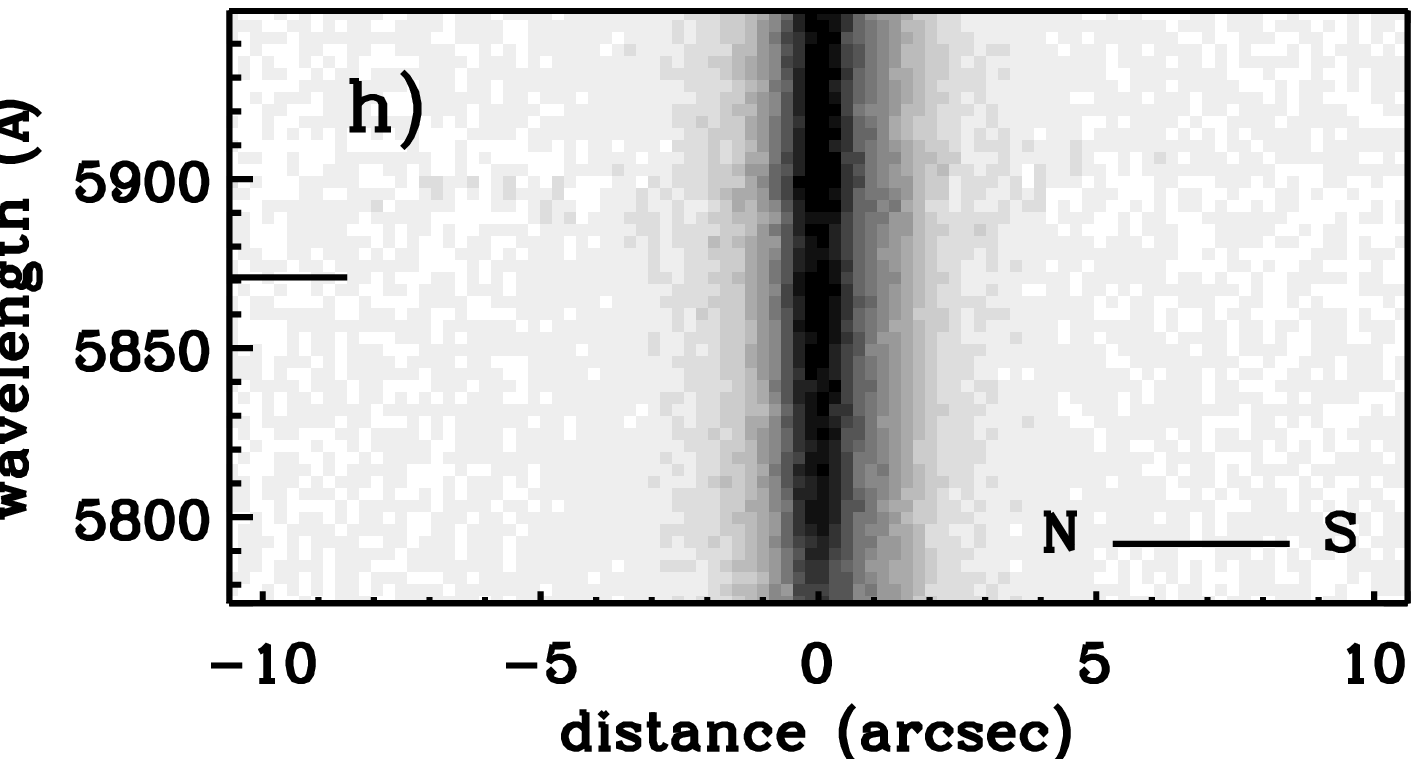}\\
\caption{\small{Panels a) and e) show the $u'$-band image of J1200+4014. Panels b) and f) show the $u'$-band  residual image after galaxy subtraction, convolved with a gaussian of $\sigma=2$ pixels for display purposes. The dashed lines indicate the position and width of the slit used to take respectively the east-west and  north-south long-slit spectra of the target. Panels c) and g) show a section of, respectively, the east-west and north-south 2D spectra of J1200+4014, with the [\ion{O}{2}] $\lambda\lambda$3727 emission line of the background galaxy at $\lambda=5871$~\AA. The spectra were skyline- and galaxy-subtracted, and convolved with a gaussian of $\sigma=1$ pixel in panel c) and $\sigma=2$ pixels in panel g) for display purposes. Panel d) and h) show the same part of the spectra after skyline subtraction only.} } 
\label{fig_1200}
\end{figure}

%FIGURE J1356+5615

\begin{figure}
\centering
\includegraphics[scale=0.4]{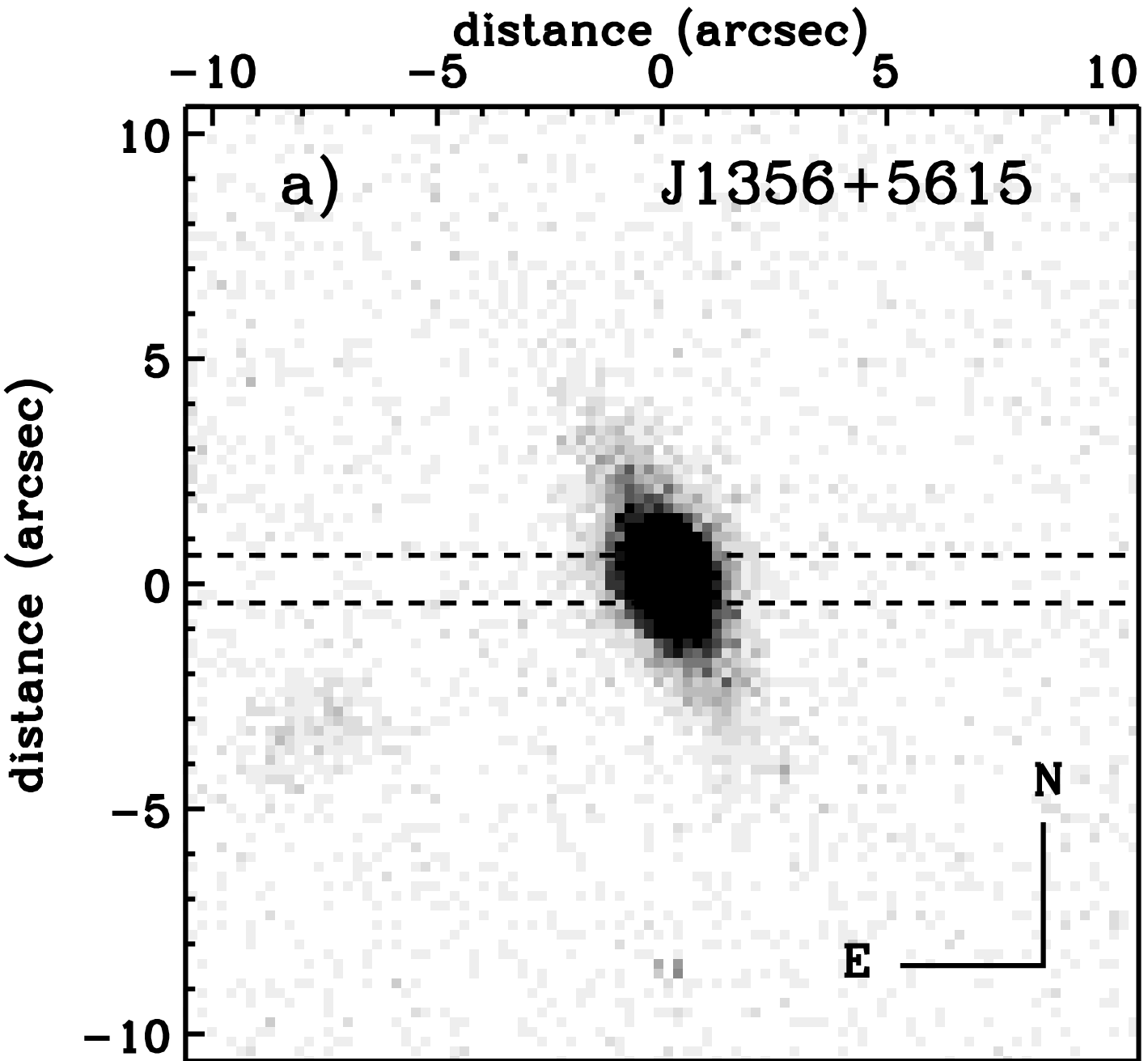} \hspace{1.cm}
\includegraphics[scale=0.4]{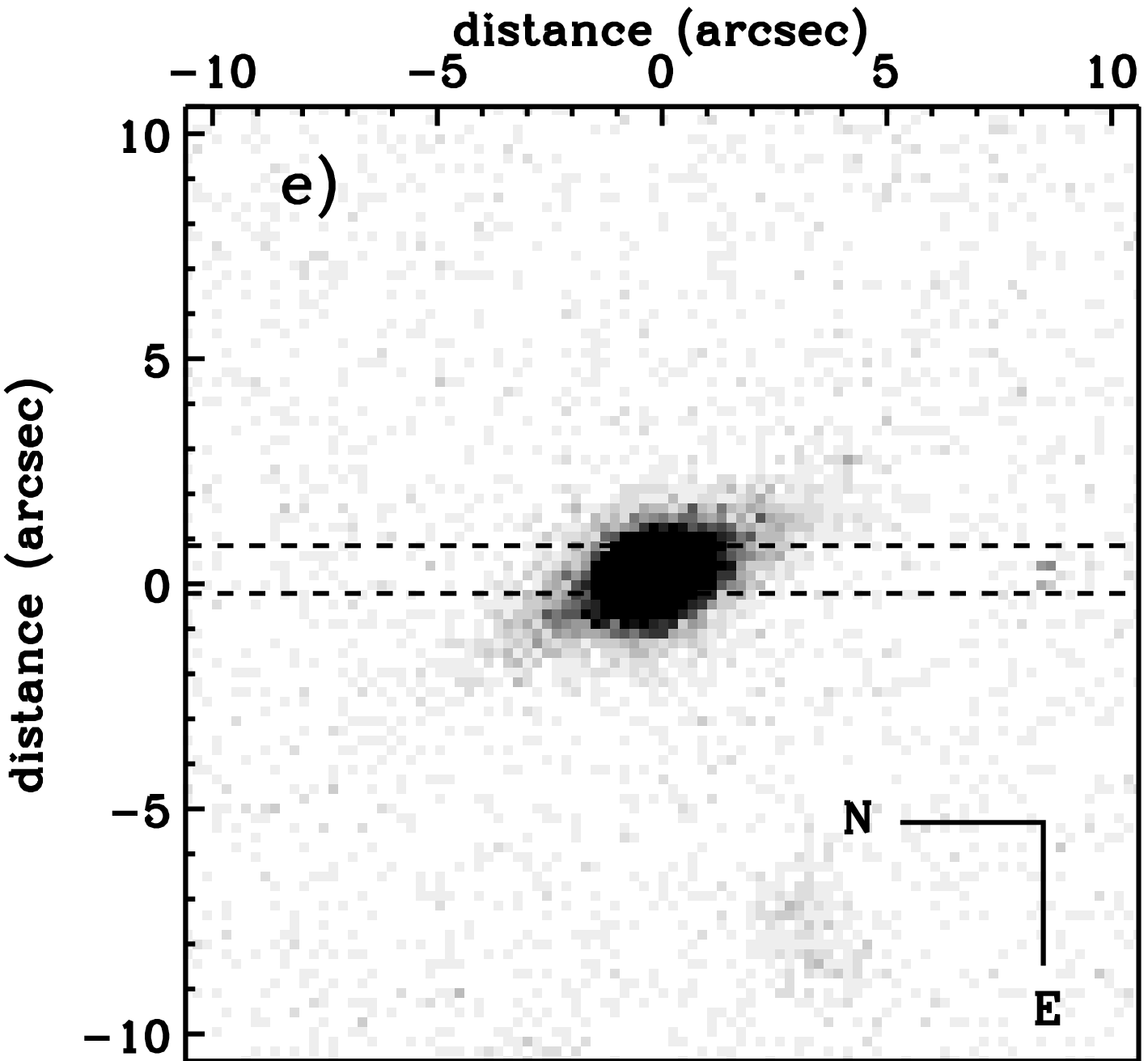}\\
\includegraphics[scale=0.4]{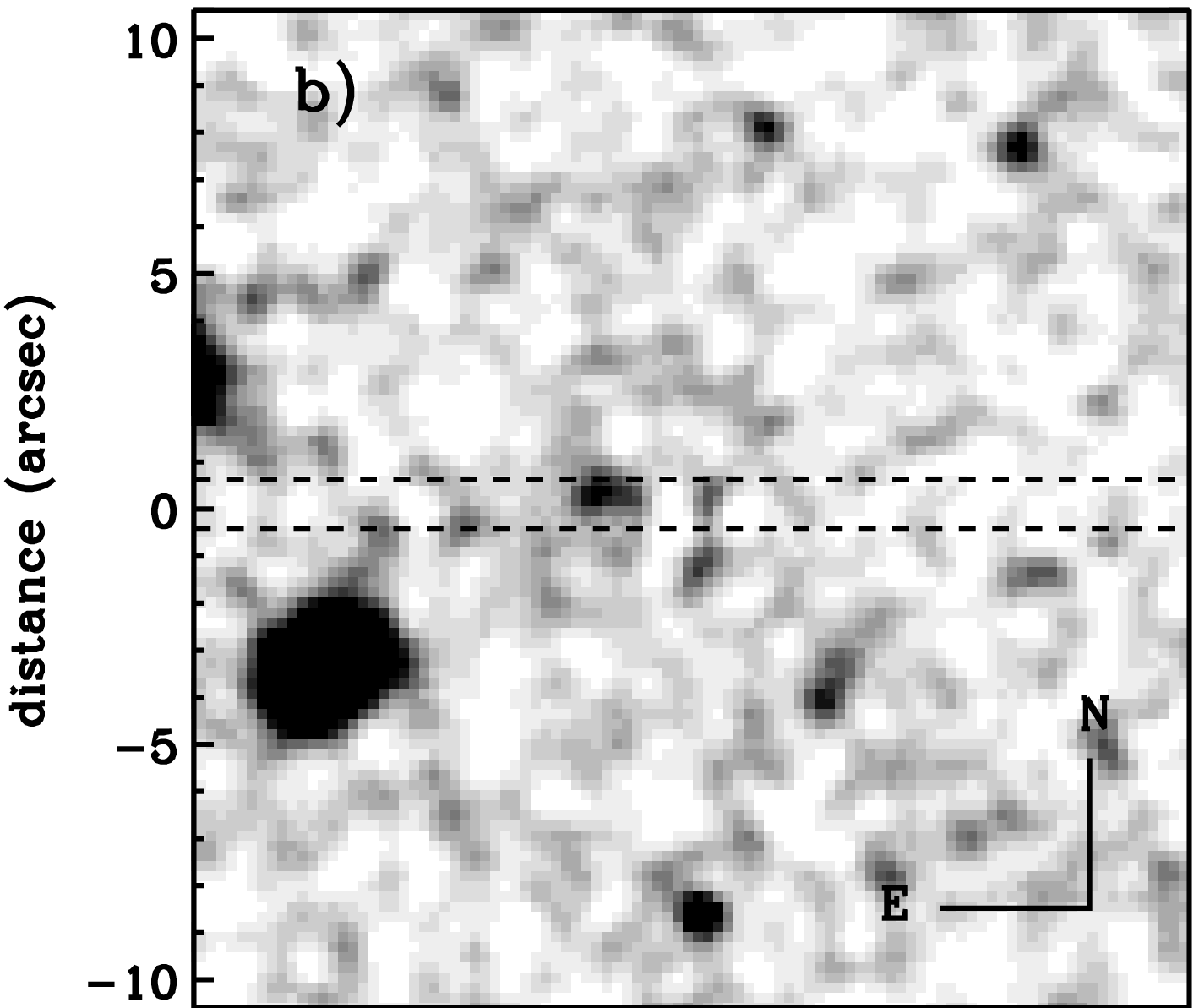} \hspace{1.cm}
\includegraphics[scale=0.4]{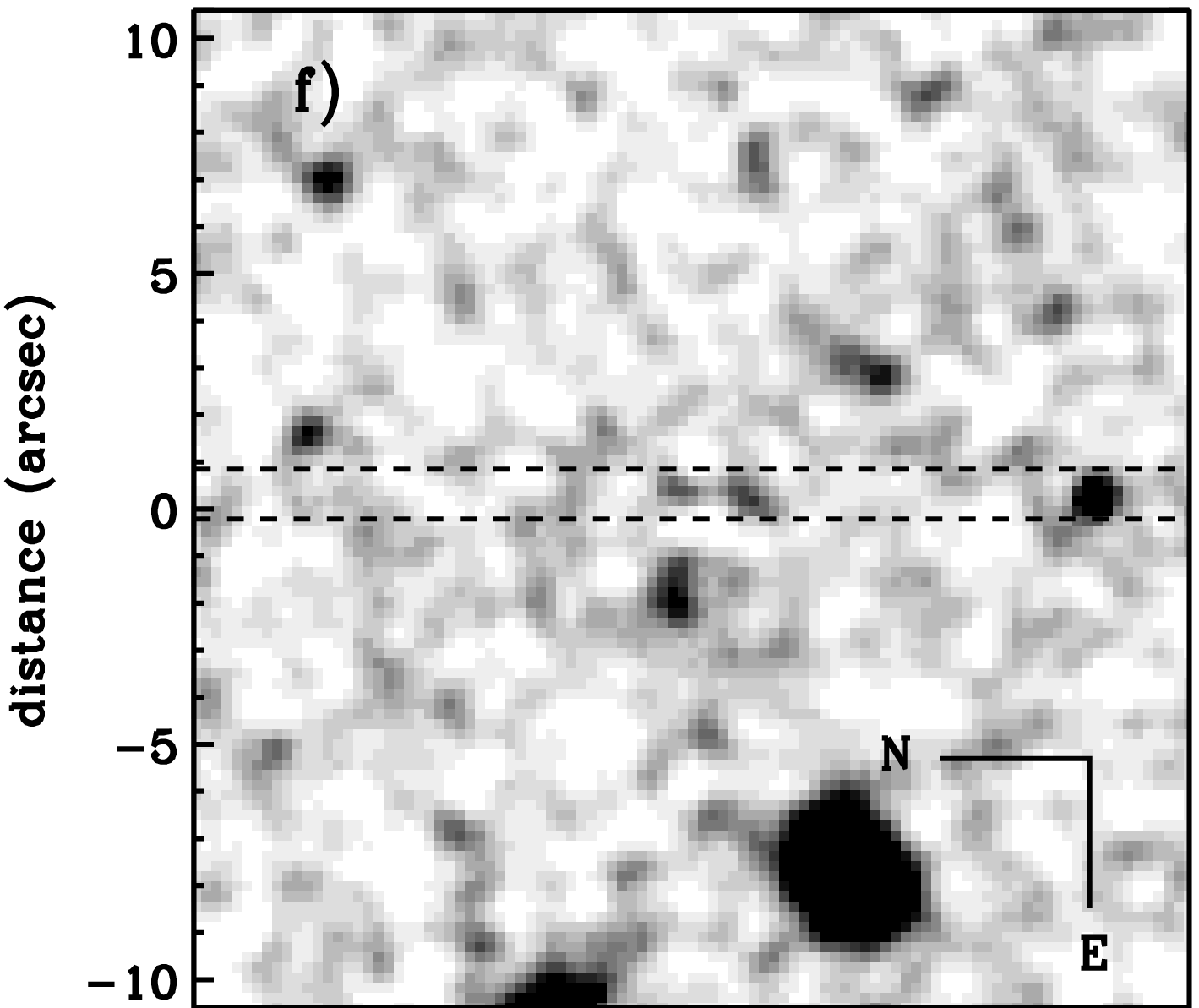}\\
\includegraphics[scale=0.4]{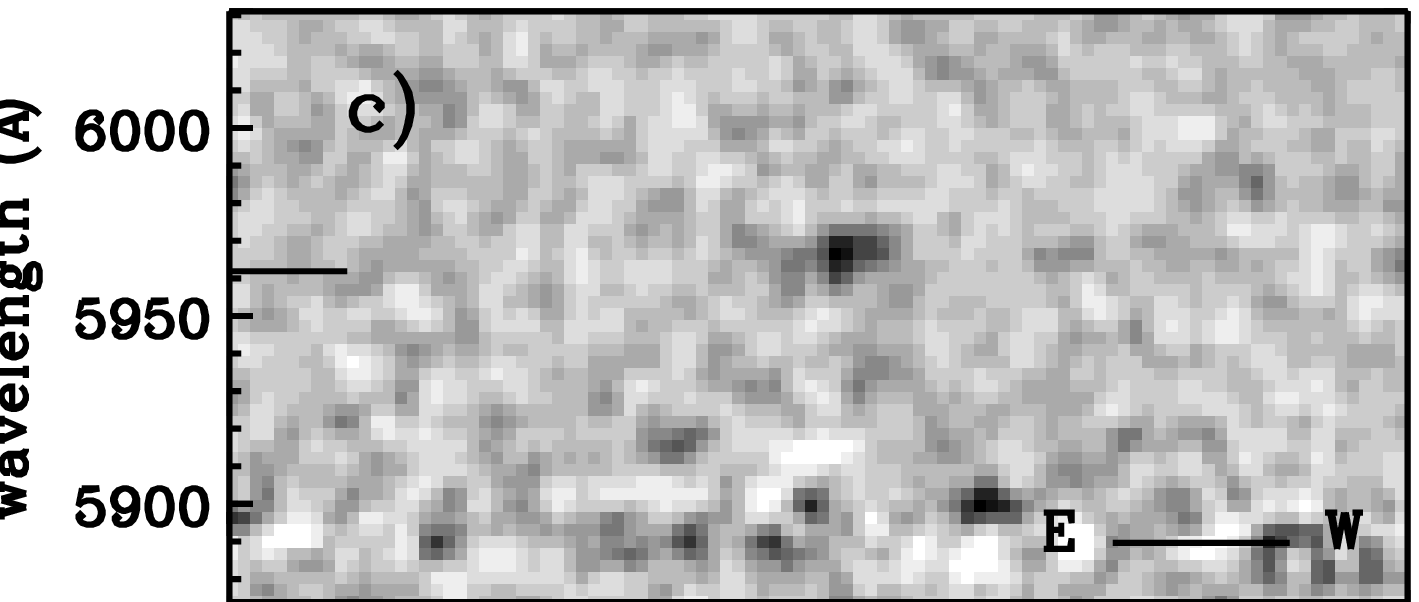} \hspace{1.cm}
\includegraphics[scale=0.4]{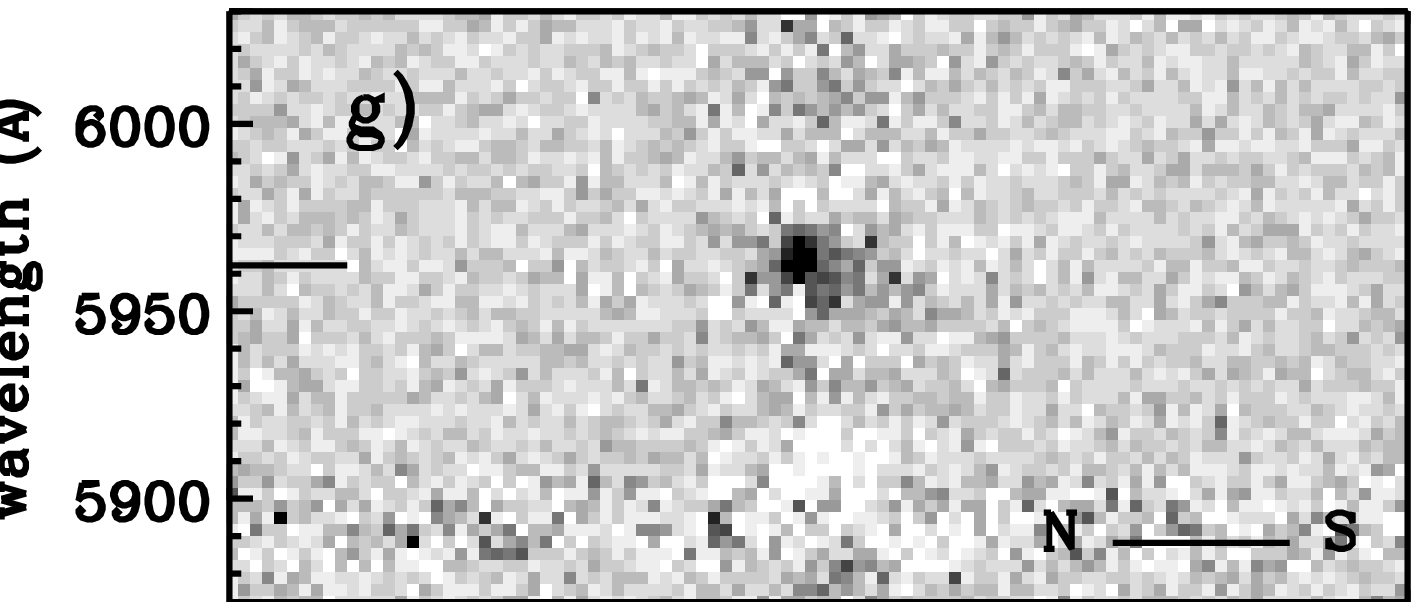}\\
\includegraphics[scale=0.4]{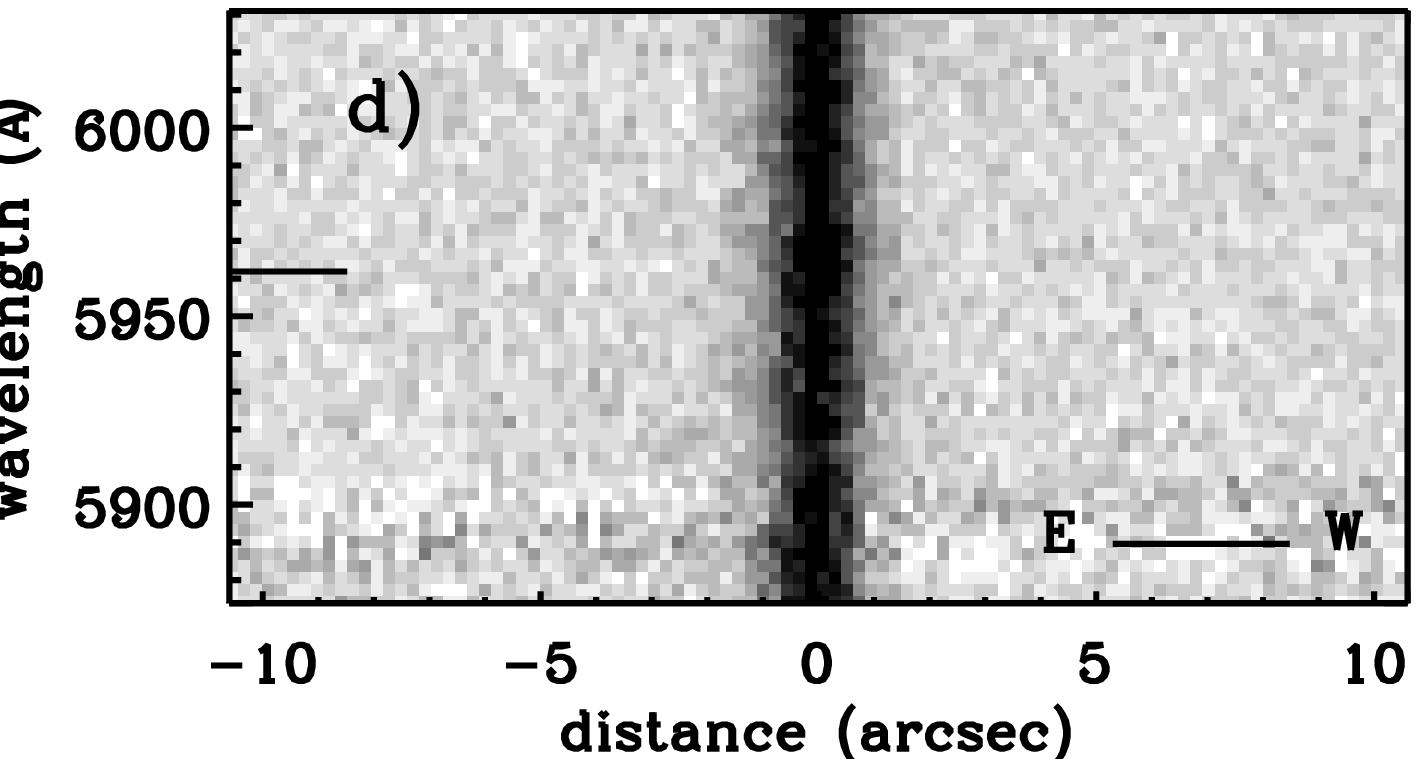} \hspace{1.cm}
\includegraphics[scale=0.4]{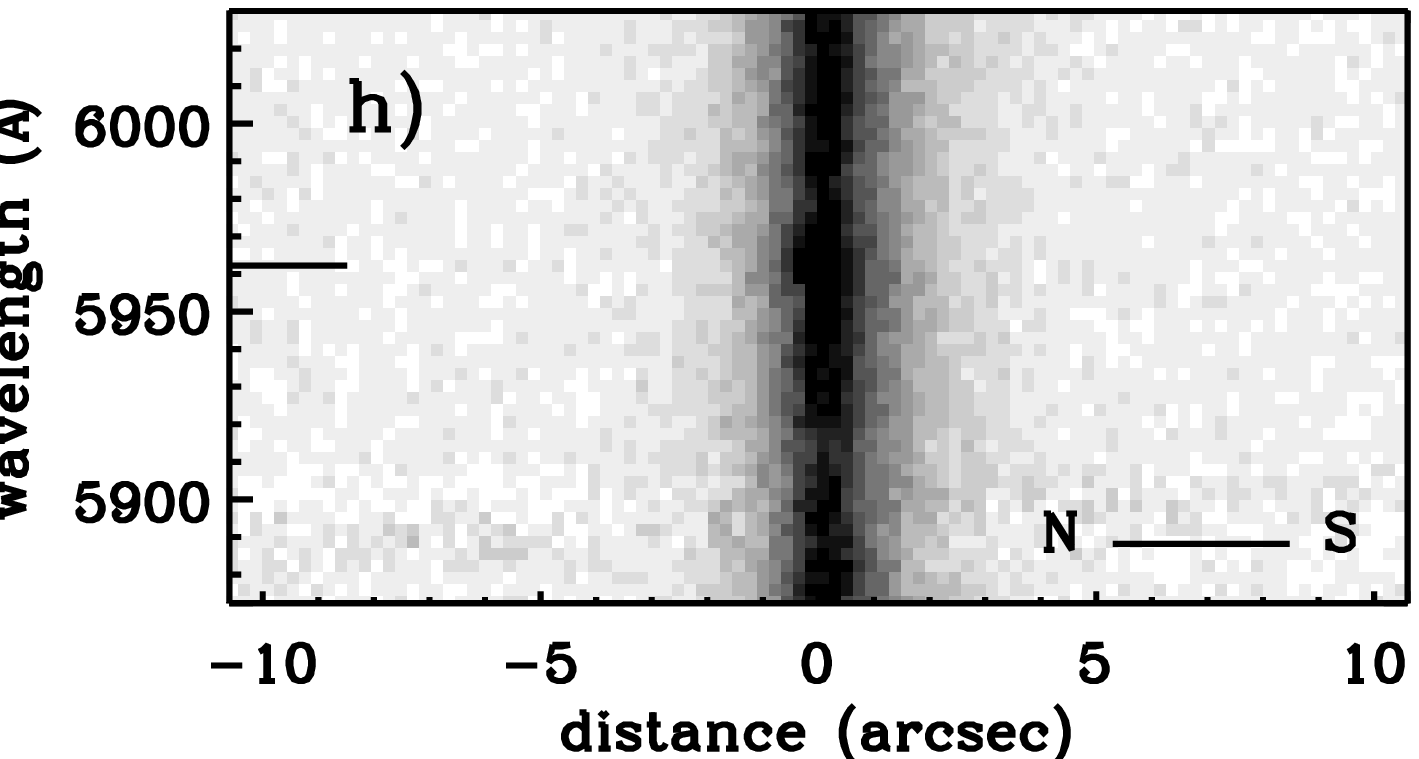}\\
\caption{\small{Panels a) and e) show the $u'$-band image of J1356+5615. Panels b) and f) show the $u'$-band  residual image after galaxy subtraction, convolved with a gaussian of $\sigma=3$ pixels for display purposes. The dashed lines indicate the position and width of the slit used to take respectively the east-west and  north-south long-slit spectra of the target. Panels c) and g) show a section of, respectively, the east-west and north-south 2D spectra of J1356+5615, with the [\ion{O}{2}] $\lambda\lambda$3727 emission line of the background galaxy at $\lambda=5962$~\AA. The spectra were skyline- and galaxy-subtracted, and the spectrum in panel c) was convolved with a gaussian of $\sigma=2$ pixels for display purposes. Panel d) and h) show the same part of the spectra after skyline subtraction only.} } 
\label{fig_1356}
\end{figure}

%FIGURE J1455+5304

\begin{figure}
\centering
\includegraphics[scale=0.4]{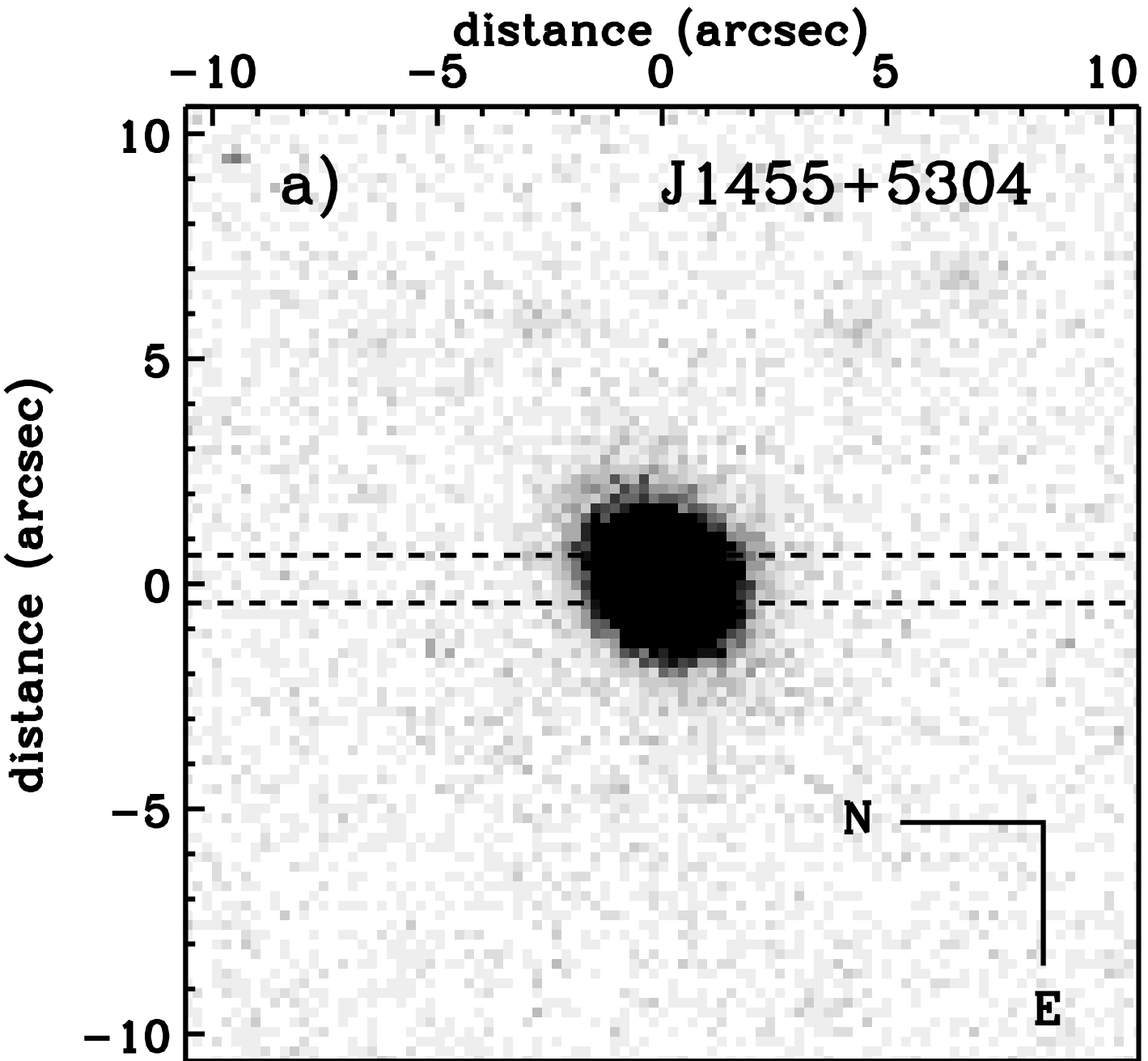}\\
\includegraphics[scale=0.4]{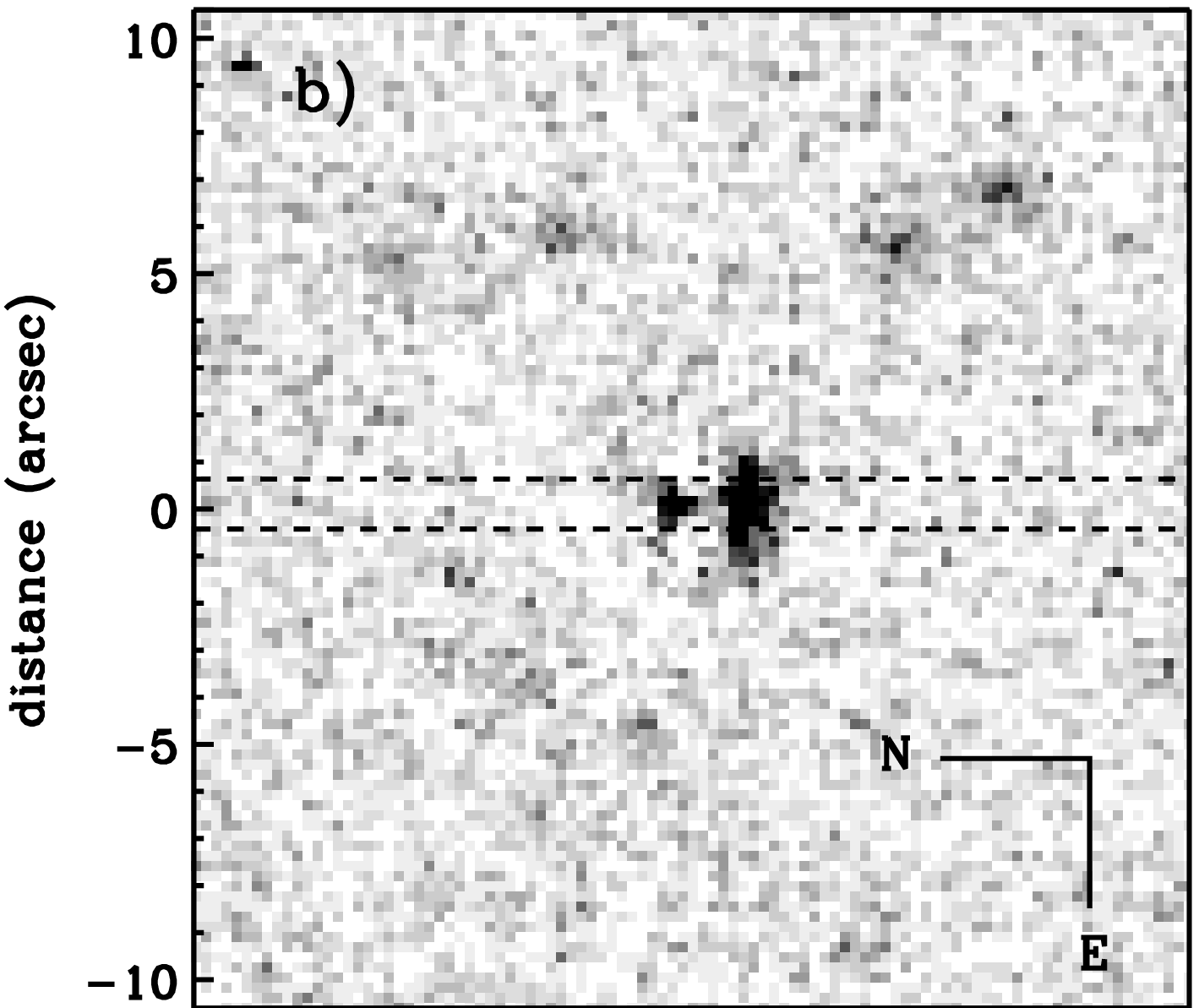}\\
\includegraphics[scale=0.4]{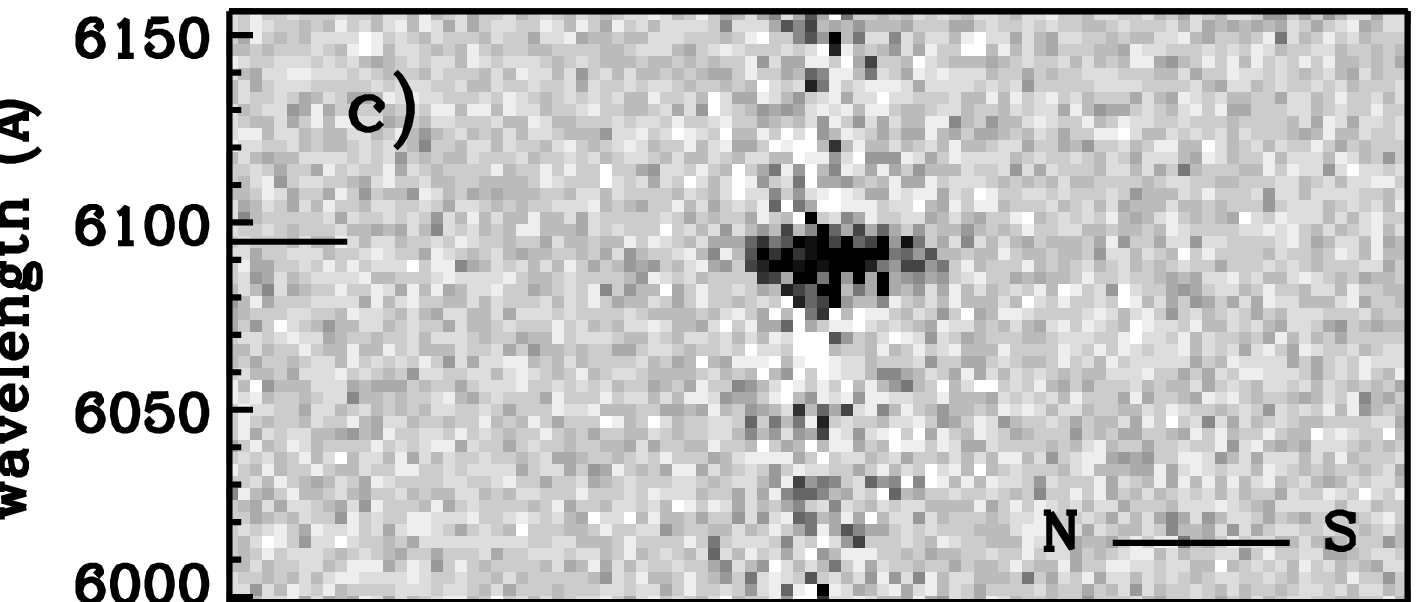}\\
\includegraphics[scale=0.4]{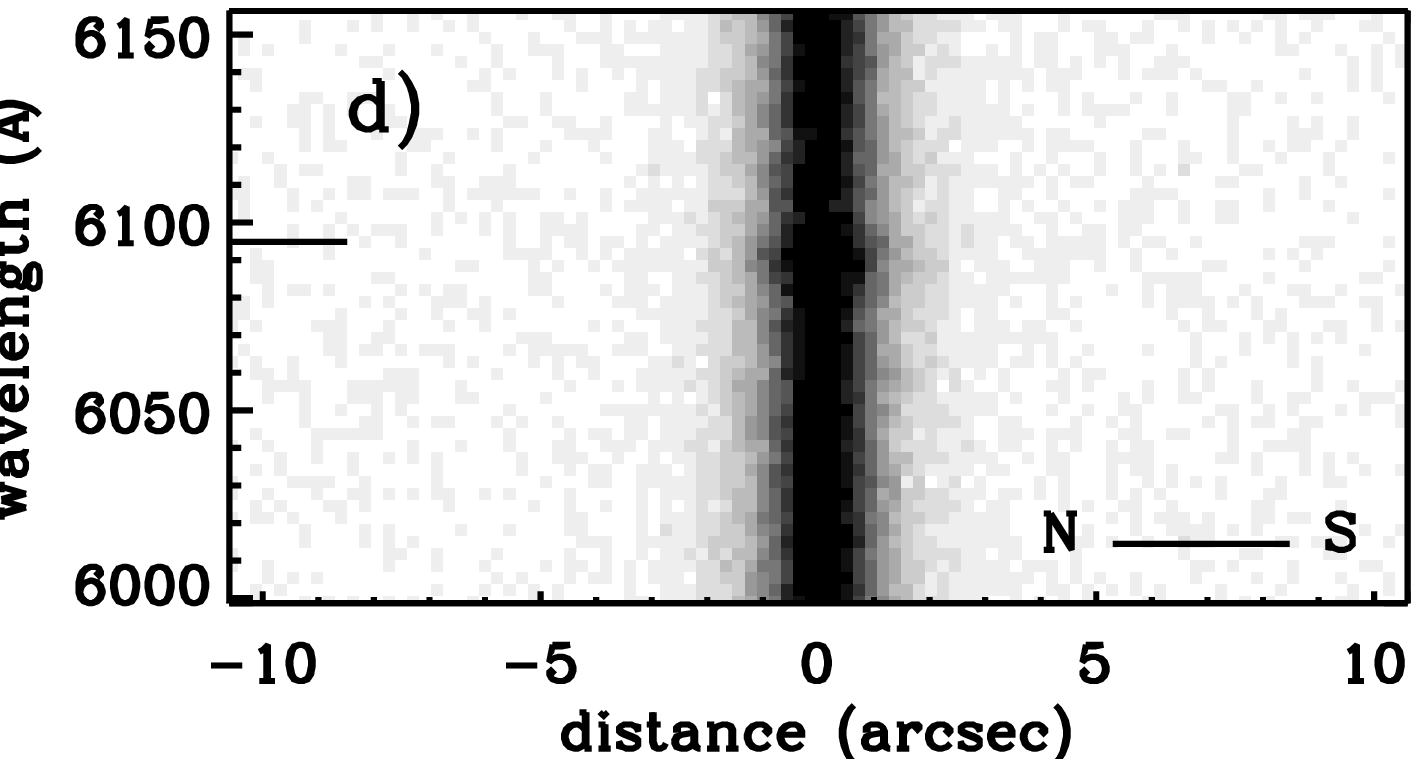}
\caption{\small{Panel a) shows the $u'$-band image of J1455+5304, and panel b) the $u'$-band  residual image after galaxy subtraction, convolved with a gaussian of $\sigma=1$ pixel for display purposes. The dashed lines indicate the position and width of the slit used to take long-slit spectra of the target. Panel c) shows a section of the 2D spectrum of J1455+5304 with the [\ion{O}{2}] $\lambda\lambda$3727 emission line of the background galaxy, at $\lambda=6095$~\AA, after skyline and galaxy subtraction; panel d) shows the same part of spectrum after skyline subtraction only.} 
\label{fig_1455}}
\end{figure}

%FIGURE J1625+2818

\begin{figure}
\centering
\includegraphics[scale=0.4]{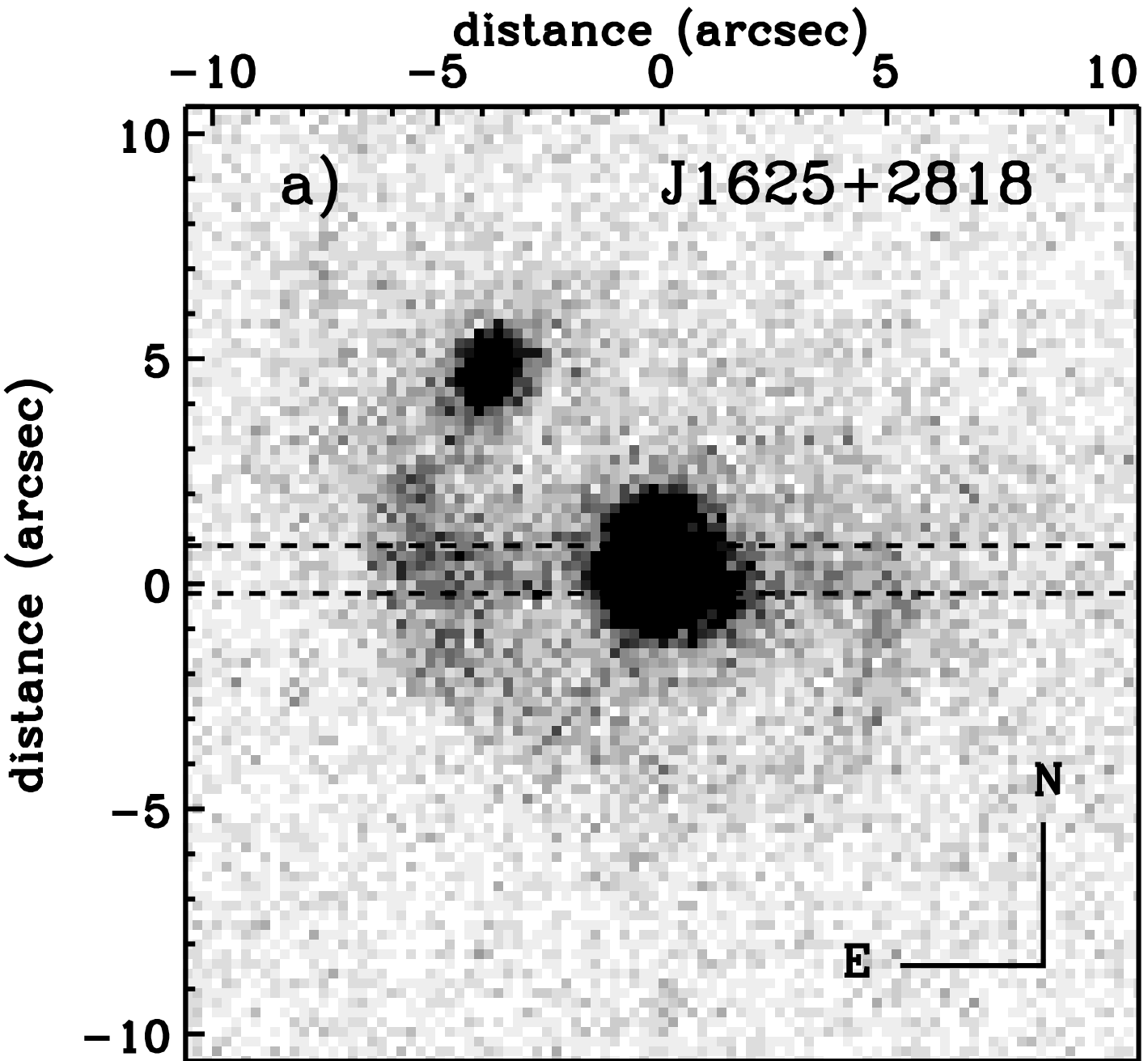} \hspace{1.cm}
\includegraphics[scale=0.4]{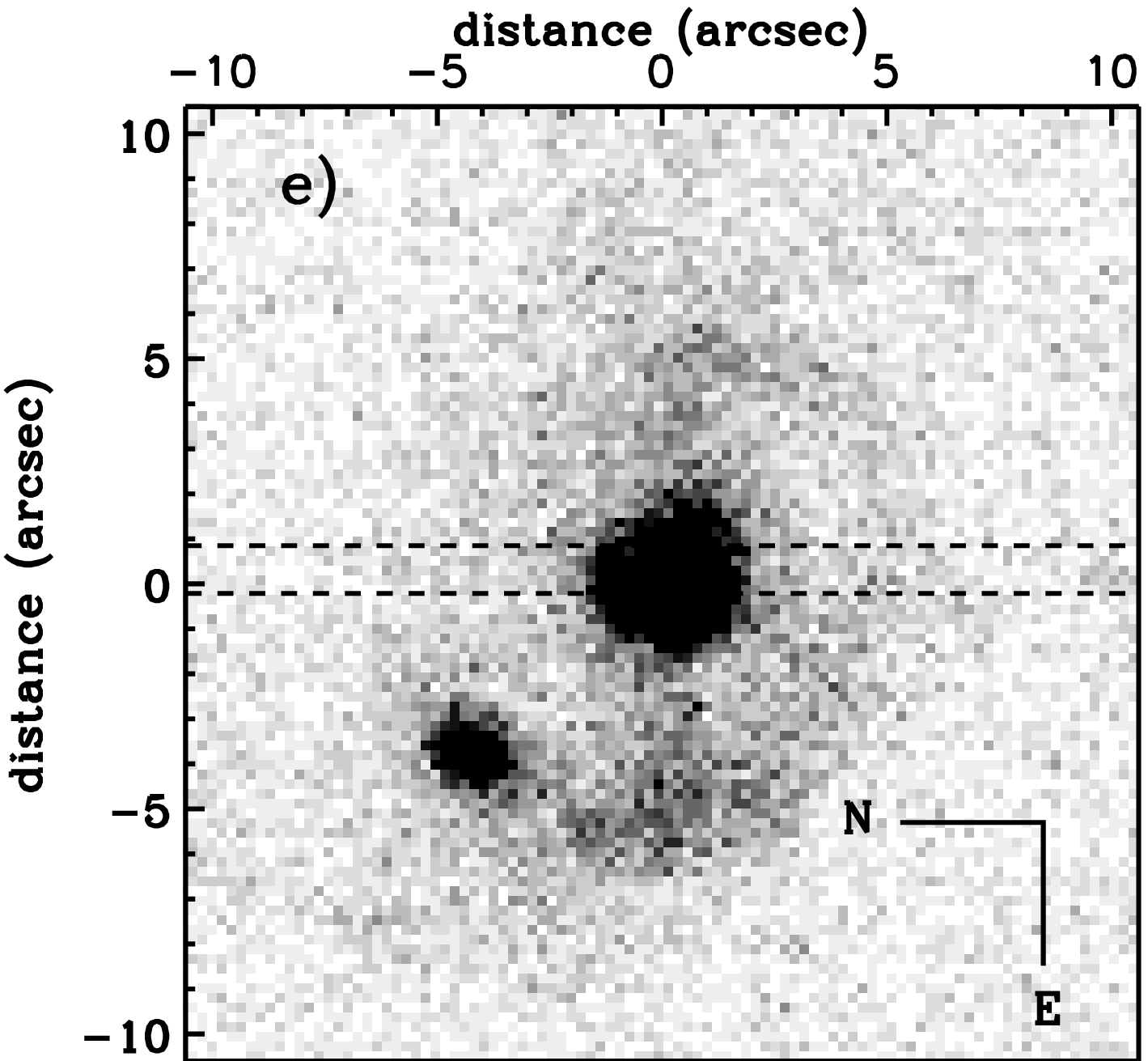}\\
\includegraphics[scale=0.4]{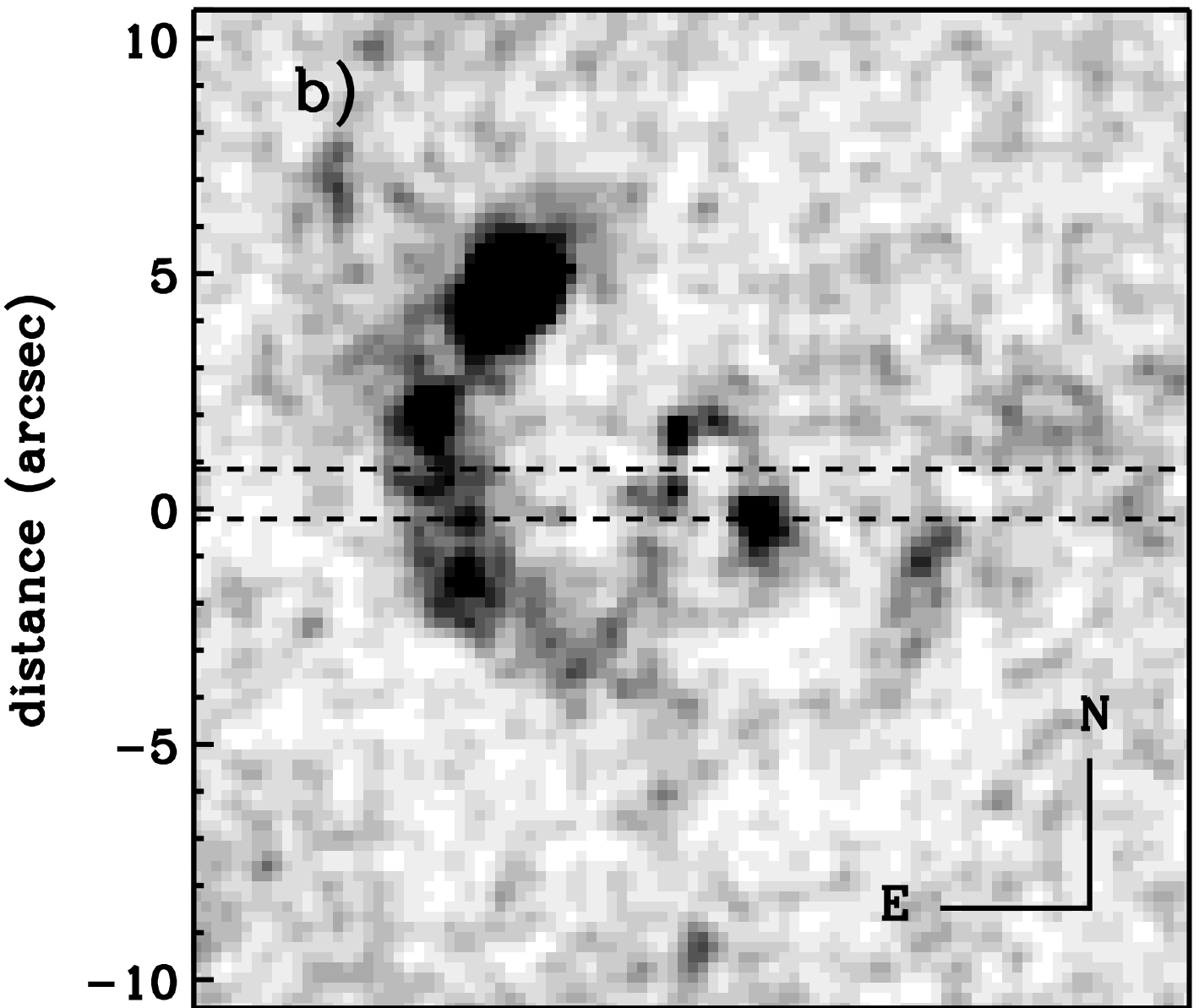} \hspace{1.cm}
\includegraphics[scale=0.4]{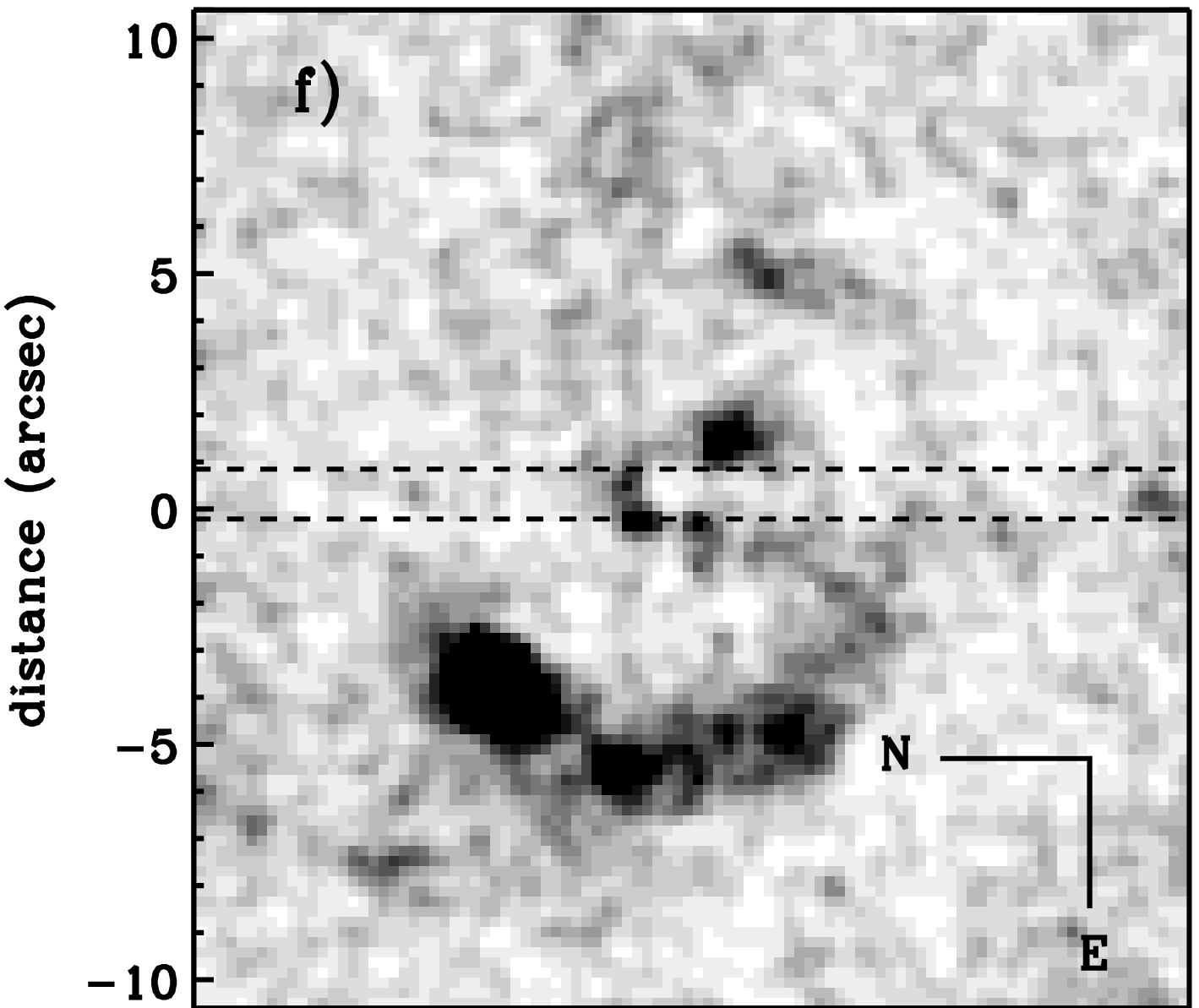}\\
\includegraphics[scale=0.4]{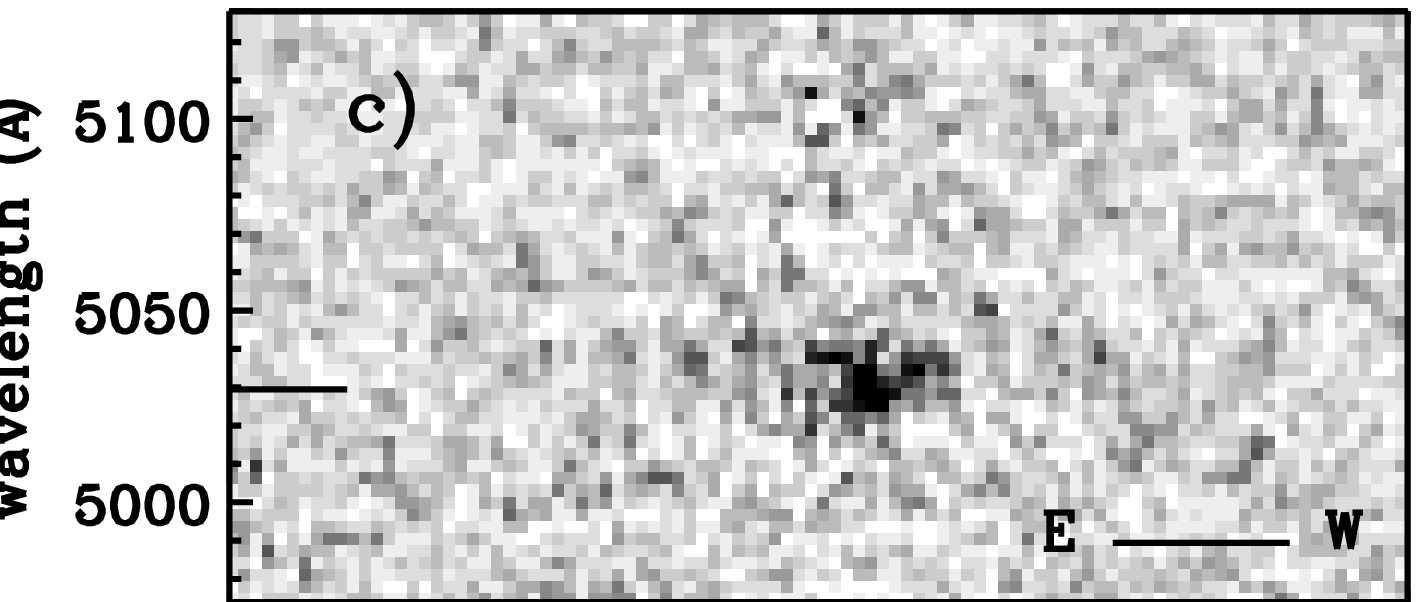} \hspace{1.cm}
\includegraphics[scale=0.4]{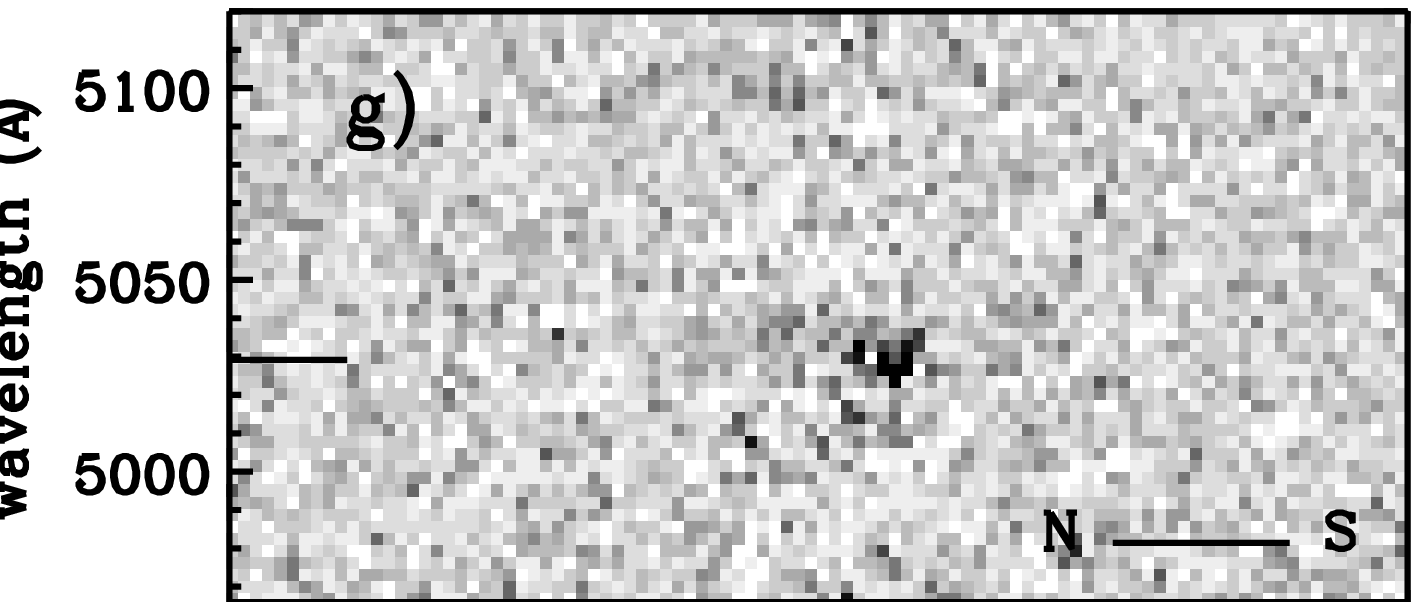}\\
\includegraphics[scale=0.4]{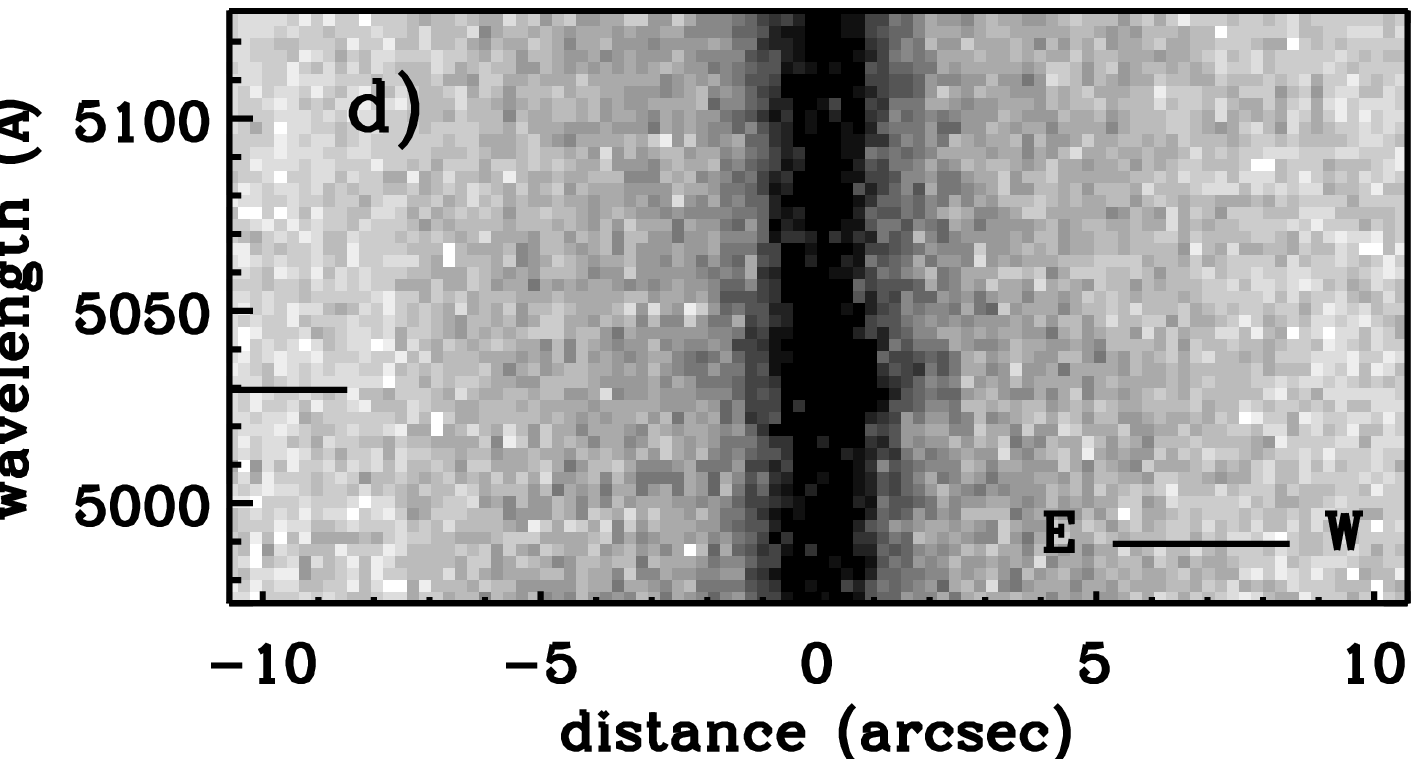} \hspace{1.cm}
\includegraphics[scale=0.4]{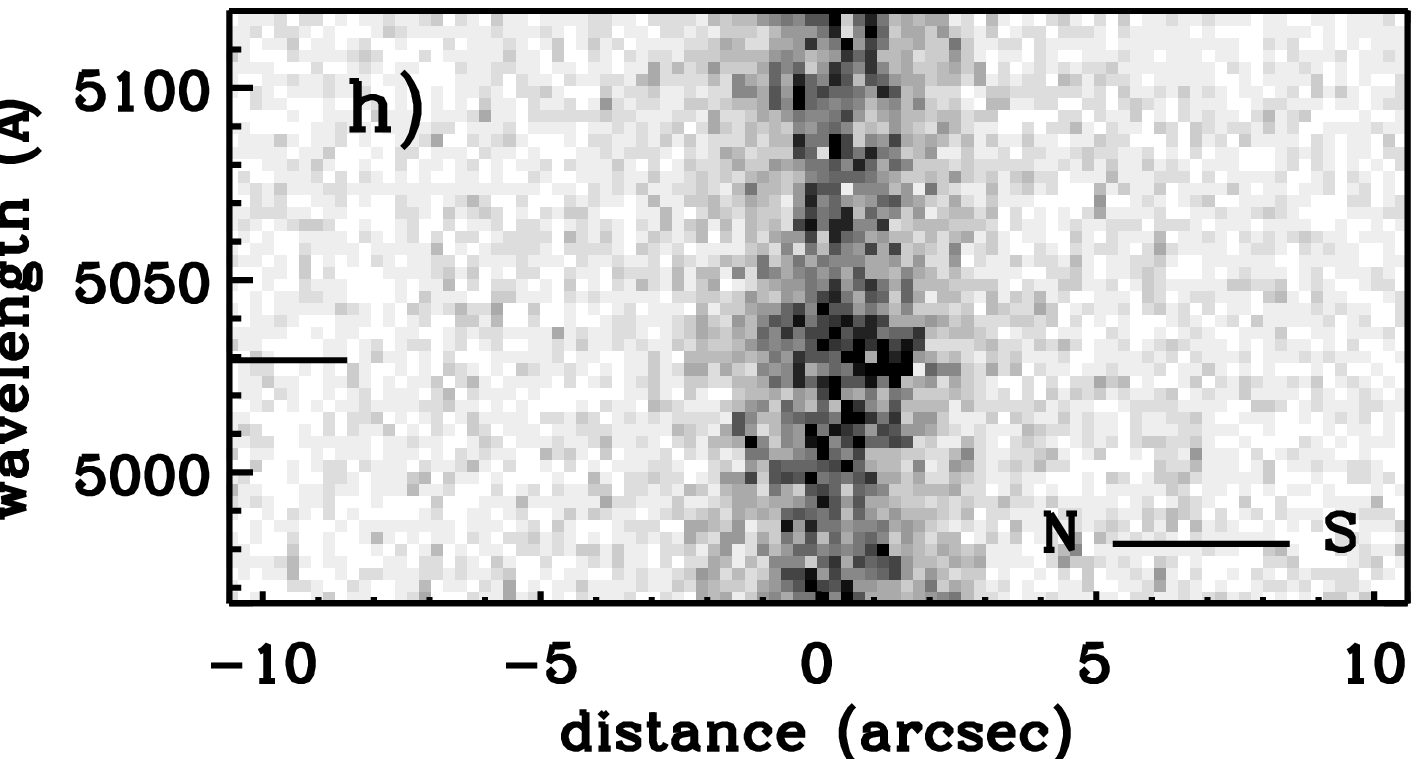}\\
\caption{\small{Panels a) and e) show the $u'$-band image of J1625+2818. Panels b) and f) show the $u'$-band  residual image after galaxy subtraction, convolved with a gaussian of $\sigma=2$ pixels for display purposes. The dashed lines indicate the position and width of the slit used to take respectively the east-west and  north-south long-slit spectra of the target. Panels c) and g) show a section of, respectively, the east-west and north-south 2D spectra of J1625+2818, with the [\ion{O}{2}] $\lambda\lambda$3727 emission line of the background galaxy at $\lambda=5029$~\AA. The spectra were skyline- and galaxy-subtracted, and convolved with a gaussian of $\sigma=1$ pixel for display purposes. Panel d) and h) show the same part of the spectra after skyline subtraction only.} } 
\label{fig_1625}
\end{figure}

%oooooooooooooooooooooooooooooooooooooooooooooooooo

\subsubsection{Galaxy modeling}\label{sec_modeling}

The surface-brightness profile of the galaxies were modeled  using the publicly available software GALFIT \citep{peng02}. We used a  S\'{e}rsic function \citep{sersic63,sersic68}, that is, a seven parameter model (x, y, orientation, ellipticity, S\'{e}rsic index $n$, scale radius and total magnitude), convolved with the point spread function, as defined from the stars in the field. The advantage of fitting a S\'{e}rsic function is to recover a large range of galaxy profiles depending on the index $n$, for example, from the de Vaucouleurs profile ($n=4$) to the exponential disk ($n=1$).

In the $u'$ band, we use from one to three S\'{e}rsic components in our models, depending on each case, in order to get a proper subtraction of the galaxy to identify the potential  lensed images in the residuals.  During this process, any strong emission from the candidate lensed background source was masked out to avoid skewing the fit to the lensing galaxy. The $u'$-band images and residual images are presented in Figs.\ \ref{fig_0812} to \ref{fig_1625} (\textit{a} and \textit{b}). The results are summarized case by case in \S~\ref{sec_results}.

The  data for the lensing galaxies in the $g'$, $r'$, and $i'$-bands were modeled by only one S\'{e}rsic component, as  here we are  only interested in the morphology of the main component. This part is treated in more detail in \S~\ref{sec_morphology}.

\subsection{Long-slit spectroscopy}

\subsubsection{Observations}

Long-slit spectroscopy of the lens candidates was obtained at the  NOT with the Andalucia Faint Object Spectrograph and Camera (ALFOSC) during 2007 May 7-11 under variable weather conditions. We used a slit width ranging from $1\arcsec$ to $1.3\arcsec$ depending on the seeing. 
We chose the grism \#4 which covers a wavelength range 3200 \AA\ to 9100 \AA\ matching that of the SDSS spectra. This was a crucial point in order to confirm the presence of the three or more emission lines detected in the SDSS spectra of our gravitational lens candidates. However, it was at the expense of the resolving power of R$\sim$710, which is lower than that used for the SDSS spectra.  The grism \#4 has a blaze angle of $-1.5887 \deg$ and a blaze wavelength of $496$ nm.

Seeing conditions during the observations are detailed in Table \ref{tab_observations}; the mean seeing during the spectroscopic run was $0.89\arcsec$.

The slit position was determined based on  the results of the galaxy subtraction presented in \S~\ref{sec_results}. For simplicity, we used only two slit positions (north-south and east-west) to cover the potential lensed images.

\subsubsection{Data reduction}

The spectroscopic data were processed using standard data reduction methods with the Image Reduction and Analysis Facility (IRAF), and using L.A.Cosmic for the cosmic-ray removal. The 2D spectra were bias subtracted, flat-fielded, and dispersion corrected using He and Ne arc spectra. The skylines were subtracted by fitting a low order cubic spline interpolation to each row and subtracting it from the 2D spectrum. The galaxy contribution was subtracted  using a similar technique along the columns of the galaxy spectrum. 1D spectra were extracted from the calibrated, skyline-subtracted 2D spectra, in order to measure with better precision the wavelength of the background source emission lines, and the redshift of the background source.

\subsection{Combined results from imaging and long-slit spectroscopy}\label{sec_results}

We present in a combined figure, for each lens candidate, the $u'$ band photometry and the long-slit spectroscopy, showing both the observations and the residuals after galaxy subtraction (see Figs.\ \ref{fig_0812} to \ref{fig_1625}). The optical images and the 2D spectra are presented with an identical spatial scale for the horizontal axis. The slit is indicated for each galaxy by two dashed lines in the $u'$-band images. In the 2D spectra, the spatial axis corresponds to the position along the slit, and  the wavelength axis corresponds to a  wavelength range around the expected [\ion{O}{2}] $\lambda\lambda$3727 emission line of the background galaxy. When spectroscopic observations of a lens candidate were obtained in two orientations  of the slit, the two spectra are presented  in the same figure, with the galaxy image rotated to present the slit horizontally, as indicated by the compass on each image. Below we discuss the lens candidates case by case, including the photometric data, spectroscopic data and combined results.

\subsubsection{J0812+5436}

\textit{Imaging}--The galaxy-subtracted $u'$-band image shows two bright residuals linked by an arc around the north of the galaxy, suggestive of lensed images (see Fig.\ \ref{fig_0812}, \textit{a} and \textit{b}). The two structures at the south  of the galaxy are difficult to identify, but the nearest to the galaxy might be a lensed image. We planned spectroscopic observations in two positions of the slit: an orientation east-west passing through the side images, and an orientation north-south to study the southern objects. Unfortunately, bad weather conditions prevented us from taking the north-south spectrum.

\textit{2D spectrum}--We study the 2D spectrum around $\lambda=5062$ \AA, the wavelength of the expected [\ion{O}{2}] $\lambda\lambda$3727 emission line. After galaxy subtraction, two bright residual images appear on each side of the galaxy at the wavelength $\lambda=5060$ \AA\ (see Fig.\ \ref{fig_0812}, \textit{c} and \textit{d}). The west image is at $0.7~\pm1.1\arcsec$ and the east image at $1.3~\pm0.6\arcsec$ from the center of the galaxy spectrum. They correspond to the positions through the slit of the images detected in the $u'$ band. Although the two images in the spectrum are not neatly separated, the faint signal between them corresponds to the arc detected in the $u'$ band, thus confirming the strong lens nature of the candidate.

\textit{Conclusions}--We confirmed that J0812+5436 is a strong gravitational lens. The Einstein radius is $r_{E}=1.8~\pm0.4\arcsec$ ($3.9~\pm0.9$ kpc), based on the half distance between the two confirmed lensed images, in the $u'$ band image. The redshift of the lensed galaxy is $z_B=0.357$, as measured from the one dimensional spectrum. However, it is not possible to detail the nature of the residual images at the south of the galaxy, as we lack spectroscopic observations for them. The object nearest to the galaxy might be a lensed image of the same background source, while the furthest object might be unrelated.

\subsubsection{J0903+5448}

\textit{Imaging}--The galaxy-subtracted $u'$-band image shows two residuals suggestive of lensed images (see Fig.\ \ref{fig_0903}, \textit{a} and \textit{b}). The north image is at $0.6~\pm0.6\arcsec$ and the south image at $0.9\pm0.3\arcsec$ from the galaxy center. We obtained spectroscopic observations with the slit in a north-south orientation, going through the two images.

\textit{2D spectrum}--We study the 2D spectrum around $\lambda=6295$ \AA, the wavelength of the expected [\ion{O}{2}] $\lambda\lambda$3727 emission line. After galaxy subtraction,   a faint and continuous signal across the galaxy width appears at  $\lambda=6298$~\AA\ (see Fig.\ \ref{fig_0903}, \textit{c} and \textit{d}), extending from $1.0\arcsec$  north to $2.0\arcsec$ south of the galaxy center. However, we do not detect distinctly an emission line from the background galaxy, nor if it is multiply imaged. The proximity of a strong skyline at $\lambda=6300$ \AA, the  subtraction of which is  bringing noise in this region of the spectrum, makes the study difficult. The [\ion{O}{3}] 5007 emission line would in theory bring better information on the lens candidate, presenting a better $S/N$ ratio (see Fig.\ \ref{fig_emlines}), but in practice the fringing in the red part of our spectra makes it even more difficult to study.

\textit{Conclusions}--The presence of a background galaxy near J0903+5448 is confirmed from the SDSS (Princeton/MIT spectroscopy) one dimensional spectrum of the target, at a redshift $z_B=0.689$. However, we do not detect properly the emission lines from the background galaxy in our own spectrum, due to the vicinity of a strong sky line near to the [\ion{O}{2}] $\lambda\lambda$3727 emission line, and the fringing in the red part of the spectrum where the other strong emission lines of the background galaxy are found. Although the two residual images appearing in the $u'$ band observation are very suggestive of lensing, better spectroscopic observations of the  [\ion{O}{3}] 5007 emission line would be necessary to confirm the lens nature of the candidate.

\subsubsection{J0942+6111}

\textit{Imaging}--The galaxy-subtracted $u'$-band image shows a residual image  at the east of the galaxy center  (see Fig.\ \ref{fig_0942}, \textit{a} and \textit{b}). We planned spectroscopic observations with the slit in the east-west orientation, passing through the potential lensed image and the center of the galaxy.

\textit{2D spectrum}--We study the 2D spectrum around $\lambda=6397$ \AA,  the wavelength of the expected [\ion{O}{2}] $\lambda\lambda$3727 emission line. After galaxy subtraction, one bright residual image appears at $1.1~\pm0.8\arcsec$ east of the galaxy center, at the wavelength $\lambda=6396$~\AA\ (see Fig.\ \ref{fig_0942}, \textit{c} and \textit{d}).

\textit{Conclusions}--We confirmed the presence of an image of a background source near J0942+6111. The redshift of the background galaxy is $z_B=0.716$, as measured from the one dimensional spectrum. We can not confirm that the system is a lens, as we detect only one image of the background source.

\subsubsection{J1150+1202}

\textit{Imaging}--The galaxy-subtracted $u'$-band image shows three bright residual images in a west, north-east and south-east configuration (see Fig.\ \ref{fig_1150}, \textit{a} and \textit{b}). We planned spectroscopic observations with east-west and  north-south orientations of the slit to cover the three potential lensed images. Unfortunately, bad weather conditions prevented us from taking the north-south spectrum.

\textit{2D spectrum}--We study the 2D spectrum around $\lambda=5710$ \AA, the wavelength of the expected [\ion{O}{2}] $\lambda\lambda$3727 emission line. After galaxy subtraction, one bright residual image appears  at $0.8~\pm0.8\arcsec$ west of the galaxy center, at the wavelength $\lambda=5704$~\AA\ (see Fig.\ \ref{fig_1150}, \textit{c} and \textit{d}). The emission line is superposed to the \ion{Mg}{1} 5175 absorption line (in white in Fig.\ \ref{fig_1150}, \textit{c}) of the foreground galaxy at redshift $z_f=0.105$. A faint extension of the [\ion{O}{2}] $\lambda\lambda$3727 emission line is seen at the east of the galaxy, which may indicate a second image of the emission line. Comparing to the $u'$ band, the bright image of the emission line corresponds to the west image, and the faint extension corresponds to the north-east image. As the west image appears brighter than the other image in the $u'$ band observations, this explains that it is less affected by the presence of the \ion{Mg}{1} 5175 absorption line.

\textit{Conclusions}--We confirmed the presence of an image of a background source near J1150+1202. The redshift of the background galaxy is $z_B=0.531$, as measured from the one dimensional spectrum. We can not confirm that the source is multiply imaged, but the presence of a faint extension to the [\ion{O}{2}] $\lambda\lambda$3727 emission line  concording with the north-east image in the $u'$-band residuals gives a strong indication for multiple images.  It is not possible to detail the nature of the residual image at the south of the galaxy, as we lack spectroscopic observations for it. However, the presence of three images in the $u'$-band residual image, concording with one and possibly two spectroscopically confirmed images of the background source, while the third image is not proven spectroscopically, indicates the system is probably a strong gravitational lens.

\subsubsection{J1200+4014}

\textit{Imaging}--The galaxy-subtracted $u'$-band image shows three residual images in a north-east, north-west and south configuration, suggestive of lensed images (see Fig.\ \ref{fig_1200}, \textit{a} and \textit{b}). Moreover, the three images seem to be linked by a faint ring, suggestive of an Einstein ring superposed to the lensed images. We obtained spectroscopic observations in the east-west and north-south orientations, covering the three potential lensed images.

\textit{2D spectrum}--We study the 2D spectra around $\lambda=5871$ \AA, the wavelength of the expected [\ion{O}{2}] $\lambda\lambda$3727 emission line. After galaxy subtraction in the east-west spectrum, two bright residual images appear on each side of the galaxy at the wavelength $\lambda=5870$ \AA\ (see Fig.\ \ref{fig_1200}, \textit{c} and \textit{d}).
In spite of the presence of a strong skyline at $\lambda=5890$ \AA, the subtraction of which induces noise in this part of the spectrum, the presence of the two residual images at the wavelength of the expected [\ion{O}{2}] $\lambda\lambda$3727 emission line gives a strong indication for multiple images of the background source.
The east image is at $1.7~\pm0.7\arcsec$ and the west image at $1.0~\pm0.6\arcsec$ from the center of the galaxy spectrum.   Comparing the position of the two images to the $u'$-band residual image, it appears these two images correspond to the north-east and north-west images. After galaxy subtraction in the north-south spectrum, we detect a faint image of the source on the north side of the galaxy at the wavelength $\lambda=5870$ \AA, but we find no evidence for the bright south image, which is therefore not belonging to the same source (see Fig.\ \ref{fig_1200}, \textit{g} and \textit{h}). 

\textit{Conclusions}--We confirmed the presence of a background galaxy near J1200+4014, with a strong indication that it is multiply imaged, and therefore, that the system is a lens. The redshift of the background galaxy is $z_B=0.575$, as measured from the one dimensional spectrum. The third image, at the south of the galaxy, is not an image of the identified background source, but probably an unrelated object which is almost superposed to the foreground galaxy. ˙

\subsubsection{J1356+5615}

\textit{Imaging}--The galaxy-subtracted $u'$-band image shows two residual images at the east and south of the galaxy (see Fig.\ \ref{fig_1356}, \textit{a} and \textit{b}). There is also a central image which might be an artifact due to the difficulty of fitting the central pixels. We obtained spectroscopic observations in the east-west and north-south orientations of the slit, covering the three potential lensed images.

\textit{2D spectrum}--We study the 2D spectra around $\lambda=5962$ \AA, the wavelength of the expected [\ion{O}{2}] $\lambda\lambda$3727 emission line. After galaxy subtraction of each of the east-west and north-south spectra, a residual image appears in each spectrum, respectively at $0.7~\pm0.5\arcsec$ west and $0.7~\pm0.3\arcsec$  north of the center of the galaxy spectrum, at the wavelength $\lambda=5962$ \AA\ (see Fig.\ \ref{fig_1356}, \textit{c}, \textit{d}, \textit{g} and  \textit{h}). These correspond to the central image in the $u'$-band residual image. We do not detect any of the two other images present in the $u'$-band residual image.

\textit{Conclusions}--We confirmed the presence of a central image of a background galaxy detected in both orientations of the slit. The redshift of the background galaxy is $z_B=0.575$, as measured from the one dimensional spectrum.  It is not possible to determine if the images in the two spectra are two detections of the same image, or if there are actually two central images of the background source, or if they form a lensing arc. Therefore we can not conclude whether J1356+5615 is a strong lens or not.

\subsubsection{J1455+5304}

\textit{Imaging}--The galaxy-subtracted $u'$-band image shows two residual images suggestive of lensed images, one on each side of the galaxy center (see Fig.\ \ref{fig_1455}, \textit{a} and \textit{b}). The north image is at $0.5~\pm0.5\arcsec$ and the south image at $1.0\pm0.4\arcsec$ from the galaxy center. We obtained spectroscopic observations with the slit in the north-south orientation, going through the two images.

\textit{2D spectrum}---We study the 2D spectrum around $\lambda=6095$ \AA, the wavelength of the expected [\ion{O}{2}] $\lambda\lambda$3727 emission line. After galaxy subtraction, a residual image extending across the galaxy width appears at the wavelength $\lambda=6090$~\AA\ (see Fig.\ \ref{fig_1455}, \textit{c} and \textit{d}). Although we observe a continuous signal, we can make the hypothesis that there are two unresolved images, and estimate their positions assuming they are separated at the center of the galaxy. This leads to a north image at $0.6~\pm0.6\arcsec$ and a south image at $0.9~\pm0.9\arcsec$ from the center of the galaxy spectrum. This is consistent with the position through the slit of the two images detected in the $u'$ band, thus confirming the strong lens nature of the candidate.

\textit{Conclusion}--We confirmed that J1455+5304 is a strong gravitational lens. The Einstein radius is $r_{E}=0.7~\pm0.4\arcsec$ ($1.4~\pm0.8$ kpc), based on the half distance between the two confirmed lensed images, in the $u'$ band image. The redshift of the background galaxy is $z_B=0.634$, as measured from the one dimensional spectrum.

\subsubsection{J1625+2818}

\textit{Imaging}--The galaxy-subtracted $u'$-band image shows residuals suggestive of lensing: one bright image at the south-west of the bulge, and three smaller images superposed to what seems an arc on the north-east side of the bulge (see Fig.\ \ref{fig_1625}, \textit{a} and \textit{b}). We obtained spectroscopic observations in the east-west and north-south orientations, covering the potential lensed images.

\textit{2D spectrum}--We study the 2D spectra around $\lambda=5029$ \AA, the wavelength of the expected [\ion{O}{2}] $\lambda\lambda$3727 emission line. After galaxy subtraction in the east-west spectrum, a residual image appears at the wavelength $\lambda=5032$ \AA, at $0.8~\pm0.3\arcsec$ west from the center of the galaxy spectrum (see Fig.\ \ref{fig_1625}, \textit{c} and \textit{d}). It corresponds to the position of the south-west image in the $u'$-band residual image. After galaxy subtraction in the north-south spectrum, a residual image appears at the wavelength $\lambda=5032$ \AA, at $0.9~\pm0.5\arcsec$ south from the center of the galaxy spectrum (see Fig.\ \ref{fig_1625}, \textit{g} and \textit{h}). When compared to the $u'$-band residual image, the position of the image in the north-south spectrum does not correspond to the position of the north-east images which the slit was intended to probe, but corresponds to the position of the south-west image at the edge of the slit. However, we can see on the $u'$ band image that the south-west image is brighter and larger than the other residual images. Moreover, the spectrum of the foreground galaxy shows many emission lines: this makes the peak of the [\ion{O}{2}] $\lambda\lambda$3727   emission line of the background galaxy difficult to detect among the peaks of the foreground galaxy (see Fig.\ \ref{fig_emlines}). This could explain that we detect only the brightest of the images of the background emission line, while the fainter images would be at the level of the noise occasioned by the foreground galaxy.

\textit{Conclusions}--We confirmed the presence of an image of a background source near J1625+2818. The redshift of the background galaxy is $z_B=0.350$, as measured from the one dimensional spectrum. We can not confirm that the source is multiply imaged, as we detect spectroscopically only the brightest image. The remaining residual images might be too faint to be detected in our spectra. However, the configuration of the residuals in the $u'$ band image points toward a good probability that the system is a strong gravitational lens.

%oooooooooooooooooooooooooooooooooooooooooooooooooooooooooo

\begin{figure}
\centering
\includegraphics[scale=0.3]{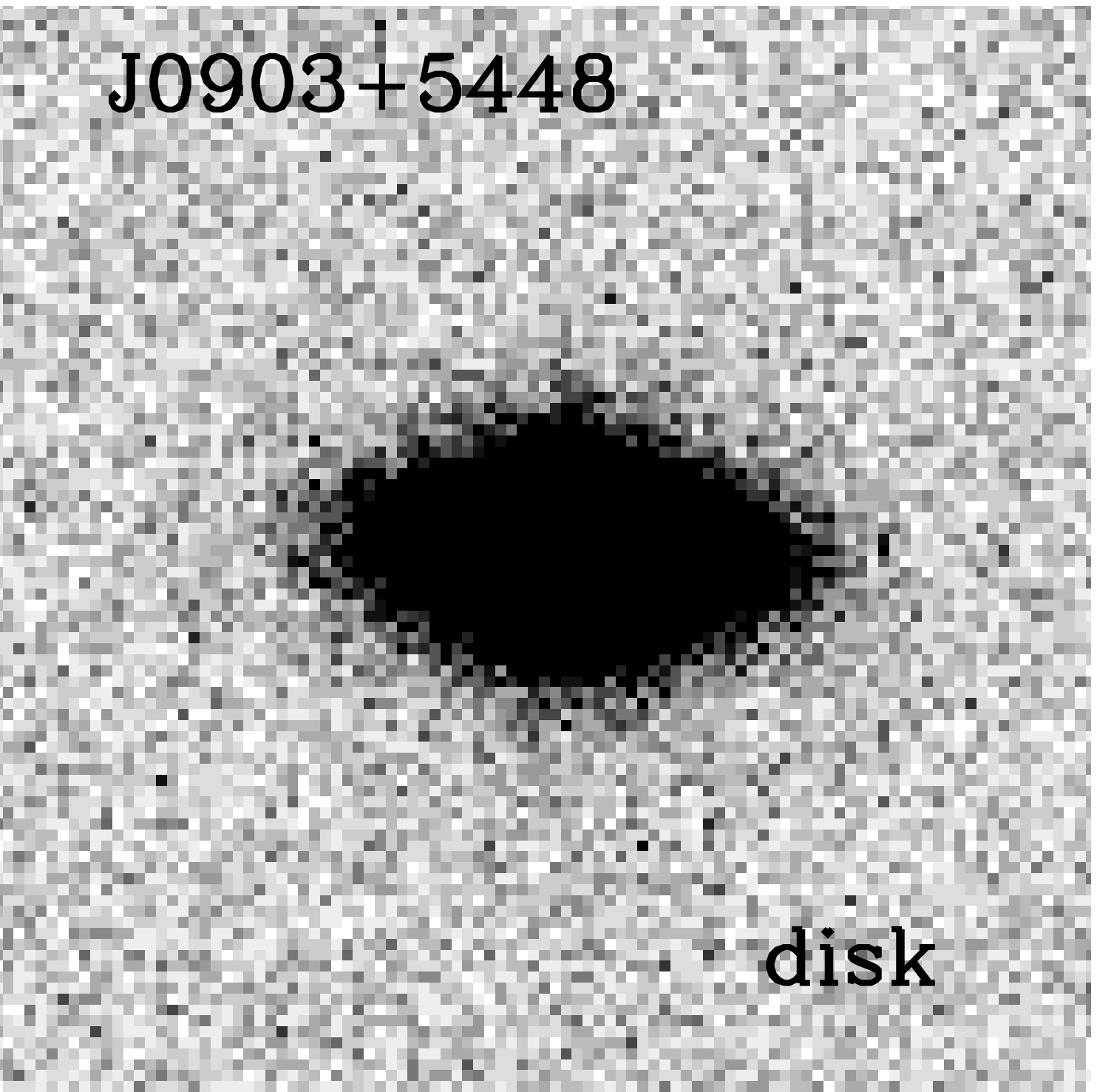}\\
\includegraphics[scale=0.3]{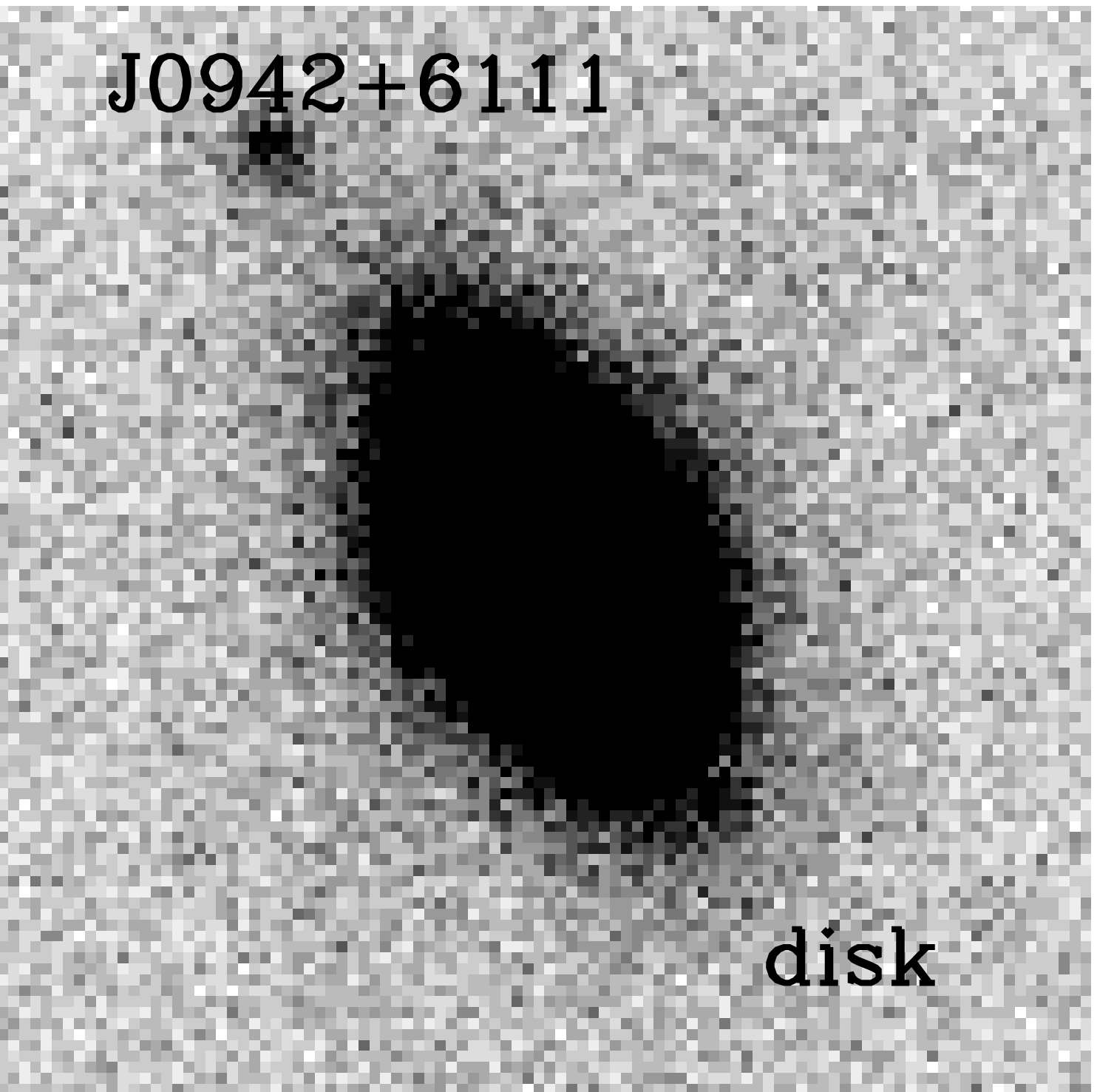}\\
\includegraphics[scale=0.3]{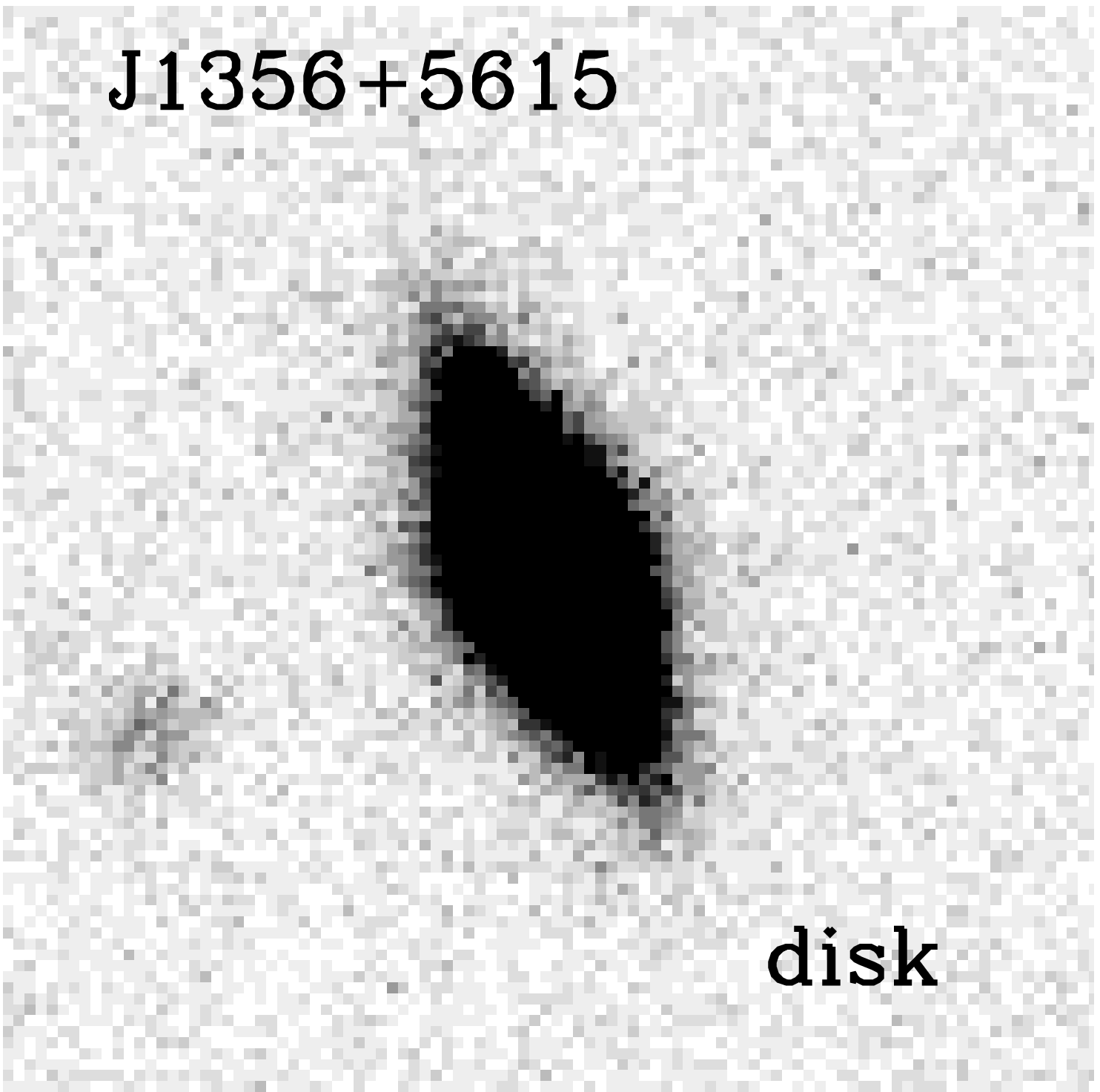}\\
\includegraphics[scale=0.3]{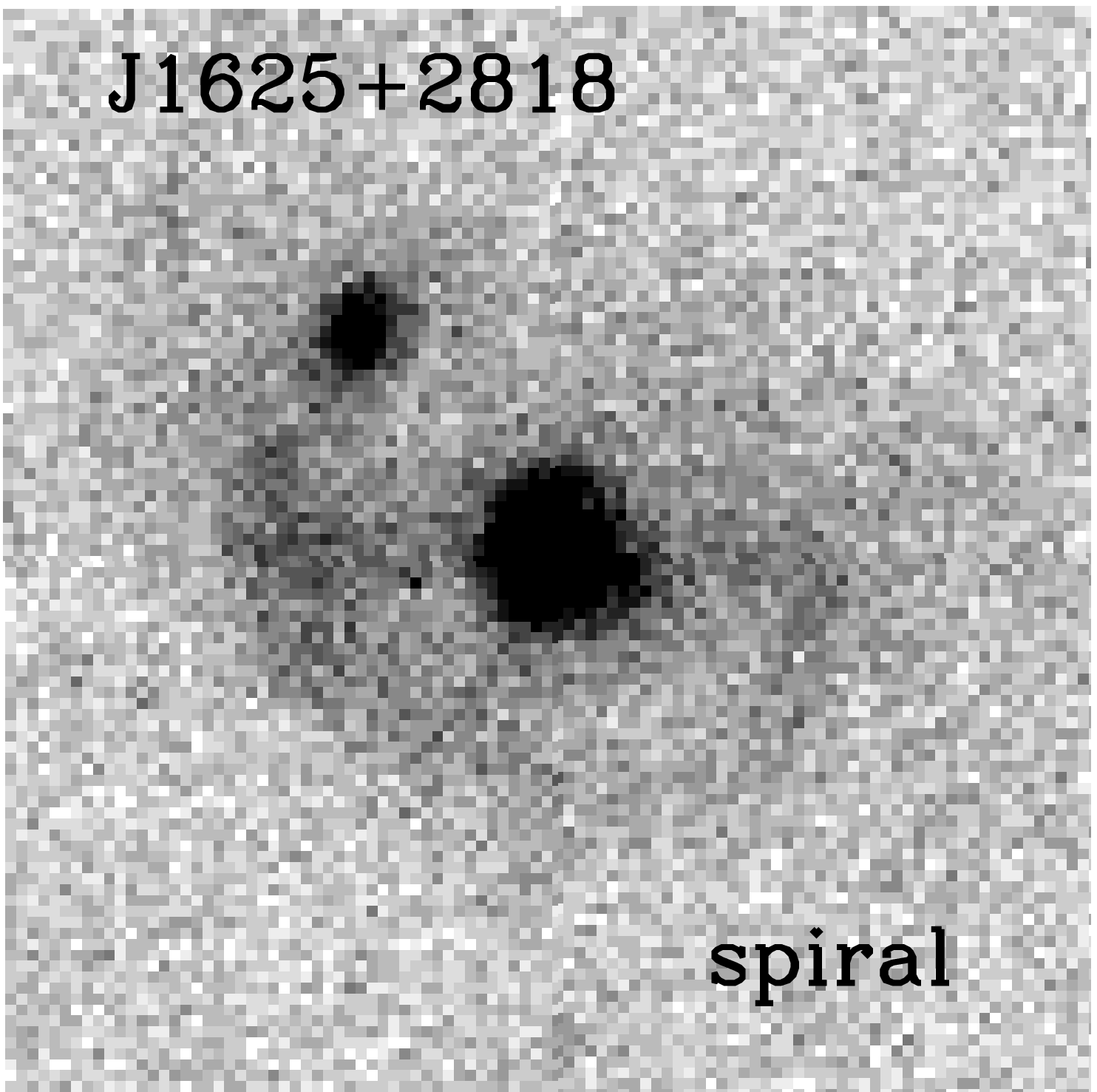}
\caption{Imaging  in the $r'$ band of the targets which are obvious disk galaxies (J1625+2818 is shown in the $u'$ band). The size of the images is $21.7\arcsec \times 21.7\arcsec$.
\label{fig_morph_disk}}
\end{figure}

\begin{figure}
\centering
\includegraphics[scale=0.3]{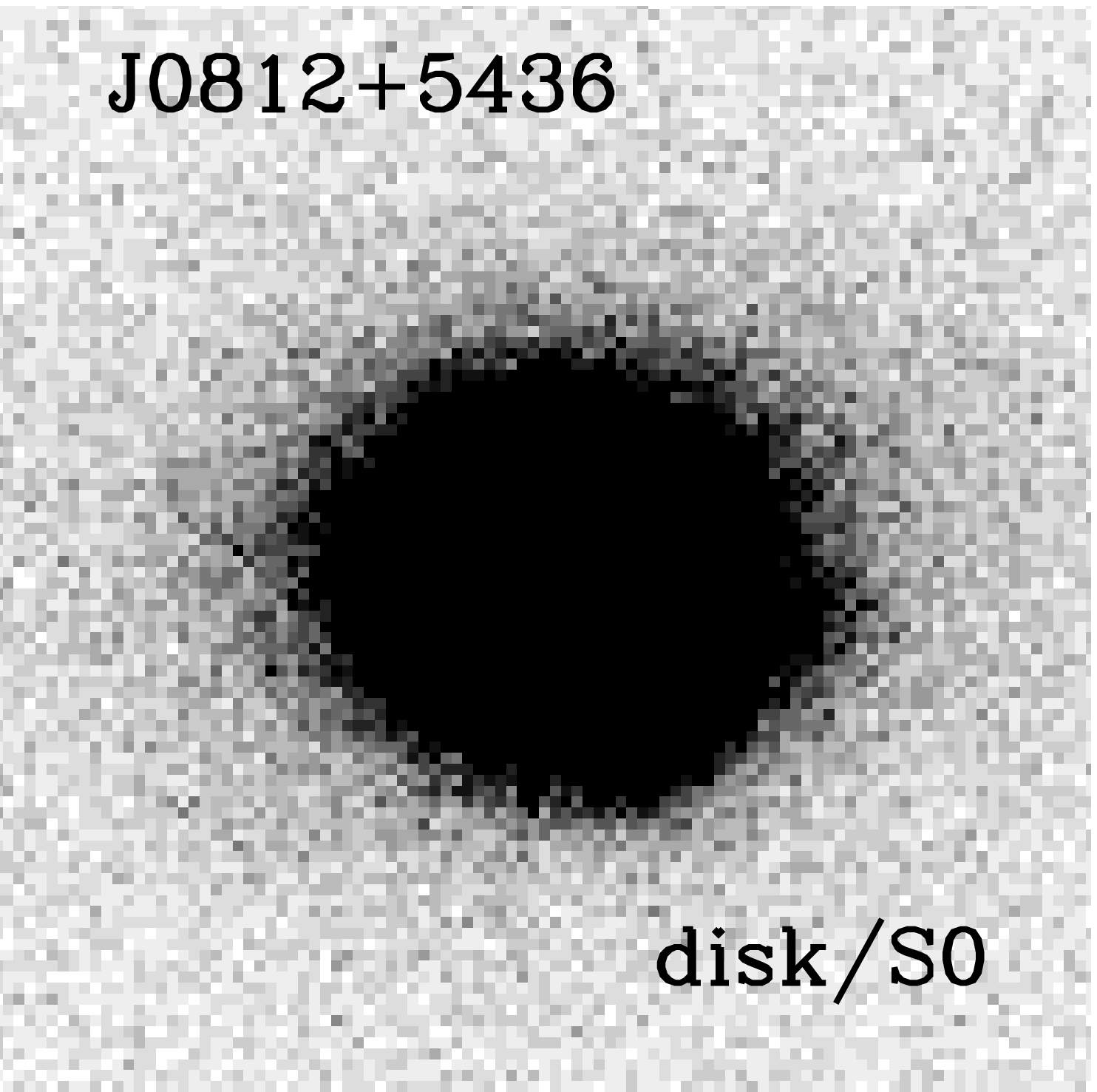}\\
\includegraphics[scale=0.3]{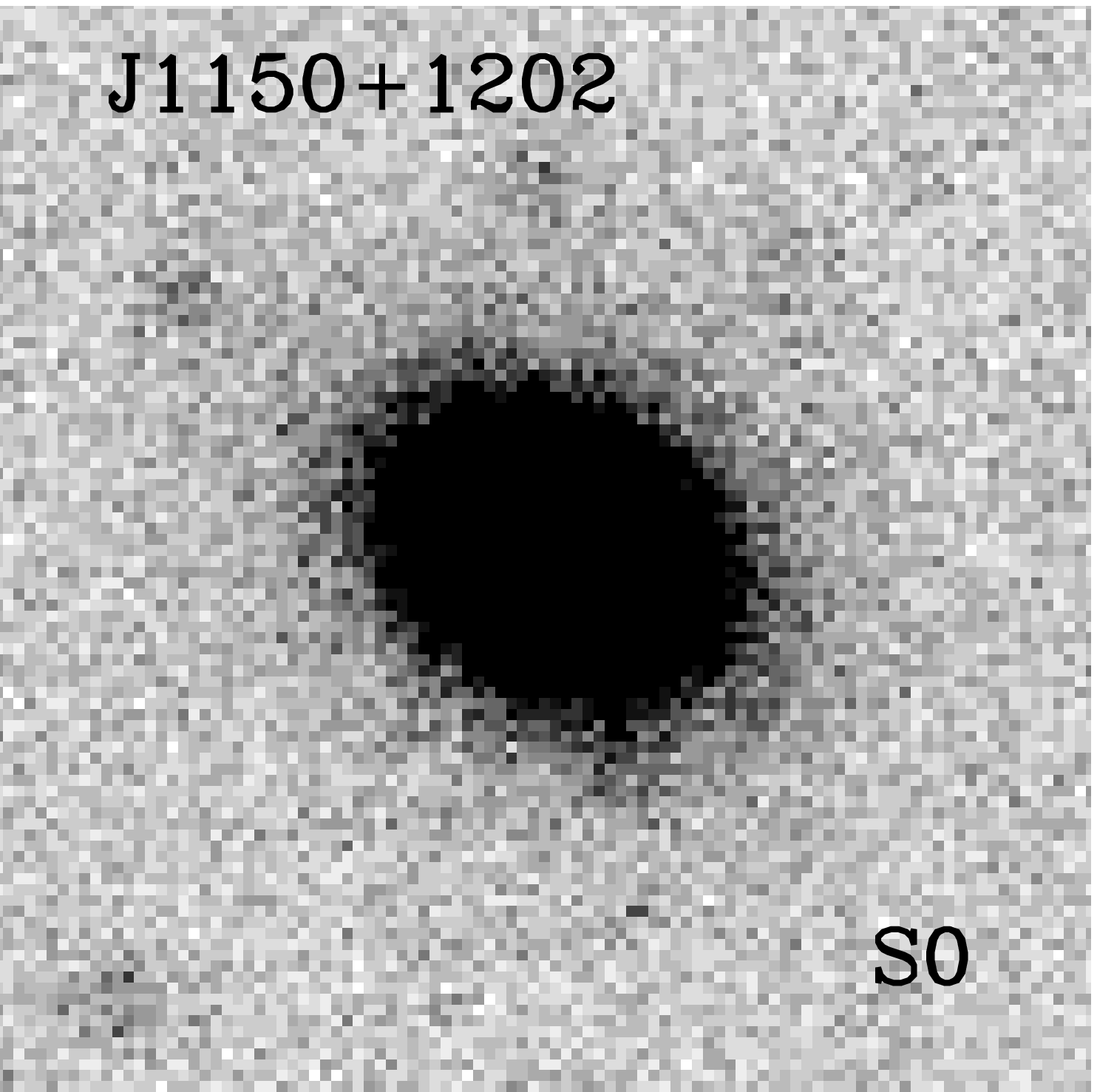}\\
\includegraphics[scale=0.3]{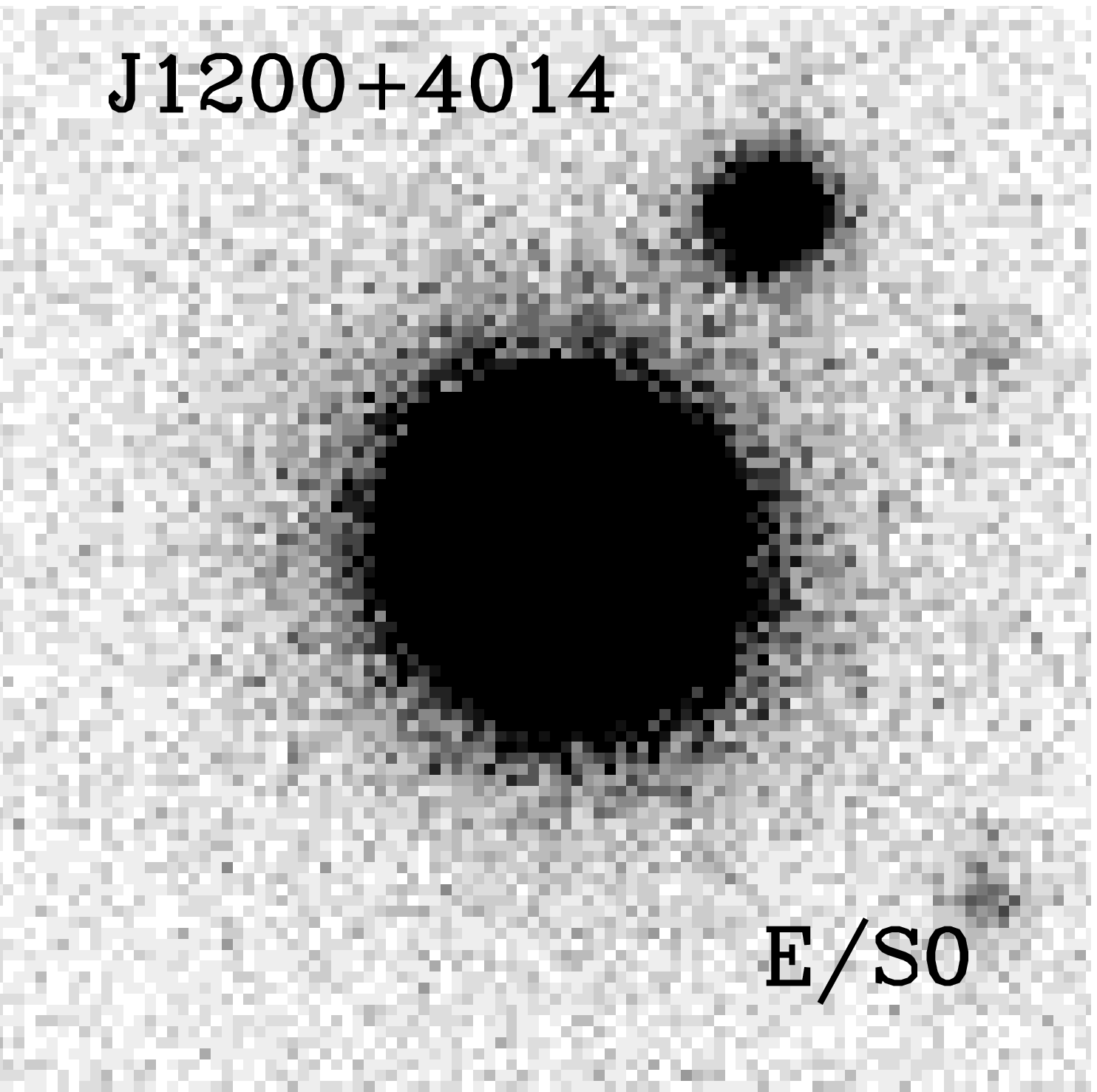}\\
\includegraphics[scale=0.3]{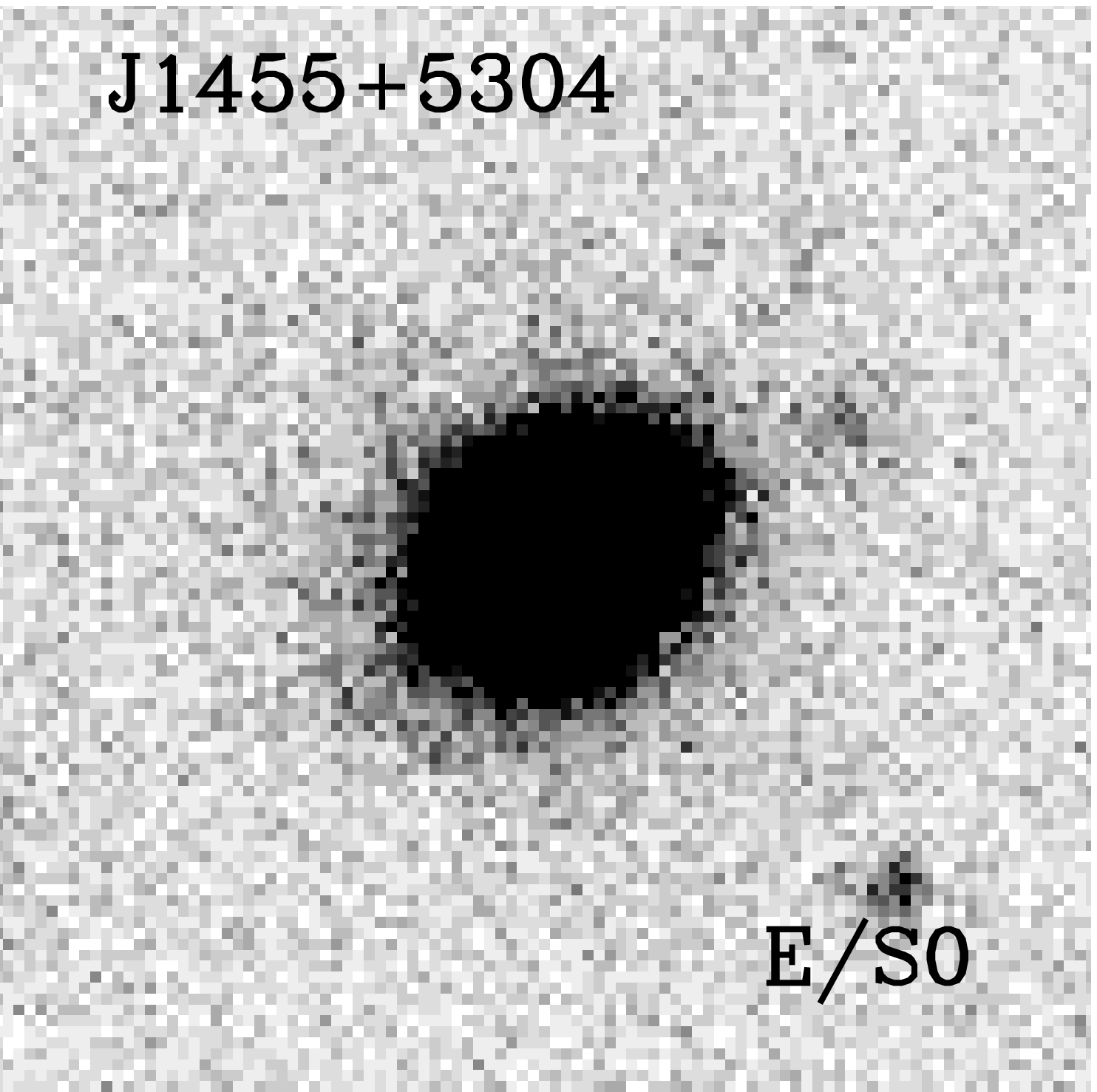}
\caption{Imaging  in the $r'$ band of the targets whose morphology can not be deduced by visual inspection and require galaxy modeling to know if they are disk galaxies. The indications of the galaxy morphology refer to the conclusions of \S~\ref{sec_morphology}. The size of the images is $21.7\arcsec \times 21.7\arcsec$.
\label{fig_morph}}
\end{figure}

%ooooooooooooooooooooooooooooooooooooooooooooooooooooooooooooooooooo

%ooooooooooooooooooooooooooooooooooooooooooooooooooooooooooooooooooooo
\begin{deluxetable}{lcccc}
\tabletypesize{\scriptsize}
\tablecaption{Morphology\label{tab_morphology}
}
\tablewidth{0pt}
\tablehead{
\colhead{ } & \colhead{$u'$} & \colhead{$g'$} & \colhead{$r'$} &
\colhead{$i'$}
}
\startdata
       &   &   J0812+5436  &   &   \\
$\chi^2_{\nu}$ for $n=1$ & 1.548 &  1.378 & 1.689 & 2.424 \\
$\chi^2_{\nu}$ for $n=4$ & 2.227 &  1.417 & 1.673 & 1.482 \\
best fit $n$ & 0.74 & 1.82 & 1.92 & 2.35  \\
\tableline \\
       &   &   J1150+1202  &   &   \\
$\chi^2_{\nu}$ for $n=1$ & 1.129 &  1.129 & 1.499 & 1.279 \\
$\chi^2_{\nu}$ for $n=4$ & 1.127 &  1.110 & 1.367 & 1.172 \\
best fit $n$ & 1.98 & 2.32 & 2.42 & 3.06  \\
\tableline \\
       &   &   J1200+4014  &   &   \\
$\chi^2_{\nu}$ for $n=1$ & 1.161 &  1.453 & 1.664 & 1.697 \\
$\chi^2_{\nu}$ for $n=4$ & 1.133 &  1.204 & 1.265 & 1.264 \\
best fit $n$  & 3.38 & 2.83 & 2.68 & 2.82  \\
\tableline \\
       &   &   J1455+5304  &   &   \\
$\chi^2_{\nu}$ for $n=1$ & 1.181 &  1.216 & 1.419 & 1.517 \\
$\chi^2_{\nu}$ for $n=4$ & 1.160 &  1.162 & 1.235 & 1.357 \\
best fit $n$  & 2.62 & 2.35 & 2.80 & 2.52  \\
\enddata
\end{deluxetable}
%oooooooooooooooooooooooooooooooooooooooooooooooooooooooooooooooooooooo

\subsection{Morphology}\label{sec_morphology}

The galaxies J0903+5448, J0942+6111 and J1356+5615 can be identified as disk galaxies directly from looking at their optical images\footnote{Fitting these galaxies would not bring more information about their morphology, as GALFIT does not include the inclination of the disk as a fitting parameter. Therefore, the S\'{e}rsic index $n$ would be increased artificially, as  the light received is more  concentrated due to the inclination of the disk.}. 
As for the galaxy J1625+2818, we can see it is a barred spiral with a dominant bulge (see Fig.\ \ref{fig_morph_disk}). However, the candidate massive disk galaxies J0812+5436, J1150+1202, J1200+4014 and J1455+5304 do not have characteristics that allow us to identify their galaxy type by simple visual inspection (see Fig.\ \ref{fig_morph}).

To gain further insight into the morphology of these galaxies, we used GALFIT to model the main component of each galaxy with a  S\'{e}rsic profile. We fitted the galaxies in each of the available photometric bands: $u'$, $g'$, $r'$ and $i'$. For each fit we used a simple method which allows us to get robust information about the morphology of the main component of the galaxies in each band: we fitted the data with  a S\'{e}rsic profile keeping $n$ fixed to $4$ (de Vaucouleurs bulge), then fitted the data with  a S\'{e}rsic profile keeping $n$ fixed to $1$ (exponential disk). The comparison of the $\chi^2$ of the two fits tells us if the morphology of the galaxy is nearer to a bulge or a disk. We also fitted the data with a S\'{e}rsic profile where $n$ is let free. 

While this method allows us to determine if the galaxy morphology is definitely a disk or a bulge, it is difficult to identify S0 galaxies, which have a S\'{e}rsic index between the two models. For most of the cases, we can only say that the galaxy is not dominated by a disk, but we can not determine clearly between S0 and elliptical galaxies. It is to note that even with {\it HST} imaging, it can be difficult to differentiate S0 galaxies from elliptical galaxies, if the disk is not a dominant feature.

Here follows a summary galaxy by galaxy of the results, which are presented in Table \ref{tab_morphology}.

\textit{J0812+5436}\\
In the $u'$ and $g'$ bands the $\chi^2$ of the fit is smaller for a disk ($n=1$) while in the $r'$ and $i'$ band the  $\chi^2$ of the fit is smaller for a bulge  ($n=4$). This indicates that the galaxy is composed of a disk and a bulge. However, due to the very bright lensed images superposed to the galaxy, the fit may not be completely reliable, although the lensed images were masked. Therefore, 
we can only conclude that J0812+5436 is a probable disk or S0 galaxy.

\textit{J1150+1202}\\
In the $u'$ band the $\chi^2$ of the fit is similar for a disk ($n=1$) and a  bulge ($n=4$), while in the $g'$, $r'$ and $i'$ bands the  $\chi^2$ of the fit is smaller for a bulge ($n=4$). In addition, the best fit values in all the bands are for $1<n<4$, which is between a disk and a bulge model.  We conclude that SDSS J1150+1202 is probably an S0 galaxy. 

\textit{J1200+4014}\\
In all bands the  $\chi^2$ of the fit is smaller for a bulge ($n=4$). We conclude that SDSS J1200+4014 is probably an S0 or an elliptical galaxy. 

\textit{J1455+5304}\\
In all bands the  $\chi^2$ of the fit is smaller for a bulge ($n=4$). We conclude that SDSS J1455+5304 is probably an S0 or an elliptical galaxy.

%oooooooooooooooooooooooooooooooooooooooooooooooooooooooooooooooo
\begin{deluxetable}{lccccc}
\tabletypesize{\scriptsize}
\tablecaption{Mass-to-light ratio within $r_E$\label{tab_mtl}
}
\tablewidth{0pt}
\tablehead{
\colhead{Name } & \colhead{$M_{\rm{tot}}/L_I$} & \colhead{$M_{\rm{tot}}/L_R$} & \colhead{$M_{\rm{tot}}/L_B$} & \colhead{$M_{\rm{tot}}/L_U$} & \colhead{$r_E$ (kpc)}
}
\startdata
J0812+5436 & $5.4~\pm1.5$ & $8.2~\pm2.0$ & $11.2~\pm2.8$ & $14.6~\pm2.3$ & $3.9~\pm0.9$ \\
J1455+5304 & $1.5~\pm0.9$ & $2.2~\pm1.3$ & $3.0~\pm1.7$ & $3.1~\pm1.9$ & $1.4~\pm0.8$ \\ 
\enddata
\end{deluxetable}
%oooooooooooooooooooooooooooooooooooooooooooooooooooooooooooooooooooooo

%oooooooooooooooooooooooooooooooooooooooooooooooooooooooooooooooo
\begin{deluxetable}{lcccc}
\tabletypesize{\scriptsize}
\tablecaption{Local galaxies around J0812+5436 (z=0.121\tablenotemark{a})\label{tab_group}
}
\tablewidth{0pt}
\tablehead{
\colhead{Name } & \colhead{Photo z (SDSS)} & \colhead{Projected distance from J0812+5436}
}
\startdata
Galaxy A & $0.166\pm0.042$ & $\sim 184$ kpc \\
Galaxy B & $0.135\pm0.030$ & $\sim129$ kpc\\ 
Galaxy C & $0.133\pm0.010$ & $\sim158$ kpc\\
Galaxy D & $0.096\pm0.008$ & $\sim219$ kpc\\
Galaxy E & $0.131\pm0.006$ & $\sim297$ kpc
\enddata
\tablenotetext{a}{Spectroscopic redshift from SDSS.}
\end{deluxetable}
%oooooooooooooooooooooooooooooooooooooooooooooooooooooooooooooooooooooo

%ooooooooooooooooooooooooooooooooooooooooooooooooooooooooooooooooooooooo
\begin{deluxetable}{lcc}
\tabletypesize{\scriptsize}
\tablecaption{Results\label{tab_results}
}
\tablewidth{0pt}
\tablehead{
\colhead{Name } & \colhead{Type} & \colhead{Morphology}
}
\startdata
J0812+5436 & strong lens & probably disk \\
J0903+5448 & unknown & disk  \\
J0942+6111 & unknown & disk \\
J1150+1202 & probably strong lens & probably S0 \\
J1200+4014 & probably strong lens & S0 or elliptical \\
J1356+5615 & unknown & disk \\
J1455+5304 & strong lens & S0 or elliptical \\
J1625+2818 & unknown & spiral \\
\enddata
\end{deluxetable}
%oooooooooooooooooooooooooooooooooooooooooooooooooooooooooooooooooooooo

\section{Mass-to-light ratios}\label{sec_ml}

We compute estimates of the total $M/L$ ratios for the gravitational lenses we have discovered.  For any circular lens, the lens potential is a function only of the distance from the lens center. The mass enclosed within the Einstein radius $r_E$ is obtained from the relation $r_E^2=\frac{4GM(<r_E)}{c^2} \frac{D_{ds}}{D_dD_s}$.  The luminosity inside the Einstein radius  is computed using a S\'{e}rsic profile \citep{graham05}. Our images are calibrated using the SDSS database magnitudes, and K-correction is applied following \citet{oke68}. The results are presented in Table \ref{tab_mtl}. However, these values are difficult to compare to usual $M/L$ ratios, as they probe the total mass and not just the stellar or baryonic matter.

In the case of the S0 or elliptical galaxy lens  J1455+5304, the Einstein radius is small  ($r_E=0.7~\pm0.4\arcsec$, that is, $1.4~\pm0.8$ kpc), so we can expect that the mass probed within is largely dominated by baryons. On this basis, the  $M_{\rm{tot}}/L_I=1.5~\pm0.9$ for J1455+5304 is consistent with the value from \citet{trott02} of $M/L_I=1.1~\pm0.2$ for the disk of the Sab spiral lens the Einstein Cross.

\begin{figure}[h]
\centering
\includegraphics[scale=0.7]{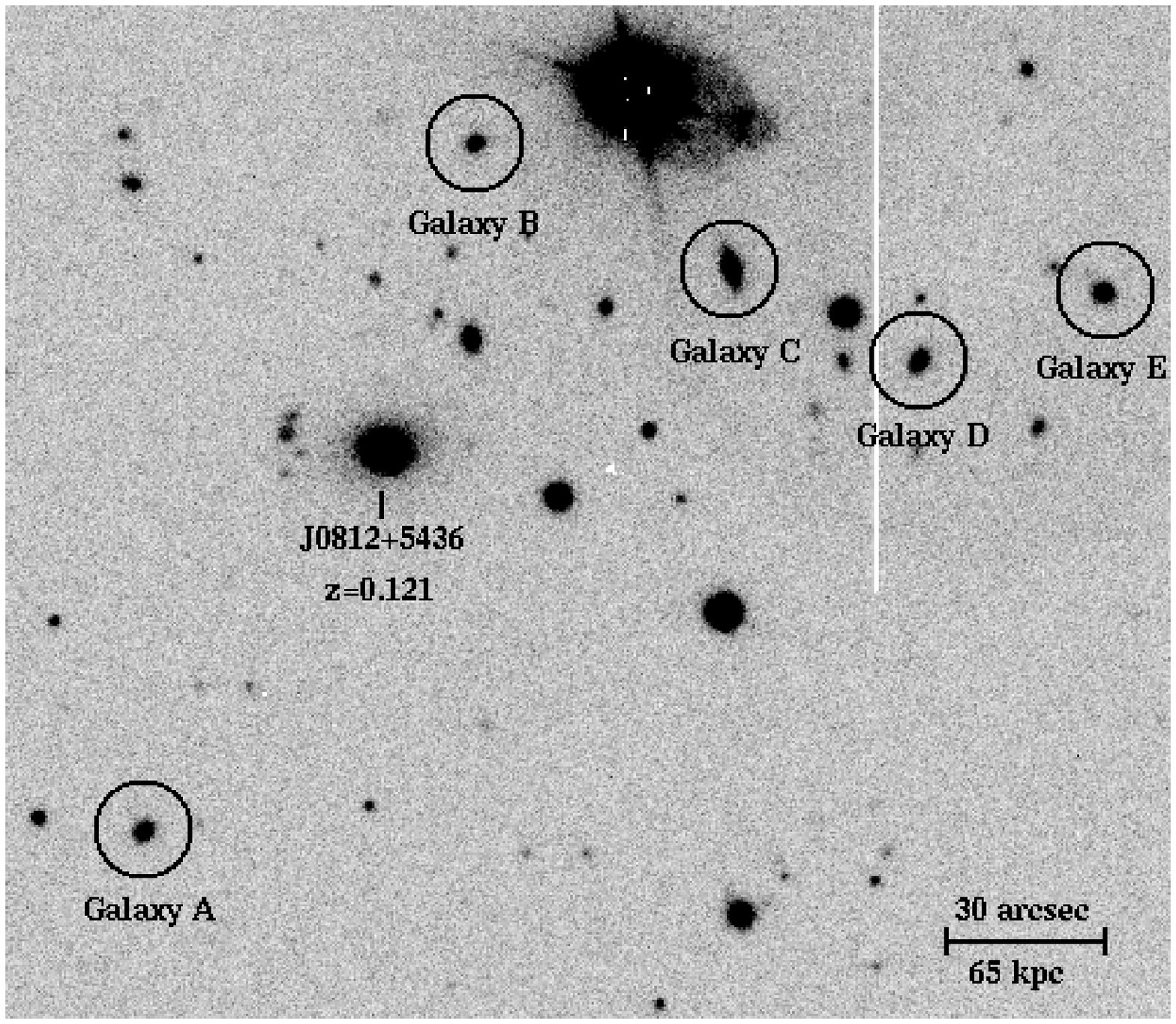}
\caption{Imaging in the $r'$ band of the field around J0812+5436. The five circled galaxies have SDSS photometric redshifts close to the redshift of J0812+5436 (see Table \ref{tab_group}). The presence of a galaxy group may account for the large Einstein radius of J0812+5436 ($r_E=1.8\pm0.4\arcsec$). At the redshift $z=0.121$, $30\arcsec$ correspond to 65 kpc.
\label{fig_group}}
\end{figure}

In the case of the (probable disk-galaxy)  lens J0812+5436, we obtain a particularly high $M/L$ ratio, with $M_{\rm{tot}}/L_I=5.4~\pm1.5$. The Einstein radius is larger than expected for galaxy-galaxy lensing systems, with $r_E=1.8~\pm0.4\arcsec$  ($3.9~\pm0.9$ kpc). The total mass within might contain a large amount of DM, although it is still in the few central kiloparsecs of the galaxy. We note that J0812+5436 is surrounded by  five galaxies at a similar redshift,  with a group radius of $\sim200$ kpc  (see Fig.\  \ref{fig_group} and Table \ref{tab_group}). This group of galaxies may add to the gravitational potential deflecting the background source. 

%ooooooooooooooooooooooooooooooooooooooooooooooooooooooooooooooooooooo
%oooooooooooooooooooooooooooooooooooooooooooooooooooooooooooooooooooooo

The limited observational precision that we obtain on measuring the Einstein radius and the surface brightness of the different components of the galaxies limits us in the study of the $M/L$ ratios of these new galaxy-galaxy lenses. Therefore, we can not gain insight on the $M/L$ ratios of the different components, nor constrain the shape of the IMF or the maximum disk hypothesis. \textit{HST}  or Adaptive Optics imaging is required to further study the $M/L$ ratio of these lens galaxies.

\section{Discussion}\label{sec_discussion}

The results of the follow-up of the  eight disk-galaxy  gravitational lens candidates are summarized in Table \ref{tab_results}.  From the optical imaging and spectroscopy presented here, we can determine clearly   that two systems are strong gravitational lenses, J0812+5436 being a probable disk or S0 galaxy, and J1455+5304 being a S0 or elliptical galaxy. 

Concerning the other candidates, we classify as probable gravitational lenses those systems in which we find indication of multiple imaging from both imaging and spectroscopy, although observations do not allow us a complete confirmation of the lens nature of the system. We find two probable gravitational lenses, J1150+1202 being probably a S0 galaxy, and J1200+4014 being a S0 or elliptical galaxy. 

Finally, we classify as unknown the systems for which we found indication of multiple imaging in optical observations, but could not confirm the presence of multiple images in the spectroscopy. These systems are good lens candidates but the quality of our observations did not allow us to conclude on their nature. In this category, we find J0903+5448, J0942+6111, and J1356+5615 which are disk dominated galaxies, as well as the spiral galaxy J1625+2818. 

We see that for most of the  gravitational lens candidates, our combined photometric and spectroscopic observations are not sufficient to determine clearly the presence of multiple images of the background source. This is due mainly to the fact that the images of the background source are very near  to the center of the foreground galaxy, the Einstein radius being expected to be of the order of $\sim 1\arcsec$. Therefore, ground-based observations suffer from seeing conditions for such a study. Moreover, some of the images might be obscured by dust in disk galaxies. 

Confirmation of the remaining lens candidates  will require either high resolution spatially-resolved spectroscopy, or high resolution imaging. Integral field spectroscopy would allow us to probe both the presence of multiple images of a background source and the configuration of the lensed images. However, because the galaxies in our study are candidate massive disk galaxies, they are not expected to present much structures in the blue part of the spectrum, which would be mistaken for lensed images. In the case of the spiral lens J1625+2818, the position of the potential lensed images around the bulge also prevented any contamination by the arms for detecting them. Therefore  \textit{HST} imaging, or Adaptive Optics \citep{marshall07,mckean07,sluse08} imaging, might be the best choice for confirming the genuine lens nature of our disk-galaxy lens candidates, confirming the disk nature of our new gravitational lenses, and studying the mass distribution of these systems.

In the following we  present suggestions for  improving the efficiency of a search and  for enlarging the  size of the survey.

1- Due to the very large number of galaxies in the SDSS Main Galaxy Sample, we covered only a part of the color parameter space corresponding to disk galaxies. We chose to limit ourselves to the region containing early-type disk galaxies, which are massive disk galaxies and therefore  better suit our purpose. A larger survey should span all  of the disk galaxy color range (with applying a suitable $r$ magnitude criterion to select massive galaxies), although this would  largely increase the size of the initial galaxy sample.\\
2- The automated method we used to select spectra  showing evidence for a background galaxy along the line of sight is based on a $S/N$ ratio per pixel emission-line detection. This was done in order to select bright peaks in the spectra.  A complete search should be based on an integrated $S/N$ ratio selection to detect and identify all background emission lines.\\
3- We did not cover the redshift range $z<0.1$, in order to preserve the efficiency of the search.  Indeed,  while half  of the  galaxies in the SDSS Main Sample lie in this region, very few lenses are expected at  such a low redshift.  However, it would be of interest for a larger survey to explore this redshift range. \\
4- Selecting lens candidates based on the presence of  three emission lines limits, in the case of the SDSS, the redshift of the background galaxy to $z<0.8$. Selection based on  one detected emission line and  its identification  as the [\ion{O}{2}] $\lambda\lambda$3727 doublet, as in the OLS-survey \citep{willis05,willis06}, would allow  us to enlarge the sample of lens  candidates. Indeed, almost all our gravitational lens candidates have a resolved [\ion{O}{2}] $\lambda\lambda$3727 doublet. However, it would also increase the false detection rate, hence reduce the efficiency of the survey.\\
5- Among the SLACS lenses, some late-type galaxy lenses were found \citep{bolton08}, although from targeting luminous red galaxies with absorption-line dominated spectra, expected to be mainly early-type galaxies. This can happen if the bulge in the  late-type galaxy is prominent, and therefore this should be taken into account when searching for the disk galaxy population in other large surveys.\\

\section{Conclusion}\label{sec_conclusion}

We have presented the first  automated search for disk-galaxy  gravitational lenses  using the SDSS database, and proved the feasibility of  such a project with the discovery of a galaxy-galaxy lens which is very probably a disk, of four interesting disk-galaxy lens candidates, and three confirmed or probable lenses which may be S0 galaxies.
This project  is the first step  in finding more disk-galaxy lenses in the SDSS, and possibly in other large surveys. 
To date, only  ten confirmed disk-galaxy lenses are known. Assembling a  larger sample of disk-galaxy lenses would open  up promising perspectives for measuring  the $M/L$ ratios of disks and bulges, and  for studying the structure of disk galaxies.
{\it HST}  or Adaptive Optics imaging will be needed to further study the systems presented in this work, particularly to probe the presence of a disk in the lens galaxy of J0812+5436 and study its mass distribution, as well as to confirm the lens nature of the interesting disk-galaxy lens candidates we discovered.

\acknowledgments

We thank the anonymous referee for many useful and constructive comments on the manuscript. It is a pleasure to thank M.\ Limousin and J.\ P.\ U.\ Fynbo for valuable discussions, and L.\ F.\ Grove for constant support during this work.  The Dark Cosmology Centre is funded by the Danish National Research Foundation. This work was supported in part by the European Community's Sixth Framework Marie Curie Research Training Network Programme, Contract No.\ MRTN-CT-2004-505183 ``ANGLES".

Based on observations made with the Nordic Optical Telescope, operated on the island of La Palma jointly by Denmark, Finland, Iceland, Norway, and Sweden, in the Spanish Observatorio del Roque de los Muchachos of the Instituto de Astrofisica de Canarias.  

 This work is using the Sloan Diagital Sky Survey database. Funding for the Sloan Digital Sky Survey (SDSS) and SDSS-II has been provided by the Alfred P. Sloan Foundation, the Participating Institutions, the National Science Foundation, the U.S. Department of Energy, the National Aeronautics and Space Administration, the Japanese Monbukagakusho, and the Max Planck Society, and the Higher Education Funding Council for England. The SDSS Web site is http://www.sdss.org/.

  The SDSS is managed by the Astrophysical Research Consortium (ARC) for the Participating Institutions. The Participating Institutions are the American Museum of Natural History, Astrophysical Institute Potsdam, University of Basel, University of Cambridge, Case Western Reserve University, The University of Chicago, Drexel University, Fermilab, the Institute for Advanced Study, the Japan Participation Group, The Johns Hopkins University, the Joint Institute for Nuclear Astrophysics, the Kavli Institute for Particle Astrophysics and Cosmology, the Korean Scientist Group, the Chinese Academy of Sciences (LAMOST), Los Alamos National Laboratory, the Max-Planck-Institute for Astronomy (MPIA), the Max-Planck-Institute for Astrophysics (MPA), New Mexico State University, Ohio State University, University of Pittsburgh, University of Portsmouth, Princeton University, the United States Naval Observatory, and the University of Washington.

{\it Facilities:} \facility{NOT (MOSCA, ALFOSC)}

\appendix

\section{Additional disk-galaxy lens candidates}

\begin{figure*}
\centering
\includegraphics[scale=0.35]{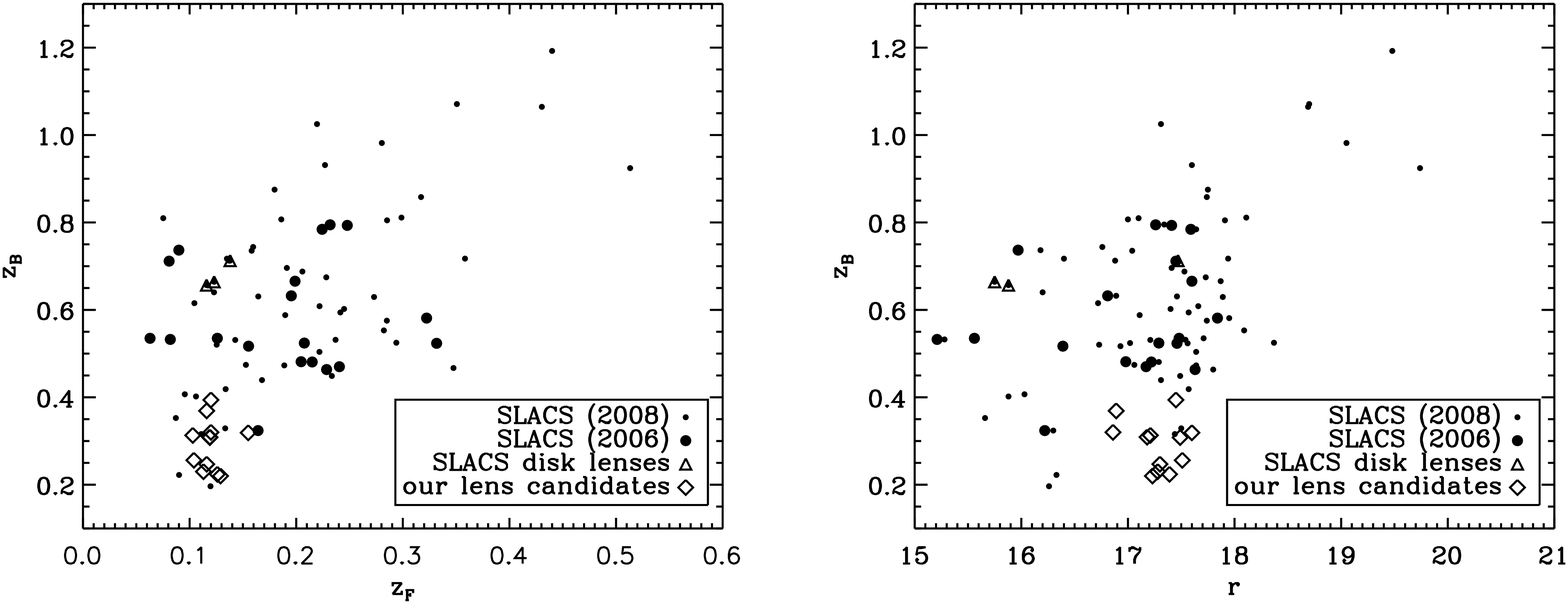}
\caption{Distribution of the foreground galaxy redshift $z_F$, the background galaxy redshift $z_B$ and the $r$ magnitude of the twelve lens candidates we did not selected for follow-up, compared to the SLACS confirmed strong lenses. The sample of SLACS strong lenses from  \citet{bolton06} was the one we actually used to select our final lens candidates. We show also the final sample of SLACS strong lenses published in \citet{bolton08}, as well as the three disk-galaxy lenses found in it. 
\label{fig_redshifts2}}
\end{figure*}

During the spectroscopic selection of the disk-galaxy lens candidates (\S~\ref{sec_selection}), we found twenty systems with evidence of a background galaxy along the line-of-sight.  Among these twenty disk-galaxy lens candidates, we found eight systems to have a similar distribution of redshifts and $r$ magnitude as the SLACS confirmed lenses (see Fig.\ \ref{fig_redshifts}), and selected these eight systems for follow-up. We present in Appendix A the remaining twelve candidates: their redshifts and $r$ magnitude distribution compared to that of the SLACS lenses is shown in Fig.\   \ref{fig_redshifts2}, and details on the candidates are presented in Table \ref{tab_lenscandidates2}. These candidates were discarded because they had both a short redshift interval between the background and the foreground galaxies, and a relatively faint $r$ magnitude for the foreground galaxy. In comparison, the SLACS lenses with such a short redshift interval show brighter $r$ magnitudes (with the $r$ magnitude being a rough tracer of the mass of the galaxy), which increases their lensing cross-section.

%oooooooooooooooooooooooooooooooooooooooooooooooooooooooooooo
%ooooooooooooooooooooooooooooooooooooooooooooooooooooooooooooooo

%ooooooooooooooooooooooooooooooooooooooooooooooooooooooooooooo
\begin{deluxetable}{lcccc}
\tabletypesize{\scriptsize}
\tablecaption{Additional lens candidates \label{tab_lenscandidates2}}
\tablewidth{0pt}
\tablehead{
\colhead{Name}  & \colhead{Plate-MJD-Fiber} & \colhead{$z_F$} & \colhead{$z_B$} &
\colhead{$r$\tablenotemark{a}}
}
\startdata
SDSS J005621.66-091201.9 & spSpec-52146-0658-393  & 0.103  & 0.313  & 17.21  \\
SDSS J090146.27+554102.0 & spSpec-51908-0450-388  & 0.116  & 0.247  & 17.30 \\
SDSS J103143.93+421859.3 & spSpec-53033-1360-415  & 0.119  & 0.308  & 17.49 \\
SDSS J110039.10+120159.1 & spSpec-53119-1603-440  & 0.129  & 0.220  & 17.23  \\
SDSS J112500.00+053604.7 & spSpec-52376-0836-536  & 0.155  & 0.319  & 17.60 \\
SDSS J114440.12+043650.5 & spSpec-52373-0839-230  & 0.104  & 0.256  & 17.51 \\
SDSS J134308.25+602754.8 & spSpec-52319-0786-236  & 0.120  & 0.320  & 16.86  \\
SDSS J145555.90+040745.3 & spSpec-52045-0588-553  & 0.126  & 0.224  & 17.39  \\
SDSS J150339.92+370728.1 & spSpec-52819-1352-472  & 0.116  & 0.369  & 16.89  \\
SDSS J171605.16+275206.3 & spSpec-52410-0977-093  & 0.119  & 0.309  & 17.18  \\
SDSS J172614.51+604142.5 & spSpec-51792-0354-151  & 0.113  & 0.230  & 17.28  \\
SDSS J235726.45-090917.7 & spSpec-52143-0650-370  & 0.120  & 0.394  & 17.45  \\
\enddata
\tablenotetext{a} { De Vaucouleurs model SDSS (AB) magnitude.}
\end{deluxetable}
%oooooooooooooooooooooooooooooooooooooooooooooooooooooooooooooooooooooo

\end{document}